\documentclass[aip,reprint,floatfix]{revtex4-2}

\usepackage{amsmath}

\DeclareMathAlphabet{\mathsf}{OT1}{phv}{b}{n}

\newcommand{\Vector}[1]{\ensuremath{\mathbf{#1}}}

\newcommand{\crossVorg}{\ensuremath{%
         \setbox0=\hbox{$V$}
        V \kern-\wd0{\raise.3ex\hbox{$\relbar$}}}}

\newcommand{\crossVxx}[2]{%
	{\setbox0=\hbox{$#1#2V$}
         \setbox1=\hbox{$#1#2$}
         \setbox2=\hbox{$#1V$}
         \dimen1=\wd0
	 \advance\dimen1-\wd1
         \raise.2\ht0\hbox{$#1#2$}\kern-.4\wd0}}

\usepackage{pifont}

\usepackage{multirow}
\usepackage{setspace}
\usepackage{enumitem}
\usepackage{verbatim}

\usepackage{amssymb}
\usepackage{amsbsy}
\usepackage[T1]{fontenc}
\usepackage{amsmath}
\usepackage{mathtools}

\usepackage{graphicx}
\usepackage{xcolor}
\usepackage{stmaryrd}
\usepackage{mathrsfs}
\usepackage[mathscr]{euscript}
\usepackage{tikz}
\usetikzlibrary{shapes}
\usepackage{cancel}
\usepackage[normalem]{ulem}

\usepackage[colorinlistoftodos]{todonotes}
\usepackage{verbatim}
\usepackage{latexsym}
\usepackage{textcomp}
\usepackage{bm}
\usepackage[Euler]{upgreek}
\usepackage{longtable}
\usepackage{subfigure}

\usepackage[colorlinks,allcolors=cyan!70!black]{hyperref}

\usepackage{epsfig}
\usepackage{dcolumn}

\usepackage{booktabs}

\newcommand{\qed}{\nobreak \ifvmode \relax \else
      \ifdim\lastskip<1.5em \hskip-\lastskip
      \hskip1.5em plus0em minus0.5em \fi \nobreak
      \vrule height0.75em width0.5em depth0.25em\fi}

\DeclareMathAlphabet\mathbfcal{OMS}{cmsy}{b}{n}
\DeclareMathAlphabet{\mathbfsf}{\encodingdefault}{\sfdefault}{bx}{sl}

\begin{document}


\title{Universal scaling and characterization of gelation in associative polymer solutions}


\author{Aritra Santra}
\affiliation{Department of Chemical Engineering, Monash University,
Melbourne, VIC 3800, Australia}
\author{B. D\"{u}nweg}
\affiliation{Max Plank Institute of Polymer Research, Ackermannweg 10, 55128 Mainz, Germany}
\affiliation{Department of Chemical Engineering, Monash University,
Melbourne, VIC 3800, Australia}
\author{J. Ravi Prakash}
\email{ravi.jagadeeshan@monash.edu}
\affiliation{Department of Chemical Engineering, Monash University,
Melbourne, VIC 3800, Australia}
 \homepage{https://users.monash.edu.au/~rprakash/}
 

\date{\today}



\begin{abstract}
A multi-particle Brownian dynamics simulation algorithm with a Soddemann-Duenweg-Kremer potential that accounts for pairwise excluded volume interactions between both backbone monomers and associating groups (stickers) on a chain, is used to describe the static behaviour of associative polymer solutions, across a range of concentrations into the semidilute unentangled regime. Predictions for the fractions of stickers bound by intra-chain and inter-chain association, as a function of system parameters such as the number of stickers on a chain, the number of backbone monomers between stickers, the solvent quality, and monomer concentration are obtained. A systematic comparison between simulation results and scaling relations predicted by the mean-field theory of Dobrynin (Macromolecules, 37, 3881, 2004) is carried out. Different regimes of scaling behaviour are identified by the theory depending on the monomer concentration, the density of stickers on a chain, and whether the solvent quality for the backbone monomers corresponds to $\theta$ or good solvent conditions. Simulation results validate the predictions of the mean-field theory across a wide range of parameter values in all the scaling regimes. The value of the des Cloizeaux exponent, $\theta_2 = 1/3$, proposed by Dobrynin for sticky polymer solutions, is shown to lead to a collapse of simulation data for all the scaling relations considered here. Three different signatures for the characterization of gelation are identified, with each leading to a different value of the concentration at the  sol-gel transition. {The Flory-Stockmayer expression relating the degree of inter-chain conversion at the sol-gel transition to the number of stickers on a chain,  modified by Dobrynin to account for the presence of intra-chain associations, is found to be validated by simulations for all three gelation signatures}. Simulation results confirm the prediction of scaling theory for the gelation line that separates sol and gel phases, when the modified Flory-Stockmayer expression is used. Phase separation is found to occur with increasing concentration for systems in which the backbone monomers are under $\theta$-solvent conditions, and is shown to coincide with a  breakdown in the predictions of scaling theory. 
\end{abstract}

\maketitle


\section{\label{sec:intro} Introduction}

Associative polymers, which are macromolecules with attractive groups~\cite{Winnik:1997hf,RubDob,RubCol}, are used in a wide variety of applications because the interactions between the attractive groups can be tuned by varying their number, strength and location on the polymer, thereby providing a means by which the physical properties of these solutions can be exquisitely controlled. For instance, they are widely used as rheology modifiers in the coating, paint, water-treatment and enhanced oil-recovery industries, since their influence on solution viscosity can be adjusted molecularly by varying the chemistry and geometry of the associations, and macroscopically by changing the temperature or concentration~\cite{RheoMod,Tripathi:2006kh}. At sufficiently high concentrations, when the suspending medium is water, associative polymers form hydrogels whose transient viscoelastic networks have found numerous applications as tissue engineering scaffolds~\cite{Yacoub}, food thickeners~\cite{FoodSc}, drug delivery carriers, soft electronics, and sensors~\cite{Rossow:2015gt,Voorhaar:2016jt,Tsitsilianis:2010kf}. Many of these uses involve the application of flow fields that influence and control the formation and duration of associations, and the evolution of the transient network structures. A fundamental understanding of the nonequilibrium dynamics of physically associative polymers is consequently essential for the rational design of these systems. Successful formulation of associative polymer systems for these various applications has largely rested on using polymer chemistry to engineer innovative polymers, followed by extensive experimental investigation to select the most suitable candidates. It is not possible currently to specify a priori the particular macromolecular architecture, the precise number, strength and location of the attractive groups, the appropriate solution temperature and concentration, and the particular flow conditions which would achieve optimal product performance. Several computational studies have been carried out aimed at making progress in this direction, i.e., towards improving our understanding of the nonequilibrium response of network structures, and deciphering the connection between molecular topology and macroscopic behaviour,  using a variety of different techniques based on coarse-grained bead-spring chain models for polymers~\cite{vandenBrule1995,HernandezCifre2003,HernandezCifre2007,Sprakel2009,GompFlo,Castillo,Omar2017,Park2017,Furuya2018}. In this work, we propose a novel alternative approach based on a muliti-particle Brownian dynamics simulation methodology that accounts for hydrodynamic interactions, and which can potentially capture both static and dynamic properties at equilibrium, along with the nonequilibrium rheological response of associative polymer solutions, across a range of concentrations that span the dilute and unentangled semidilute regimes. In order to validate the proposed simulation methodology, we report here the results of a detailed comparison of its predictions of equilibrium static properties with the analytical predictions of the seminal scaling theories of Semenov and Rubinstein~\cite{RnSstatics} and Dobrynin~\cite{Dob}. As will be seen, these results set the stage for a subsequent study of the equilibrium and nonequilibrium dynamics of associative polymer solutions within a systematic and coherent framework. 

An essential feature of physically associative polymer solutions is the prevalence of \emph{intra-chain}  and \emph{inter-chain} associations between the attractive groups on the chains, which lead to the formation of micelles and network structures. A central prediction of scaling theories~\cite{RnSstatics,Dob} is the dependence of the fraction of attractive groups that are stuck through {intra-chain} ($p_1$) and {inter-chain} ($p_2$) associations, on the various parameters that define the system, such as the number of attractive groups (stickers) on a chain ($f$), the number of monomers between two stickers (or the spacer length $\ell$), the strength of association between the stickers ($\epsilon_{st}$), the monomer concentration ($c$), and the solution temperature ($T$). Apart from a preliminary Monte Carlo study~\cite{SanatK}, these predictions have, to our knowledge,  so far not been thoroughly tested through simulations. The formulation adopted in the present work enables a careful examination of the predictions of these theories for the scaling dependence of $p_1$ and $p_2$ on system parameters.

With increasing monomer concentration, associative polymer solutions undergo a transition from the sol to the gel phase with the appearance of an incipient system spanning network. Dobrynin has pointed out that within the framework of a mean-field lattice based theory for associative polymer solutions, it is not possible to identify the location of the sol-gel transition, since chains are not distinguished as belonging to the sol or the gel phase~\cite{Dob}. As a result, in order to describe the phase behaviour of associative polymer solutions within mean-field theory, \citet{Dob}\! assumes a modified form of the Flory-Stockmayer expression~\cite{FloryBook,Stock2} 
\begin{equation}
p_2 = \frac{1}{(1-p_1)f-1}
\label{FS-D}
\end{equation}
which relates the degree of inter-chain conversion $p_2$ at the gelation threshold,  to the number of stickers on a chain available for inter-chain association, $(1-p_1)f$. In the original Flory-Stockmayer theory it is assumed that the fraction of intra-chain associations $p_1$ is zero (which is expected to hold at high concentrations), leading to following well known simple relation at the location of the sol-gel transition
\begin{equation}
p_2 = \frac{1}{f-1}
\label{FS}
\end{equation}
It should be noted that the Flory-Stockmayer theory assumes that the gel network is a treelike structure and prohibits the formation of loops~\cite{RubCol}. Semenov and Rubinstein~\cite{RnSstatics}, on the other hand, show that Eq.~(\ref{FS}) can be formally derived if one assumes that the sol-gel transition coincides with the monomer concentration at which the concentration of free chains in the system (i.e. those with no inter-chain associations) undergoes a maximum. In real polymer networks, one expects that the formation of loops is a common occurrence. In the formalism adopted in the present work, the formation of cyclic structures is not prohibited, and as a result, we are able to examine both the validity of Eq.~(\ref{FS-D}), and the assumption of Semenov and Rubinstein~\cite{RnSstatics} regarding the coincidence of the sol-gel transition with the free chain concentration maximum. 

Interestingly, from an equilibrium statics point of view (as opposed to a rheological characterization~\cite{Winter:2000gw,Li:1997cu, Li:1997gi}), there does not appear to be a commonly agreed definition of the concentration at which the sol-gel transition occurs. Descriptions of gelation based on percolation models define the sol-gel transition as the concentration at which the first system spanning network appears~\cite{RubCol}. Alternatively, the sol-gel transition is also identified as the concentration at which the probability distribution of chain cluster sizes becomes bimodal~\cite{SanatK}. In this interpretation, it is expected that in the sol phase the probability of finding a cluster with $m$ chains decreases monotonically with increasing $m$, while the appearance of a second peak in the probability distribution, at a non-zero value of $m$, signals the onset of gelation. It is not clear if the three definitions of the sol-gel transition, namely, the appearance of the system spanning network, the appearance of bi-modality in the chain cluster size probability distribution, or the occurrence of a maximum in the free-chain concentration, are all located at the same monomer concentration, and if the degree of inter-chain conversion $p_2$ is related to the number of stickers on a chain available for inter-chain association by Eq.~(\ref{FS-D}), in all the three definitions. These questions are examined in the present work, and we show that while the three different definitions are located at different monomer concentrations, the dependence of $p_2$ on  $p_1$ and $f$ is given by Eq.~(\ref{FS-D}) in all three cases, for sufficiently long chains.

The outline of this paper is as follows. In section~\ref{sec:simulation_details}, the proposed multi-particle Brownian dynamics algorithm that accounts for hydrodynamic interactions, and which is capable of simulating associative polymer solutions across a range of concentrations, is described. Also discussed in this section is the adoption of the Soddemann-Duenweg-Kremer (SDK) potential~\cite{SDK,Aritra2019} to model the pair-wise interactions between both the backbone and sticker monomers, which is a key aspect of the suggested methodology. In section~\ref{sec:scaling}, a brief summary of the predictions of scaling theories~\cite{RnSstatics,Dob} for the degrees of intra-chain and inter-chain conversions as a function of system parameters, is given. These predictions provide a basis for identifying the quantities that need to be evaluated by simulations, and a framework for the interpretation of simulation results. Essentially, the theories identify three different regimes of scaling behaviour depending on the solvent quality of the backbone monomers, the monomer concentration and the density of stickers along the backbone. Section~\ref{sec:simdet} discusses the choice of various simulation parameter values that enables the exploration of these different scaling regimes. Simulation predictions for the dependence of the degrees of conversion on spacer length and concentration, at constant temperature and sticker strength, are discussed in subsection~\ref{sec:ell_and_c}, while subsection~\ref{sec:gss} first examines the influence of temperature and sticker strength, before combining the dependencies on all system parameters together in master plots. A comparison of the scaling of radius of gyration with concentration between homopolymer and sticky polymer solutions is carried out in subsection~\ref{sec:Rg2vc} and the behaviour of a sticky polymer solution in which the chains as a whole are under $\theta$-solvent conditions, is considered in subsection~\ref{sec:thetasticky}. Section~\ref{sec:gel} considers the sol-gel transition and the various definitions that are used to find its location, and the validity of the modified Flory-Stockmayer expression at the gelation threshold (Eq.~(\ref{FS-D})) is examined. An interesting correlation observed between the breakdown of scaling predictions and the occurrence of phase separation is highlighted in section~\ref{sec:phase}. Finally, the key results of the present work are summarised in the concluding section. 

\section{\label{sec:simulation_details} Brownian dynamics of associative polymer solutions}

Although several previous computational studies of associative polymer solutions have been based on Brownian dynamics as the simulation technique, they differ from each other in a number of different aspects. For instance, while in some studies entire micelles are coarse-grained to single particles~\cite{Sprakel2009,Park2017}, others represent individual polymer chains as bead-spring dumbbells~\cite{vandenBrule1995,HernandezCifre2003,Cass2008}. Whereas in some recent investigations of the shear flow of associative polymer solutions, bead inertia has been taken into account in the context of Langevin dynamics of bead-spring chains~\cite{Omar2017,Furuya2018}, in earlier enquiries, associative polymers in shear flow have been modelled as non-interacting dumbbells~\cite{HernandezCifre2003}, or non-interacting bead-spring chains~\cite{HernandezCifre2007}, with beads switching between associated and dissociated states. None of these previous investigations, however, have taken hydrodynamic interactions into account. 

Hydrodynamic interactions have been successfully incorporated over the past several decades in computational studies of polymer solution dynamics in the \emph{dilute} concentration regime~\cite{Aust19995660,Petera19997614,Kairn2004,Kroger2004,Todd2007, Ermak1978,Fixman1981,Rey1989,zylka89,Iniesta1990,Zylka1991,jendrejack,Jendrejack20027752,Prabhakar2002,PraRavi04,Larson2005,Schroeder20051967,Shaqfeh2005,KailashJCP2018,Prakash2019,KailashPRR2020}. These studies have established beyond doubt that the inclusion of hydrodynamic interactions is essential for accurately capturing dynamic properties not only in the equilibrium and linear viscoelastic regimes, but also rheological material functions in the far from equilibrium non-linear regime~\cite{Larson2005,Shaqfeh2005,Prakash2019}. Accounting for hydrodynamic interactions in the \emph{semidilute} regime of concentration is more challenging since both intra and inter-molecular interactions need to be taken into consideration, particularly in the case of the latter since hydrodynamic interactions are long-ranged in space. Significant advances have been made over the last decade in our capacity to simulate semidilute polymer solutions due to the development of a variety of mesoscopic simulation techniques based on coarse-grained bead-spring chain models for polymer molecules~\cite{BD1999,Stoltz,kapral2008,BurkhardLadd,Gompper2009,JainPRE,SaadatSemi,JainCES,Yong2016,Dyer2017,Sing2018,Duenweg2018}. These algorithms have recently been successfully employed to examine a number of different problems in the semidilute regime~\cite{BD1999,JainCES,Huang2010,Jain2012,Fedosov2012,Chandi2017,Phan-Thien2018,Theers2018,Prakash2019,Qi2020}. 

The recent numerical investigations of associative polymer solutions by the J\"ulich group using multiparticle collision dynamics (MPCD)~\cite{GompEq,GompFlo}, and by \citet{Castillo} using nonequilibrium molecular dynamics (NEMD), automatically account for hydrodynamic interactions through the exchange of momentum between the beads on polymer chains and solvent molecules, since the latter are simulated explicitly. By implementing an attractive interaction potential between selected beads on the chain to model the association between sticker monomers, these pioneering studies have essentially extended the framework for studying semidilute polymer solutions to one that is capable of describing associative polymer solutions. In this work we introduce an alternative approach for describing associative polymer solutions that accounts for hydrodynamic interactions and is based on Brownian dynamics simulations. As in recent extensions of MPCD and NEMD, the proposed methodology is an extension of an algorithm developed previously to study semidilute polymer solutions, but in contrast to these methods, treats the solvent implicitly~\cite{JainPRE}. We hasten to add that while the formalism includes hydrodynamic interactions, the focus in this work is on the prediction of equilibrium static properties, which we consider as a first step towards exploring the predictive capabilities of the proposed algorithm. In the Supporting Information, however, we briefly consider how even though hydrodynamic interactions have no effect on equilibrium static properties, they do have a significant influence on the time scales in which equilibration occurs, and presage their fundamental role in determining dynamic properties.

\subsection{\label{sec:gov_eqs} Governing equations for sticky polymer solution dynamics}
Sticky polymers have been modelled here as a linear sequence of $N_b$ coarse-grained beads connected by $N_b-1$ entropic springs~\cite{Bird}, with the chain configuration specified at any time $t$ by the set of bead position vectors $\textbf{r}_{\nu}  (t)\, (\nu = 1, 2, ..., N_b)$. Each polymer is a multi-sticker chain with $f$ equispaced stickers positioned along the backbone (except at the chain ends where there are no stickers) separated by $\ell$ spacer (or backbone) monomers. A sticker is assumed to associate with only one other sticker (i.e. with functionality $\varphi =1$). Such systems can be designed experimentally\cite{Kornfield1,Kornfield2,Guo2005}. In general, while the proposed methodology can support any value for $\varphi$, the specific choice of $\varphi =1$ is made here in order to compare simulation predictions with the analytical predictions of Semenov and Rubinstein~\cite{RnSstatics} and Dobrynin~\cite{Dob}, where this constraint on sticker functionality has been chosen for the sake of simplicity. Note that once $\ell$ and $f$ are fixed, the number of beads in a chain can be calculated from,
\begin{equation}\label{Eq:Nb_calc}
 N_b = (f+1)\ell+f
\end{equation}

An associative polymer solution is modelled as an ensemble of such bead-spring chains, immersed in an incompressible Newtonian solvent. A total of $N_c$ chains are initially enclosed in a cubic and periodic cell of edge length $L$, giving a total of $N = N_b \times N_c$ beads per cell at a bulk
monomer concentration of $c = N/V$, where $V = L^3$ is the volume of the simulation cell. The evolution of bead positions in Brownian dynamics simulations is governed by an It{\^o} stochastic differential equation for the vectors $\textbf{r}_{\mu}$. The Euler integration algorithm for the non-dimensional form of this equation is given by,
\begin{widetext}
\begin{align}\label{GovEqn}
\begin{aligned}
\mathbf{r}_\mu(t + \Delta t) =\, & \mathbf{r}_\mu(t) + \left(\bm{\kappa}\cdot\textbf{r}_{\nu}(t)\right)\Delta t + \frac{\Delta t}{4} \sum\limits_{\nu=1}^N\mathbf D_{\mu\nu}\cdot(\mathbf F_\nu^{s}+ \mathbf F_\nu^{\textrm{SDK}}) +\frac{1}{\sqrt{2}}\sum\limits_{\nu=1}^N \mathbf B_{\mu\nu}\cdot\Delta \mathbf W_\nu
\end{aligned}
\end{align}
\end{widetext}
Here the length and time scales are non-dimensionalised with $l_H=\sqrt{k_BT/H}$ and $\lambda_H=\zeta/4H$, respectively, where $T$ is the absolute temperature, $k_B$ is the Boltzmann constant, $H$ is the spring constant, and $\zeta=6\pi\eta_s a$ is the Stokes friction coefficient of a spherical bead of radius $a$ where $\eta_s$ is the solvent viscosity. The quantity $\bm{\kappa} = (\bm{\nabla v})^{T}$ is a $3\times 3$ tensor, with $\bm{v}$ corresponding to the unperturbed solvent velocity field. For the static property predictions considered here, this term is set to zero. $\Delta\pmb W_\nu$ is a non-dimensional Wiener process with mean zero and variance $\Delta t$. The components of $\bm{\Delta}\Vector{W}_{\nu}$ are obtained here from a real-valued Gaussian distribution with zero mean and variance $\Delta t$. $\pmb{B}_{\mu\nu }$ is a non-dimensional tensor whose evaluation requires the decomposition of the diffusion tensor $\pmb D_{\mu\nu}$, defined as $\pmb D_{\mu\nu} = \delta_{\mu\nu} \pmb \delta + \pmb \Omega_{\mu\nu}$, where $\delta_{\mu\nu}$ is the Kronecker delta, $\pmb \delta$ is the unit tensor, and $\pmb{\Omega}_{\mu\nu}$ is the hydrodynamic interaction tensor. Defining the matrices $\mathcal{D}$ and $\mathcal{B}$ as block matrices consisting of $N \times N$ blocks each having dimensions of $3 \times 3$, with the $(\mu,\nu)$-th block of $\mathcal{D}$ containing the components of the diffusion tensor $\pmb{D}_{\mu\nu }$, and the corresponding block of $\mathcal{B}$ being equal to $\pmb{B}_{ \mu\nu}$, the decomposition rule for obtaining $\mathcal{B}$ can be expressed as $\mathcal{B} \cdot {\mathcal{B}}^{\textsc{t}} = \mathcal{D}\label{decomp}$. Since hydrodynamic interactions do not affect equilibrium static properties, all the static simulation results reported here are carried out with ${\pmb{\Omega}_{\mu \nu}} =\mathbf{0}$. For the few simulations carried out with hydrodynamic interactions, the regularized Rotne-Prager-Yamakawa (RPY) tensor is used to compute hydrodynamic interactions. Details and simulation results for this case are reported in the Supporting Information. 

The bonded interactions between the beads are represented by a spring force, $\mathbf F_\nu^{s}$, arising from a spring potential which is assumed here to be a finitely extensible nonlinear elastic (FENE) potential, $U_{\textrm{FENE}}$, between adjacent beads,
\begin{align}\label{eq-fraenkel}
U_{\textrm{FENE}}= -\frac{1}{2}Q_0^2\ln \left( 1-\frac{r^{2}}{Q_0^2}\right)
\end{align}
Here $Q_0$ is the dimensionless maximum stretchable length of a single spring, and $k_B T$ is used to non-dimensionalise energy. All the simulations reported in this work use a value of $Q_0^2 = 50.0$. Note the quantity $Q_0$ used here is identical to the square root of the more commonly used FENE $b$-parameter. The large value of $Q_0$ used here indicates a soft spring potential, which could result in chain-crossing. This would be problematic when the  prevention of chain-crossing is important, such as when examining the dynamics of entangled systems. However, in the present study of static properties, and planned future studies of dynamics, we restrict our attention to the unentangled regime, where topological constraints are not relevant. In fact, by allowing chain-crossings, the phase space is expected to be explored faster, which is advantageous. 

\subsection{\label{sec:sdk} Modelling excluded volume interactions in sticky polymers}

The only quantity in Eq.~(\ref{GovEqn}) that remains to be defined is $\mathbf F_\nu^{\textrm{SDK}}$, that describes the short-ranged excluded volume force on a bead $\nu$ due to its pair-wise interactions with other beads in its neighbourhood, which could be either stickers or spacer monomers that belong to the same chain or neighbouring chains. The strength of the interaction depends on the nature of both the interacting beads. In homopolymer solutions, pair-wise excluded volume interactions are frequently modelled with the Lennard-Jones potential, which is able to capture polymer conformations in poor, $\theta$ and good solvents, depending on the value chosen for the potential's well-depth. In the case of sticky polymer solutions, the introduction of sticky groups on polymer chains leads to a decrease in the size of the chain due to the relative affinity of sticky groups for each other, making the chains more collapsed or less swollen at a given temperature compared to the corresponding homopolymer of the same molecular weight, and resulting in the whole phase diagram being modified due to the presence of stickers~\cite{RnSstatics,Dob}.

In a recent publication~\cite{Aritra2019}, we have examined the collapse transition of dilute sticky polymer solutions and found it convenient to use the SDK potential~\cite{SDK} as an alternative to the Lennard-Jones potential, to model the pair-wise interactions between both the backbone and sticker monomers. The reasons for this choice have been elaborated in Ref.~\citenum{Aritra2019}. In this section, we give details of the potential, and briefly summarise some of the key results of our earlier publication that are relevant to the current work. This is necessary because the predictions of the scaling theories of Semenov and Rubinstein~\cite{RnSstatics} and Dobrynin~\cite{Dob}, with which we plan to compare the results of simulations, depend on which of three different scaling regimes the sticky polymer solution belongs to. 

The pair-wise excluded volume force, $\mathbf F_\nu^{\textrm{SDK}}$,  on a bead $\nu$ due to interactions with either stickers or spacer monomers is modelled here by the potential, $U_{\text{SDK}}$, proposed by Soddemann, D\"unweg and Kremer,~\cite{SDK}
\begin{align}\label{eq:SDK}
U_{\textrm{SDK}}=\left\{
\begin{array}{l l l}
&4\left[ \left( \dfrac{\sigma}{r} \right)^{12} - \left( \dfrac{\sigma}{r} \right)^6 + \dfrac{1}{4} \right] - \epsilon;  & r\leq 2^{1/6}\sigma \vspace{0.5cm} \\
& \dfrac{1}{2} \epsilon \left[ \cos \,(\alpha \left(\dfrac{r}{\sigma}\right)^2+ \beta) - 1 \right] ;& 2^{1/6}\sigma \leq r \leq r_c \vspace{0.5cm} \\
& 0; &  r \geq r_c
\end{array}\right.
\end{align}
The potential has a minimum at $r=2^{1/6}\sigma$, and the quantity $\epsilon$ is the attractive well-depth of the potential. The repulsive part of the SDK potential is modelled by a truncated Lennard-Jones (LJ) potential while the attractive contribution is modelled with a cosine function. Unlike the LJ potential, which has a long attractive tail, the short ranged attractive tail of the SDK potential approaches zero smoothly at a finite cut-off distance $r_c$, which leads to increased simulation efficiency~\cite{SDK}. When $\epsilon = 0$, the SDK potential corresponds to a  purely repulsive Weeks-Chandler-Anderson (WCA) potential. With increasing values of well-depth $\epsilon$, the solvent quality reduces from athermal to poor, and  the complete range of solvent qualities can be explored by simply varying $\epsilon$, since it changes the attractive component of the SDK potential without affecting the repulsive force (in contrast to the LJ potential). {In the present study, backbone-backbone (spacer-spacer) monomer interactions are denoted by $\epsilon_{bb}$, sticker-sticker monomer interactions are denoted by $\epsilon_{st}$, and spacer-sticker monomer interactions are assumed to be the same as spacer-spacer interactions, i.e., equal to $\epsilon_{bb}$. The value of the non-dimensional distance $\sigma$ (non-dimensionalised with $l_H$) is taken to be 1}. The constants $\alpha$ and $\beta$ are determined by applying the two boundary conditions, $U_{\text{SDK}} = 0$ at $r=r_c$, and $U_{\text{SDK}}=-\epsilon$ at $r=2^{1/6}\sigma$. 

For homopolymer solutions, \citet{Aritra2019} have shown that when the SDK potential is used in conjunction with Brownian dynamics simulations, the expected asymptotic scaling behaviour, in all regimes of solvent quality, is obtained with $r_c = 1.82 \,\sigma$ (for which  $\alpha= 1.5306333121$ and $\beta= 1.213115524$). Since only backbone monomer-monomer interactions exist for homopolymers, the well-depth $\epsilon$ is equal to $\epsilon_{bb}$ for all bead pairs. It is found~\cite{Aritra2019}, both from the scaling of the radius of gyration with chain length and from an estimation of the second virial coefficient, that $\theta$-solution conditions are reproduced for $\epsilon_{bb} \coloneqq \epsilon_\theta = 0.45$. As a consequence, the choice $\epsilon_{bb} < 0.45$ leads to good solvents, while $\epsilon_{bb} > 0.45$ leads to poor solvents. In particular, by defining the solvent quality $z = k_{\text{SDK}} \, \hat{\tau} \sqrt{N_b}$, where the parameter $\hat{\tau}$ is defined in terms of the potential well-depth by, 
\begin{equation}\label{Eq:zHP}
 \hat{\tau} =  \left(1-\frac{\epsilon_{bb}}{\epsilon_{\theta}}\right)
\end{equation}    
\citet{Aritra2019} show that by an appropriate choice of the constant $k_{\text{SDK}}$, simulation predictions of the swelling ratio $\alpha_g$, which is the ratio of the radius of gyration $R_g$ in a good solvent to that in a $\theta$-solvent, can be collapsed on to the universal swelling curve of $\alpha_g$ versus $z$ that describes the thermal crossover between $\theta$ and good solvents~\cite{schafer}. Note that $\hat{\tau}$ corresponds physically to $(1-T_\theta/T)$, so that in accordance with its definition in the experimental literature,~\cite{schafer,Hayward19993502,Pan2014a,Pan2014b} the solvent quality $z \sim (1-T_\theta/T) \sqrt{M}$, where $T_\theta$ is the $\theta$-temperature, and $M$ is the molecular weight.

In the case of the sticky polymer solutions considered here, the attractive strength $\epsilon$ of the SDK potential for a pair of monomers $\mu$ and $\nu$, is given by
\begin{equation} 
\label{Boolean}
\epsilon = (1- q_{\mu\nu})\, \epsilon_{bb} + q_{\mu\nu} \,\epsilon_{st} 
\end{equation} 
where $q_{\mu\nu} \in \{0,1\}$ is a Boolean variable such that for each pair of monomers $\mu$ and $\nu$,  $q_{\mu\nu}$ is zero whenever at least one of the two monomers is a backbone monomer, while for a pair of sticker monomers, $q_{\mu\nu}$ is zero if no bond exists between the two stickers, and $q_{\mu\nu} = 1$ for a bonded pair of sticker monomers. Typically, in all the simulations considered here, $ \epsilon_{bb} \le  \epsilon_\theta \le \epsilon_{st}$. At each time step, the simulation algorithm updates the variables $q_{\mu\nu}$ for stickers according to the following simple rules:
\begin{enumerate}
\item Whenever two stickers $\mu$ and $\nu$ come within the cutoff radius of the SDK potential, $r_c$, the value of $q_{\mu\nu}$ is changed from zero to one, provided that both stickers are not bonded to other stickers.
\item If three stickers are within the interaction range $r_c$, the decision regarding which pairs of beads stick together is made according to a scheme that depends on the bead number label of each of the stickers. Thus if stickers $\zeta$, $\delta$ and $\rho$, with $\zeta < \delta < \rho$ are within interaction range $r_c$, then $\zeta$ and $\delta$ form a pair with $q_{\zeta\delta} = 1$, and the sticker $\rho$ remains unbound with $q_{\zeta\rho} = q_{\delta\rho} = 0$. When more than three stickers are within the interaction range, the same scheme is implemented by treating each bead pair in turn and considering their respective bead number labels. While the choice of which pairs to stick based on bead number labels may seem arbitrary, it turns out that the scheme is effectively equivalent to picking the sticking pairs at random when three or more stickers are within the interaction range. Since the probability of three and higher body interactions amongst stickers is very low, and since we have considered a large ensemble of chains distributed randomly in a simulation box (implying random labelling of the stickers), the two schemes effectively produce the same results. This is explicitly demonstrated in the Supporting Information by comparing the predictions of different equilibrium static properties when the two different sticking rules are implemented.

\item As soon as the distance between two stickers $\mu$ and $\nu$ becomes greater than $r_c$, $q_{\mu\nu}$ is reset to zero, and new bondings may occur. 
\end{enumerate}

A knowledge of the monomer coordinates and the values of $q_{\mu\nu}$ is clearly sufficient to calculate the interaction energy of the system uniquely. For the sake of simplicity, we have implemented a rule where bond formation or breakage is determined purely by whether sticker pairs are within or outside the cut-off radius. Typically, bond formation or breakage is determined by implementing a Monte-Carlo scheme based on a Boltzmann weight~\cite{Hoy2009,Sing2011}. In a sense, the current rule can also be considered as a special case of a method based on a Boltzmann weight, where the activation energy barrier for binding is zero, such that the probability of bond formation is unity whenever a sticker pair is within the cut-off radius, while the activation energy for unbinding is infinite, implying that bonds break only when the relative distance between previously stuck pairs is larger than the cut-off radius. 

\begin{figure}[ptbh]
 \begin{center}
 {\includegraphics[width=80mm]{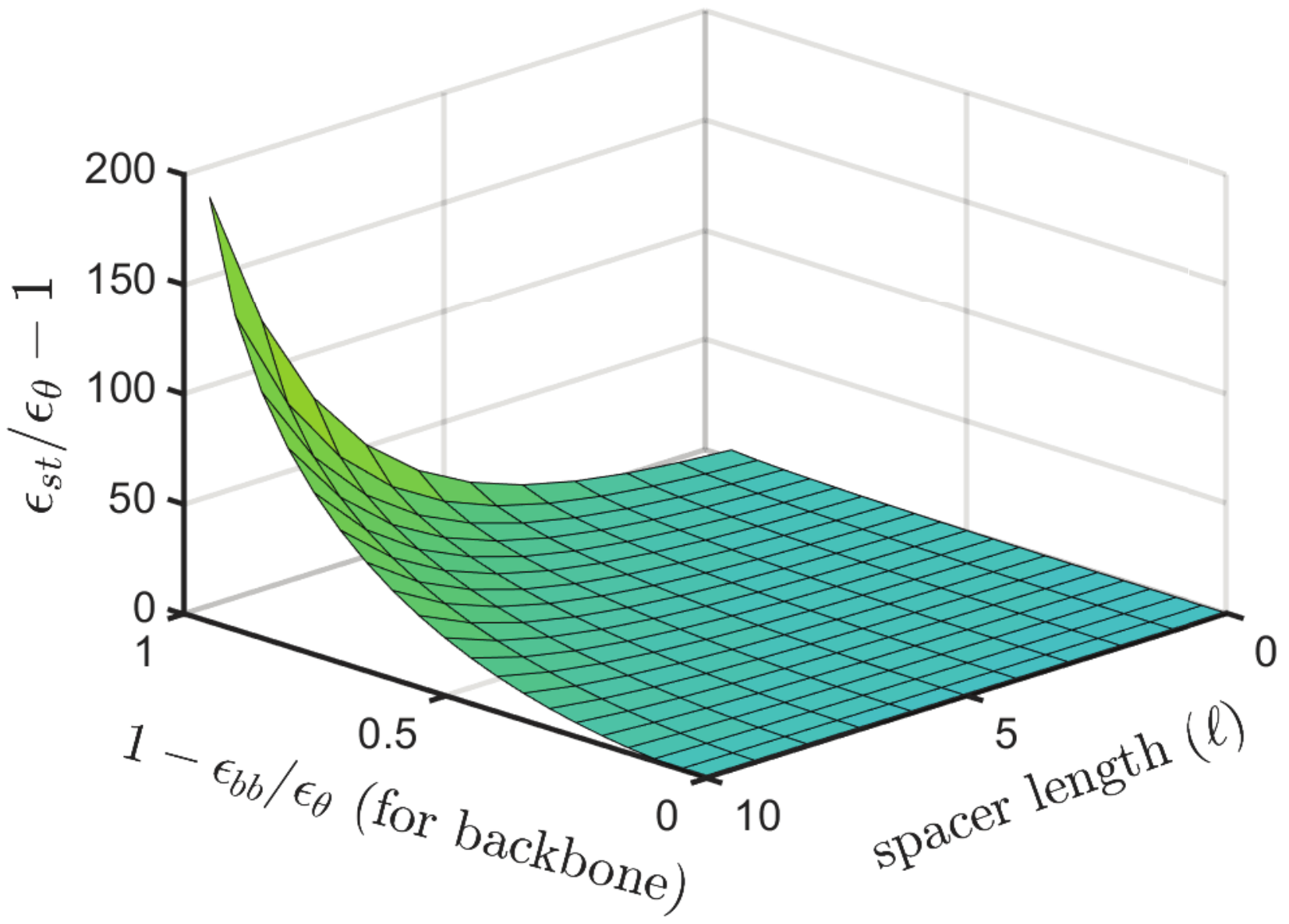}}
 \end{center}
 \vskip-15pt
\caption{Schematic of the $\theta$-surface for sticky polymer solutions in terms of the scaled variables $(\epsilon_{st}/\epsilon_{\theta}-1)$ plotted against the scaled backbone solvent quality, ($1-\epsilon_{bb}/\epsilon_{\theta}$), and spacer length, $\ell$. Points below the surface represent sticky polymer solutions under good solvent conditions while points above the surface indicate solutions under poor solvent conditions. Reproduced from Ref.~\citenum{Aritra2019} with permission from The Royal Society of Chemistry.}
\label{fig:thetasurface}
\vskip-10pt
\end{figure} 

As mentioned earlier, the introduction of stickers on chains alters the solvent quality of a polymer solution. In particular, \citet{Aritra2019} show that the well-depth of the SDK potential, $\epsilon_{st}^{\theta}$, at which $\theta$-solution conditions are observed in sticky polymer solutions is different from that for homopolymer solutions ($\epsilon_\theta$), and that it depends on the backbone well-depth $\epsilon_{bb}$ and spacer length $\ell$. A schematic representation of this dependence, reproduced from their paper, in shown in Fig.~\ref{fig:thetasurface}, where the two-dimensional surface corresponds to values of $\epsilon_{st}^{\theta} (\ell, \epsilon_{bb})$ that separate good and poor solvent regions. Since $\epsilon_{bb}< \epsilon_{\theta}<\epsilon_{st}$, chain conformations are a result of a competition between backbone-backbone repulsion and sticker-sticker attraction. As indicated in Fig.~\ref{fig:thetasurface}, the value of $\epsilon_{st}^{\theta}$ keeps increasing: (i) as the backbone solvent quality gets better at a given value of $\ell$, and (ii) with increasing spacer length, at a given value of $\epsilon_{bb}$.

As will become evident in Secs.~\ref{sec:simdet} and \ref{sec:deg_of_conv}, the behaviour of dilute sticky polymer solutions summarised here is very helpful for estimating sticky chain parameters and the well-depths of the SDK potential that lead to simulation results in the precise scaling regimes defined in the theory of Dobrynin~\cite{Dob}, thereby enabling a direct comparison between them.

\section{\label{sec:scaling} Scaling relations for fractions of associated stickers}

The phase behaviour of physically associative polymer solutions has been described theoretically by a number of different analytical approaches.~\cite{TanakaPRL1989,IshTan,RubDob,RnSstatics,Erukhimovich2001,Dob,tanaka2011polymer,Ozaki2020} The majority of these studies only treat the presence of inter-chain associations and neglect the formation of intra-chain associations, which is a reasonable approximation at sufficiently high polymer concentrations. The theory developed by~\citet{Dob}, on the other hand, explicitly accounts for the presence of intra-chain associations as well. In the present work, since the simulations can predict both intra and inter-chain degrees of conversion, we compare our results with the predictions of Dobrynin's theory.  It should be noted, however, that the scaling prediction by~\citet{RnSstatics} are identical to that of~\citet{Dob} for the fraction of stickers bonded by inter-chain associations. 

An expression for the free energy of an associative polymer solution has been derived by~\citet{Dob} with the help of a lattice based mean-field theory, combined with blob scaling arguments to describe polymer chain conformations in semidilute solutions. By minimising the free energy with respect to both intra and inter-chain degrees of conversion, equations for the dependence of the equilibrium degrees of conversion $p_1$ and $p_2$ on system parameters, such as $T$, $c$, $\ell$, $\epsilon_{st}$, etc., are obtained. Though \citet{Dob} also estimates the phase diagram of associative polymer solutions in the temperature and concentration plane, we do not attempt to map out the entire phase diagram with simulations in the present work. In section~\ref{sec:gel}, however, we briefly consider the relationship predicted by simulations between the monomer concentration, $c$, and the number of stickers, $f$, along the gelation line that separates the sol and gel phases, and compare with the prediction of scaling theory. 

\begin{figure}[t]
    \centerline{
    \begin{tabular}{c}
   {\includegraphics[width=50mm]{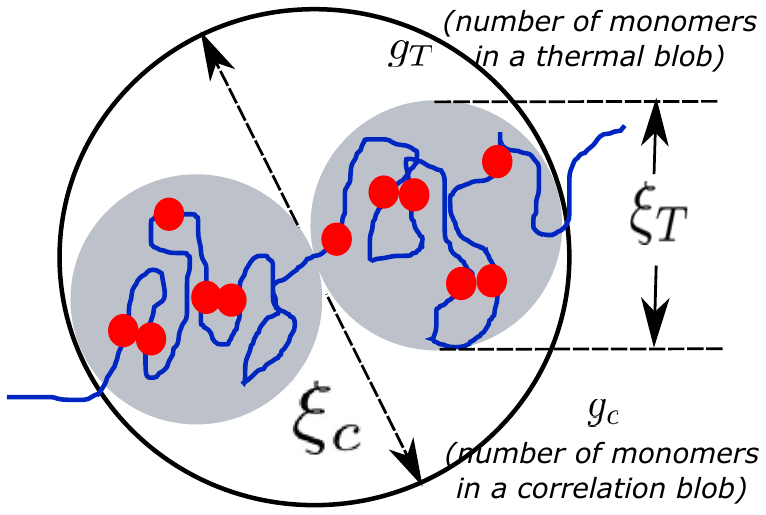}} \\
    { (a) }\\[5pt]
         \includegraphics[width=43mm]{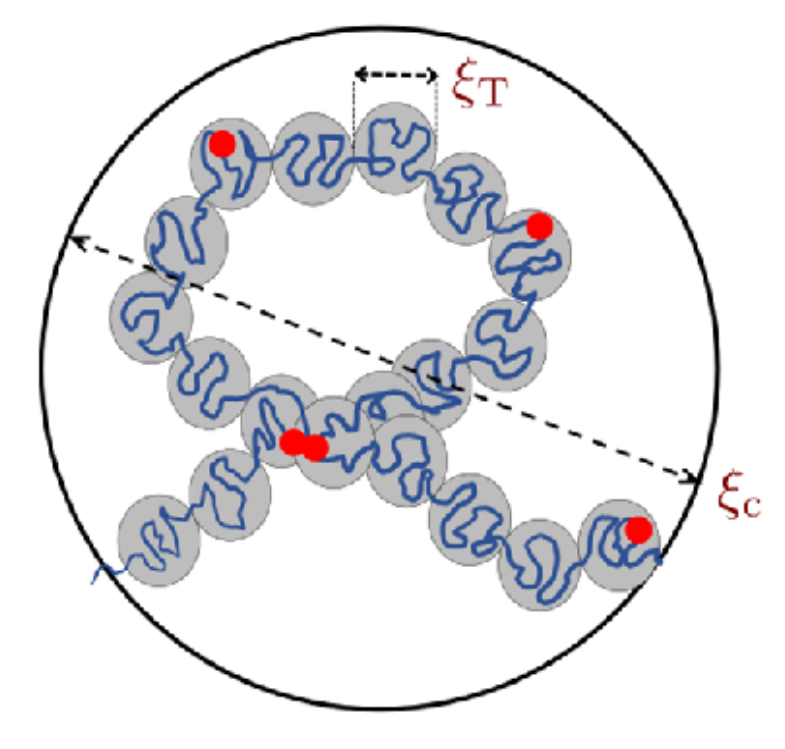}  \\
    { (b) }\\[5pt]
   \includegraphics[width=45mm]{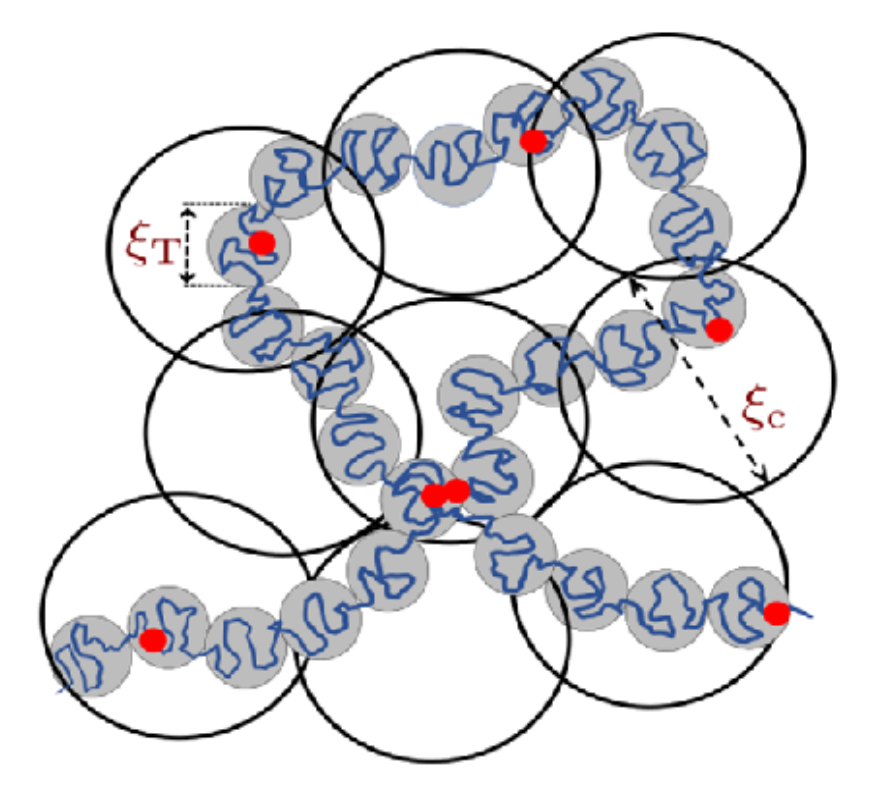} \\
  (c) \\
    \end{tabular}
    }
\caption{Three scaling regimes when the chain of backbone monomers is in a good solvent. Stickers are indicated by red circles. (a) $\ell<g_T<g_c$ (regime I), (b) $g_T<\ell<g_c$ (regime II) and (c) $g_T<g_c<\ell$ (regime III).}
\label{fig:GoodSolBlobs}
\vskip-10pt
\end{figure}

\citet{Dob} derives separate sets of relations for the fractions $p_1$ and $p_2$, depending on the quality of the solvent with respect to the backbone monomers (i.e., whether they are in $\theta$ or good solvents), and the number of monomers between two stickers ($\ell$). In particular, when the chain of backbone monomers is under good solvent conditions, three separate regimes are identified that are best understood in terms of the schematic representation in Fig.~\ref{fig:GoodSolBlobs}, which is inspired by a similar figure in Ref.~\citenum{Dob}. 

As is well known, semidilute polymer solutions that lie in the double crossover region of solvent quality and concentration can be described in terms of thermal and correlation blobs, which represent the dependence on solvent quality $z$ and scaled concentration $c/c^*$~\cite{Jain2012,Prakash2019}. The size of a thermal blob is denoted by $\xi_T$, with $g_T$ numbers of monomers in it, while the size of a correlation blob is denoted by $\xi_c$, with $g_c$ numbers of monomers in it. The three regimes defined by~\citet{Dob} depend on the relative magnitudes of $\ell$, $g_T$ and $g_c$, as indicated in Fig.~\ref{fig:GoodSolBlobs}.

A central element in Dobrynin's theory is the estimation of the probability of two stickers coming together to form a bond. Under $\theta$-solvent conditions this is straightforward to determine since a polymer chain obeys random walk statistics and the probability of contact between two monomers on a chain is proportional to the probability of loop formation between two ends of a Gaussian chain segment~\cite{RubCol,Dob}. For a segment of chain with $\ell$ monomers between stickers, this implies that the probability of sticker contact is $p_\theta (\ell)\sim\ell^{-3/2}$. The situation is more complicated in a good solvent, and depends on which of the three regimes in Fig.~\ref{fig:GoodSolBlobs} is relevant. In regime~I  (Fig.~\ref{fig:GoodSolBlobs}~(a)), since there are many stickers in a thermal blob, and a chain segment within a thermal blob obeys random walk statistics, the sticker association probability is governed by the same physics as for a chain under $\theta$-solvent conditions. On the other hand, in regime~II (Fig.~\ref{fig:GoodSolBlobs}~(b)), the calculation of the probability that two stickers come into contact involves two steps. First two thermal blobs must come into contact, followed by two stickers within these thermal blobs coming together to form a bond. Since thermal blobs follow self-avoiding walk statistics, their probability of contact is equivalent to the probability that two internal monomers of a chain in a good solvent come into contact. This problem was first solved using renormalisation group methods by~\citet{desClo1980}, who derived the following expression for the probability of contact between two internal monomers on a self-avoiding walk chain,
\begin{equation}\label{Eq:theta2}
p_{\text{good}} \sim \left[\frac{\delta}{r(n)}\right]^{3+\theta_2}
\end{equation}
where $\delta$ is the spatial distance between the two monomers, $r(n)$ is the root mean squared end-to-end distance between the two monomers, which are considered to be separated by $n$ monomers along the chain backbone, and $\theta_2$ is a geometrical exponent, the so-called des Cloizeaux exponent~\cite{Redner80,Witten82,Duplantier89,Hsu2004}. The numerical value of $\theta_2$ will be discussed shortly below. Once the thermal blobs are in contact, the probability that two stickers within them come into contact is estimated using the same expression as for two monomers on a segment of a chain under $\theta$-solvent conditions. Using similar arguments, \citet{Dob} also derives the probability of two stickers coming into contact when the good solvent conditions for the chain backbone correspond to those represented by regime~III.

\begingroup
\renewcommand{\arraystretch}{1.5}
\begin{table*}[t]
\centering
\resizebox{\textwidth}{!}{\begin{tabular}{|c|c|c|c|}
\hline
\textbf{(a)}  & {$\displaystyle\frac{p_1(1-p_1/2)}{(1-p)^2}$} & {$\displaystyle\frac{p_2}{(1-p)^2}$} & {$\displaystyle\frac{p_1(1-p_1/2)}{p_2}$}  \\[10pt]
 \hline 
 \hline  
Regime I & $g_{ss}\,\ell^{-{3}/{2}}$ & $ g_{ss} \, \ell^{-1} \, g_T^{-1/2} \left({g_T}/{g_c}\right)^{\nu(3+\theta_2)-1}$  & ${g_T}^{{1}/{2}} \, {\ell}^{-{1}/{2}} \left({g_c}/{g_T}\right)^{\nu(3+\theta_2)-1}$ \\
 \hline
Regime II &$g_{ss}\,\ell^{-{3}/{2}}\left({g_T}/{\ell}\right)^{\nu(3+\theta_2)-\frac{3}{2}}$ & $ g_{ss} \, \ell^{-1} \, g_T^{-1/2} \left({g_T}/{g_c}\right)^{\nu(3+\theta_2)-1}$ &$\left({g_c}/{\ell}\right)^{\nu(3+\theta_2)-1}$ \\
 \hline
Regime III  &$g_{ss}\,\ell^{-{3}/{2}}\left({g_T}/{g_c}\right)^{\nu(3+\theta_2)-\frac{3}{2}}$  & $ {g_{ss} \, \ell ^{-1} \, g_T^{-{1}/{2}}} \left({g_T}/{g_c}\right)^{\nu(3+\theta_2)-1}$ & $\left({g_c}/ {\ell}\right)^{{1}/{2}}$ \\
 \hline    
  \hline 
\addlinespace[10pt]
  \multicolumn{4}{c}{$ \nu = \tfrac{3}{5}  \, ; \quad \theta_2 = \tfrac{1}{3} \, ; \quad g_T =  \hat{\tau}^{-2} \, ; \quad g_c =  \hat{\tau}^{-\tfrac{6\nu -3}{3\nu -1}}  \, c^{-\tfrac{1}{3\nu -1}}  =   \hat{\tau}^{-\tfrac{3}{4}} \, c^{-\tfrac{5}{4}} $  }  \\
\addlinespace[10pt]  
\hline   
\textbf{(b)}  & {$\displaystyle\frac{p_1(1-p_1/2)}{(1-p)^2}$} & {$\displaystyle\frac{p_2}{(1-p)^2}$} & {$\displaystyle\frac{p_1(1-p_1/2)}{p_2}$}  \\[10pt]
 \hline 
 \hline  
Regime I & $g_{ss}\,\ell^{-{3}/{2}}$ & $g_{ss} \, \hat{\tau}^{-1/4} \, \ell^{-1}  \, c^{5/4} $  & $ \hat{\tau}^{1/4} \, \ell^{-1/2} \, c^{-5/4}  $ \\
 \hline
Regime II & $ g_{ss} \, \hat{\tau}^{-1}  \, \ell^{-2} $ & $ g_{ss} \,  \hat{\tau}^{-1/4}   \, {\ell}^{-1} \,c^{5/4} $  & $ \hat{\tau}^{3/4} \, {\ell}^{-1}   \, c^{-5/4} $ \\
 \hline
Regime III  &$g_{ss}\, \hat{\tau}^{-5/8} \,  \ell^{-{3}/{2}} \, c^{5/8}  $  & $ g_{ss}  \, \hat{\tau}^{-1/4}  \, {\ell}^{-1} \, c^{5/4} $ & $ \hat{\tau}^{-3/8} \, \ell^{-1/2} \,  c^{-5/8}  $ \\
 \hline
 $\theta$-solvent & $g_{ss}\,\ell^{-{3}/{2}}$ & $g_{ss} \, \ell^{-1}  \, c $  & $ \ell^{-1/2}  \, c^{-1}  $ \\
 \hline
 \hline           
\end{tabular}}
\caption{Relations for the intra-chain and inter-chain association fractions, predicted by~\citet{Dob}. Table~(a) corresponds to the three scaling regimes that arise when the backbone monomers are under good solvent conditions. The expressions are in terms of the spacer length $\ell$, the number of monomers in a thermal blob $g_T$, the number of monomers in a correlation blob $g_c$, and the function $g_{ss}$, which depends on the \emph{effective} sticker strength. Note that $p = p_1+ p_2$ is the total fraction of associated stickers. Table~(b) gives the simplified forms of the relations for good solvents when $g_T$ and $g_c$ are expanded in terms of the backbone solvent quality parameter $\hat{\tau}$, the monomer concentration $c$, and the specific choices $\nu = 3/5$ and $\theta_2 = 1/3$ are made, along with the corresponding relations for the case when the backbone monomers are under $\theta$-solvent conditions.  \label{tab:Scaling_rel} }
\end{table*}
\endgroup

With this background, the relevant relations for the degrees of intra and inter-chain conversion derived by~\citet{Dob} are displayed in Table~\ref{tab:Scaling_rel}. Note that $p = p_1+ p_2$ is the total fraction of associated stickers. Table~\ref{tab:Scaling_rel}~(a) displays the most general form of the relations when the backbone monomers are under good solvent conditions, for the three different scaling regimes that have been identified in Fig.~\ref{fig:GoodSolBlobs}. {The relations are in terms of $\ell$, $g_T$, $g_c$, and $g_{ss}$. The function $g_{ss}$ is assumed to depend exponentially on an \textit{effective} associating energy $\tilde \epsilon_{a}$, which is a combination of the interaction energy between stickers, $\epsilon_{st}$, and the Flory-Huggins interaction energy between monomer and solvent molecules on adjacent lattice sites, $\epsilon_{ps}$. In the context of scaling theory, where the solvent is treated explicitly, the magnitude of $g_{ss}$ can be chosen independently from $\ell$. When the spacer segment length $\ell$ is changed while keeping $\epsilon_{st}$ fixed, it is possible to control the influence of $g_{ss}$ on chain statistics independently by appropriately tuning $\epsilon_{ps}$. The situation is more subtle in the model adopted in the present work, since the solvent is treated implicitly rather than explicitly. It is not possible to keep $\epsilon_{st}$ and $\epsilon_{bb}$ fixed, and vary only $\ell$ without also simultaneously influencing chain statistics, since as exemplified by the schematic representation in Fig.~1, the effective interaction energy between stickers, which determines the conformations of polymer chains in a sticky polymer solution, is a complex function of  $\epsilon_{st}$, $\epsilon_{bb}$, and $\ell$. As a consequence, the function $g_{ss}$ depends on all three of these variables, and in general cannot be varied independently of spacer length $\ell$. The nature of this dependence is discussed in more detail in section~V~B. } Table~\ref{tab:Scaling_rel}~(b) gives the simplified form of the equations for good solvent conditions that are used in the current work, along with the corresponding relations for the case when the backbone monomers are under $\theta$-solvent conditions. Before discussing the derivation of these simplified relations, however, it is worth making a few remarks about the des Cloizeaux exponent $\theta_2$. 

The value $\theta_2 = 0.71$ was derived by~\citet{desClo1980} approximately using renormalised field theory. Subsequently, it was shown by~\citet{Witten82} and~\citet{Duplantier89} that $\theta_2$ could be related analytically to critical exponents that characterise star polymers.  {The critical exponents for star polymers with up to 80 arms have been obtained extremely accurately by~\citet{Hsu2004}, using Monte Carlo simulations with the PERM algorithm}. Based on the expression connecting $\theta_2$ to the critical exponents of stars derived by~\citet{Duplantier89}, and using the values computed for these exponents by~\citet{Hsu2004}, one can determine that $\theta_2 = 0.8142(17)$. This is probably the most refined value of the des Cloizeaux exponent that has been estimated to date. In addition to reporting the value for the exponent $\theta_2$ derived by~\citet{desClo1980}, \citet{Dob} also estimates a value for $\theta_2$ using an alternative argument. Essentially, by equating the probability of binary contact between monomers within a correlation blob (in the context of the mean field theory) to the contact probability given by Eq.~(\ref{Eq:theta2}), \citet{Dob} obtains $\theta_2 = 1/3$, which is considerably different from the value of the des Cloizeaux exponent derived from combining analytical arguments with Monte Carlo simulations. However, the latter value has been obtained for a self-avoiding walk chain in the dilute limit. The value of $\theta_2$ in the context of associative polymer solutions at finite concentrations, where both Flory screening and attractive interactions between stickers is present, is currently unknown. We will show subsequently that using $\theta_2 = 1/3$ leads to excellent collapse of simulation data under a wide range of conditions. 

The simplified form of Dobrynin's relations can be obtained by expanding $g_T$ and $g_c$ in terms of the backbone solvent quality parameter $\hat{\tau}$, and the monomer concentration $c$. 
{Within the blob scaling ansatz, for a semidilute solution in the double crossover region, the number of thermal blobs $\mathcal{N}_T$, and the number of correlation blobs $\mathcal{N}_c$ on a chain, are determined solely by the solvent quality $z$, and the scaled concentration $c/c^*$, respectively~\cite{Jain2012}
\begin{equation}
\begin{aligned}
\label{Eq:NTNc}
\mathcal{N}_T & = z^{2} \\ 
\mathcal{N}_c & = \left(\frac{c}{c^*}\right)^{\frac{1}{3\nu-1}} 
\end{aligned}
\end{equation} 
where $\nu$ is the Flory exponent, and the overlap concentration $c^*$ is defined by,
\begin{equation}
\label{Eq:c*}
c^{\ast} = \frac{N_b}{({4 \pi}/{3} ){R_{g0}}^3}
\end{equation} 
Here, $R_{g0}  \coloneqq \sqrt{\langle R_{g0}^2 \rangle}$ is the radius of gyration of a homopolymer chain of backbone monomers in the dilute limit, where $\langle R_{g0}^2 \rangle$ is given by,
\begin{equation}
\langle R_{g0}^2 \rangle = \frac{1}{2 N_b^2}
\sum_{\mu = 1}^{N_b} \sum_{\nu = 1}^{N_b} \langle  r_{\mu \nu}^2\rangle
\end{equation}
with angular brackets representing ensemble averages, and $r_{\mu \nu} = \vert \textbf{r}_{\nu} - \textbf{r}_{\mu} \vert$ being the inter-bead distance. It follows that,
\begin{equation}
\begin{aligned}
\label{Eq:gTgc}
g_T & = \frac{N_b}{\mathcal{N}_T} \sim \hat{\tau}^{-2} \\ 
g_c & = \frac{N_b}{\mathcal{N}_c} = N_b \left(\frac{c}{c^*}\right)^{- \frac{1}{3\nu-1}} 
\end{aligned}
\end{equation} 
}
In a good solvent, since a homopolymer  is a self-avoiding walk of thermal blobs, $R_{g0} = \xi_T \, (N_b/g_T)^\nu$, where $\xi_T = b \, g_T^{1/2}$, and $b$ is the size of a monomer. It follows that, $c^* \sim  N_b^{1-3\nu} g_T^{3\nu-3/2}$, and from Eq.~(\ref{Eq:gTgc}),
\begin{equation}
\label{Eq:gc}
{g_c \sim  \hat{\tau}^{-\tfrac{6\nu -3}{3\nu -1}}  \, c^{-\tfrac{1}{3\nu -1}} }
\end{equation}
Substituting for $g_T$ and $g_c$ from Eqs.~(\ref{Eq:gTgc}) and~(\ref{Eq:gc}) into the general scaling relations in Table~\ref{tab:Scaling_rel}~(a), and setting $\nu = 3/5$ and $\theta_2 = 1/3$, leads to the expressions displayed in Table~\ref{tab:Scaling_rel}~(b). The choice of simulation parameters that enable the validation of these scaling predictions, {and the details of the simulation algorithm are} discussed in the next section.

\section{\label{sec:simdet} Choice of parameters and details of simulations}

{In order to establish the validity of scaling laws, one would ideally vary independent variables such as the concentration and temperature, and parameters such as the number of stickers, spacer length and so on, over a very wide range of values so as to capture not only the asymptotic regimes, but also the crossover behaviour from one regime to another. In the present instance, this goal is constrained due to several factors. Firstly, the different scaling regimes are not obtained by varying the magnitude of a single variable. This is unlike, for instance, in the case of homopolymer solutions where one can go from the dilute to the concentrated entangled regime via the semidilute unentangled, semidilute entangled and concentrated unentangled regimes, by just varying the concentration~\cite{RubCol}. Secondly, the need to remain in the good solvent regime of the sticky polymer solution (i.e., below the $\theta$-surface shown schematically in Fig.~\ref{fig:thetasurface}) in order to avoid phase separation imposes constraints on the choice of parameter values. Finally, the use of the Brownian dynamics simulations methodology, which has the advantage of accurately predicting dynamic properties due to the incorporation of hydrodynamic interactions, makes the computations very intensive (even in the absence of HI). Each of these points are discussed in greater detail in section~\ref{subsec:simcon} below, while the particulars of the current algorithm are given in sections~\ref{subsec:sims} and~\ref{subsec:clus}.
}

{ \subsection{\label{subsec:simcon} Simulation contraints}} 

{In Dobrynin's scaling theory~\cite{Dob}, the different scaling regimes depend on the relative magnitudes of the spacer length $\ell$, the number of monomers in a thermal blob $g_T$, and the number of monomers in a correlation blob $g_c$ (as shown schematically in Fig.~\ref{fig:GoodSolBlobs}). As a result, in order to traverse from Regime~I ($\ell < g_T < g_c$) to Regime~II ($g_T<\ell<g_c$) it is necessary to change either $\ell$ or the solvent quality parameter $\hat{\tau}$ (which affects $g_T$). Changing the concentration (which would change $g_c$) would have no effect, provided both $g_T$ and $\ell$ were maintained less than $g_c$. On the other hand, in order to traverse from Regime~II ($g_T<\ell<g_c$) to Regime~III ($g_T<g_c<\ell$), one can either change $\ell$ or the concentration $c$. Changing $\hat{\tau}$ would have no effect provided $g_T$ was always maintained the smallest of the three magnitudes. Note that Dobrynin's scaling theory does not consider the case where $g_c < g_T$, which would occur for concentrations $c > c^{**}$, where $c^{**}$ represents the concentration at which $\xi_c = \xi_T$. These considerations imply that it is not possible to move all the way from Regime~I to Regime~III through the change of a single variable, such as the concentration. }

{Since $\hat{\tau}$ is given by Eq.~(\ref{Eq:zHP}), the values of $\hat{\tau}$ are in the range, $ 0 \le \hat{\tau} \le 1$, for $0 \le \epsilon_{bb} \le \epsilon_{\theta}$ (which follows from the requirement that the backbone monomers must be in a good solvent), and consequently, $1 \le g_{T} \le \infty$ (setting all unknown pre-factors equal to 1). Since the values of concentration are in the range $ 0 \le {c}/{c^*} \le {c^{**}}/{c^*}$, we have, $N_b \left({c^{**}}/{c^*}\right)^{- \frac{1}{3\nu-1}} \le g_c \le \infty$. The lower bound is always satisfied provided $g_T < g_c$. While conceptually, both $g_c$ and $g_T$ can be greater than $N_b$, the spacer length $\ell$, which is an input parameter in the simulations, must satisfy, $1 \le \ell \le N_b$. Since there must be at least one or more thermal blobs in a chain for good solvent conditions, $g_T$ must be less than $N_b$ in Regime~I. For this reason, and in order to satisfy the constraint with regard to $\ell$, $g_T$ must be less than $N_b$ in Regimes~II and~III, while $g_c$ cannot be greater than $N_b$ in Regime~III.} 

{With this background, we can now consider the constraints that exist in each of the regimes with respect to the choice of parameters. Consider Regime~I, where $\ell  < g_T < g_c $. Since a reasonable length of spacer segment must be chosen to enter the scaling regime, the value of $\ell$ cannot be too small. The value of $g_T$ increases rapidly from 1 as $\epsilon_{bb}$ increases from $0$ to $\epsilon_{\theta}$ (since $g_T =  [1- (\epsilon_{bb}/\epsilon_{\theta})]^{-2}$). While it is consequently not difficult to achieve $g_T > \ell$, its value cannot be too large, since the need to maintain $ N_b \ge g_T$ would lead to  excessive computational cost from simulating long chains (as discussed in greater detail below). The requirement that the scaled concentration $c/c^*$ must be such that $g_c > g_T$, connects the range of variation of $c/c^*$ to the choice of $\epsilon_{bb}$. As can be seen from Eq.~(\ref{Eq:gTgc}), a large value of $c/c^*$ implies  choosing a large value of $N_b$ in order to satisfy the constraint on $g_c$. Clearly, for a given chain length $N_b$, there is a limit to how closely $\epsilon_{bb}$ can approach $\epsilon_{\theta}$ (since $g_T \le N_b$), and also an upper bound on the choice of $c/c^*$ (since $g_c > g_T$).} 

{The value of $\epsilon_{bb}$ also affects the choice of sticker strength $\epsilon_{st}$. It is clear from Fig.~\ref{fig:thetasurface} that for a given value of $\ell$, the value of $\epsilon_{st}^\theta$ decreases with increasing $\epsilon_{bb}$. As will be seen later, in order to avoid phase separation with increasing concentration, it is necessary for $\epsilon_{\theta} < \epsilon_{st} < \epsilon_{st}^\theta$, which  corresponds to good solvent conditions for the sticky polymer solution as a whole. At the same time, it is necessary to sufficiently separate the values of $\epsilon_{bb}$ and $\epsilon_{st}$ in order to achieve a reasonable frequency and duration of association between stickers, and to distinguish between sticky and homopolymer solutions.} 

{Of the three scaling regimes, it is relatively easiest to simulate Regime~II ($g_T<\ell<g_c$), since $g_T$ can be chosen to have a small value by choosing $\epsilon_{bb}$ close to zero. This also leads to a fairly wide range of permissible values for $c/c^*$. Nevertheless, for large values of $\ell$, maintaining $g_c > \ell$ would require larger and larger values of $N_b$ for increasing values of $c/c^*$ (as can be seen from  Eq.~(\ref{Eq:gTgc})), leading to excessive computational cost. }

{The smallest value of the scaled concentration permissible in Regime~III is  $c/c^* =1$. This is because,  for this value (from Eq.~(\ref{Eq:gTgc})) $g_c = N_b$, and Regime~III requires that $g_c < \ell \,\, (\le N_b)$. As discussed above, it is straightforward to achieve a small value of $g_T$ by choosing $\epsilon_{bb}$ close to zero. However, both the requirements that $c/c^* \ge 1$ and $\ell > g_c$, lead to significant computational costs since they imply an increase in the number of monomers in a simulation cell, associated with an increase in the number of chains in a box, and large values of $N_b$, respectively.}

{It is clear from the discussion above that for given values of $\epsilon_{st}$ and $N_b$, simulations in any one of the three different scaling regimes can be carried out by appropriately choosing the values of $\ell$, $\epsilon_{bb}$ and $c/c^*$. The range of values of these parameters that can be explored is, however, very dependent on the chain length $N_b$. Provided that large enough values of $N_b$ are used, it would in principle be possible to thoroughly examine both the asymptotic scaling behaviour in all the different regimes, and the crossover between them. The computational intensity of the current Brownian dynamics algorithm, however, places quite stringent restrictions on the range of values that can be explored. The majority of the results reported here have been carried out on Australia's peak research supercomputer based at the National Computational Infrastructure. Details of the machine, and the computational cost estimates for simulating chains of various lengths $N_b$, spacer lengths $\ell$, and concentrations $c/c^*$ have been given in Table~S1 in the supplementary material. It suffices here to say that while $N_b=34$, $\ell=6$, $c/c^*=2.0$ requires roughly 3 hours of CPU time, $N_b=79$, $\ell=15$, $c/c^*=6.5$ requires around 3 days and 9 hours of CPU time for obtaining data at these individual concentrations, from a typical simulation. This computational intensity has implied that we have only been able to explore a limited range of parameter values. Nevertheless, as will be seen from the results presented below, the simulations are adequate to reach clear conclusions regarding the validity of the scaling predictions in the different regimes, and to tease out some aspects of the crossover behaviour. It is hoped that future improvements in the BD algorithm will enable a more complete examination of the predictions of the scaling theory.} 

{Table~\ref{tab:parameters} summarises all the values of parameters (with $g_T$ and $g_c$ rounded to the nearest integer) used in the current simulations in order to explore the different scaling regimes. }

\begingroup
\begin{table}[t]
    \centering
\resizebox{\columnwidth}{!}{\begin{tabular}{c|c|c|c|c|c|c}
        \hline
         & ($N_b,\,\ell,\,f$) & $\epsilon_{bb}$ & $\epsilon_{st}$ & $g_T$ & $c/c^*$ & $g_c$  \\
        \hline
        \hline 
  	\multirow{6}{*}{$\theta$}    & ($24,\,4,\,4$) & No EV  & $5.0$ & -- & $0.1 - 1.6$ & --\\
           & ($34,\,4,\,6$) & No EV & $5.0$ & -- & $0.1 - 0.5$ & -- \\
               & ($34,\,6,\,4$) & No EV & $2.5$ & -- & $0.1 - 0.6$ & -- \\
            & ($29,\,4,\,5$) & 0.45  & $5.0$ & -- & $0.2 - 0.8$ & -- \\
           & ($34,\,4,\,6$) & 0.45  & $5.0$ & -- & $0.2 - 1.2$ & -- \\
           & ($34,\,6,\,4$) & 0.45  & $2.5$ & -- & $0.1 - 0.6$ & -- \\
               \hline    
            & ($29,\,4,\,5$) & $0.3$ & $2.5$  & $9$ & $0.7 - 2.0$ & $45 - 12$ \\ 
        Regime  I      & ($34,\,4,\,6$) & $0.3$ & $2.5$  & $9$ & $0.8 - 1.6$ &   \multirow{2}{*}{$45 - 19$}  \\
  ($\ell<g_T<g_c$)     &  ($34,\,6,\,4$) & $0.3$ & $2.0$ & $9$ & $0.8 - 1.6$ &    \\
                      &  ($34,\,6,\,4$) & $0.35$ & $2.5$ & $20$ & $1.0 - 1.5$ &  $34 - 20$  \\ 
            \hline
 &  ($24,\,4,\,4$) &  $0.0$  & $5.0$ & 1 &  \multirow{8}{*}{${0.5} - 1.9$} & $57 - 11$  \\
                 & ($29,\,5,\,4$) &  $0.0$  & $5.0$ & 1 & & $69 - 13$ \\ 
                 & ($34,\,4,\,6$) &  $0.0$  & $5.0$ & 1 & & \multirow{3}{*}{$81 - 15$}\\
 Regime  II    & ($34,\,6,\,4$) &  $0.0$  & $5.0$ & 1 & $ $ &  \\
  ($g_T<\ell<g_c$)   & ($34,\,6,\,4$) &  $0.0$  & $4.0$ & 1 & \\
   & ($39,\,4,\,7$) &  $0.0$  & $5.0$ & 1 &  & $93 - 17$\\
       & ($44,\,4,\,8$) &  $0.0$  & $5.0$ & 1 &  & 105 - 20  \\                         
    & {($64,\,12,\,4$)} &  {$0.0$}  & {$5.0$} & {1} & & 152 - 29 \\    
                 \hline 
        Regime III  & ($64,\,12,\,4$) & \multirow{2}{*}{$0.0$}  & \multirow{2}{*}{$5.0$} & \multirow{2}{*}{$1$} & \multirow{2}{*}{${4.0} - 6.5 $} & $11 - 6$ \\  
     ($g_T<g_c<\ell$)  &($79,\,15,\,4$) & & & & & $14 - 8$\\
    \hline
    Sticky $\theta$ chain & ($34,\,6,\,4$) & {$0.35$}  & {$3.6$} & {$20$} & {$0.2 - 6.0$} & {$254 - 4 $} \\  
 \hline
  \hline
    \end{tabular}
}     
\caption{Parameter sets used to simulate the different scaling regimes of associative polymer solutions that arise when the backbone monomers are under $\theta$ and good solvent conditions, and when the sticky chain as a whole is under $\theta$ solvent conditions. {Note that $g_T$ must be less than $N_b$ in Regime~I in order for the chain to be under good solvent conditions. It must also satisfy the constraint $g_T < \ell < N_b$ in Regimes~II and~III. On the other hand it is possible for  $g_c >N_b$  in Regimes~I and~II, but must satisfy $g_c < \ell < N_b$ in Regime~III. }}
\label{tab:parameters}
\vskip-10pt   
\end{table}
\endgroup

{\subsection{\label{subsec:sims} Simulation details}}

The protocol described below is followed with regard to the selection of parameter values {listed in Table~\ref{tab:parameters}}. For any choice of values of $N_b$ and $\epsilon_{bb}$, single chain simulations are carried out to determine $R_{g0}$ and the end-to-end vector $R_{e0}$. Note that the finite extensibility parameter is set to $Q_0 = \sqrt{50}$ in all the simulations reported here. The size of the simulation box is then fixed at $L = 2 R_{e0}$ to ensure that chains do not overlap with themselves in the periodic cell. Once $L$ is determined, the monomer concentration $c$ (and consequently $g_c$) can be adjusted by choosing the number of chains $N_c$ in a simulation cell, since $c = (N_c \times N_b)/L^3$. The scaled concentration $c/c^*$ (with $c^*$ defined in Eq.~(\ref{Eq:c*})) can also then be estimated. Finally, the choice of the number of stickers $f$ per chain determines the number of spacer monomers $\ell$ between stickers. In this manner, the relative magnitudes of $\ell$, $g_T$ and $g_c$ can be varied to probe each of the three scaling regimes that arise when the backbone monomers are under good solvent conditions. In the case when the backbone monomers are under $\theta$-solvent conditions, two different procedures are followed here. In the first, we set $\epsilon_{bb} = \epsilon_\theta = 0.45$ in the SDK potential to reproduce $\theta$ conditions for the backbone, and in the second, we neglect excluded volume (EV) interactions altogether, i.e., we simulate ghost chains that can cross themselves and each other.  

Once the parameter choices are made, a typical simulation consists of a pre-equilibration run of about $3$ to $4$ Rouse relaxation times for a system of chains with only backbone monomers and no stickers, followed by the introduction of stickers and an equilibration run of about $5$ to $8$ Rouse relaxation times. Finally, sampling is carried out over a production run of about $5$ Rouse relaxation times. Time averages, from each independent trajectory, are calculated during the production run, from a set of data collected at intervals of 1000 to 5000 non-dimensional time steps between sampling points. Ensemble averages and error of mean estimates of different equilibrium properties are then computed over a collection of such independent time averages, evaluated from 64 to 128 independent trajectories. All simulations have been carried out with a non-dimensional time-step $\Delta t=0.001$. In the absence of hydrodynamic interactions, the CPU time for the BD algorithm used here to determine all the static properties, scales linearly with system size $N$, for a fixed simulation box size $L$. It should be noted, however, that when the box size is increased, for instance to accomodate chains with a larger number of beads $N_b$, there is a large change in the pre-factor for the calculation of CPU time, due to various changes in bookkeeping, such as neighbour lists and so on.

{\subsection{\label{subsec:clus} Computation of clusters}}

The estimation of the fraction of associated intra and inter-chain stickers (required for the validation of scaling relations), and the enumeration of the number of chains in a cluster (required for the identification of the gelation transition), are both carried out here with the help of the cluster computation algorithm proposed by~\citet{SevOtt1988} A brief description of the application of the algorithm in the context of sticky polymer solutions is given here. 

To compute the intra-chain and inter-chain associations between stickers, a connectivity matrix for sticky beads is constructed such that, for any pair of stickers $i$ and $j$, the corresponding element in the connectivity matrix has a value equal to 1 for direct contact ($r_{ij} \le r_c$) and 0 otherwise. Clearly, in general, there can also be stickers which are not in direct contact but still belong to the same cluster through indirect contacts. The~\citet{SevOtt1988} algorithm also takes this into account and generates a reduced connectivity matrix, where each linearly independent column of the matrix represents a cluster of stickers which are either in direct or indirect contact. The total number of such independent columns gives the number of clusters in the system. In the simulations carried out here, however, there are no indirect contacts between stickers since they always associate in pairs (the functionality of stickers has been chosen to be one). All the necessary information regarding the state of intra-chain or inter-chain association, of every sticker in the system, is recovered by appropriately labelling the non-zero elements in each independent column of the reduced connectivity matrix. 

A similar connectivity matrix is also constructed for entire chains to determine whether they are either directly or indirectly connected to other chains via at least one sticky bead. Note that in this case there can be indirect contacts between chains, since there is typically more than one sticker per chain. The information on the number of chains in a cluster, or the spatial span of a cluster of chains, can be obtained from the columns of the chain connectivity matrix.
\vskip10pt

\section{\label{sec:deg_of_conv} Validation of scaling relations for degrees of conversion}

\begin{figure*}[t]
    \centerline{
    \begin{tabular}{cc}
       \includegraphics[width=66mm]{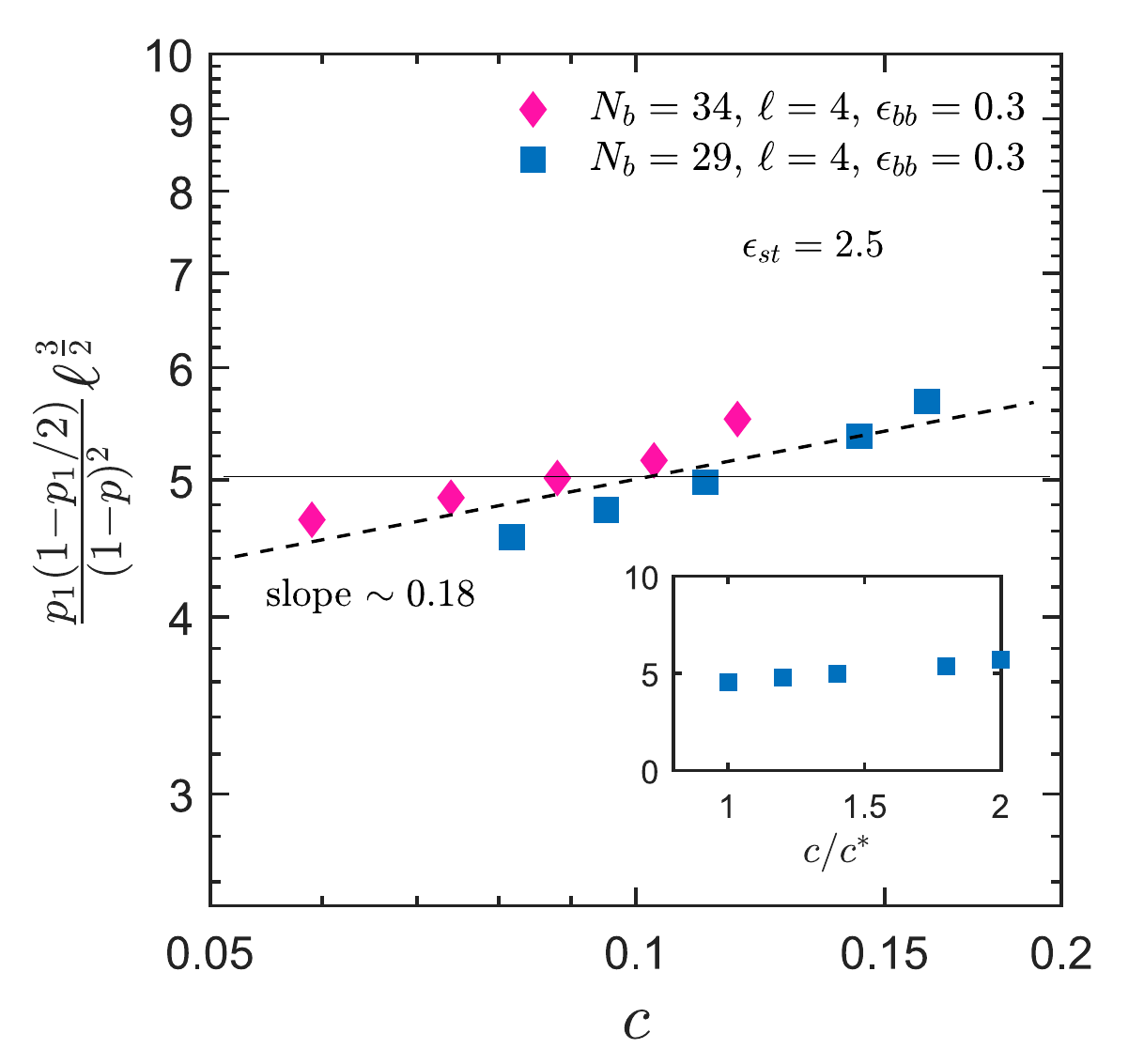} 
       &         \includegraphics[width=65mm]{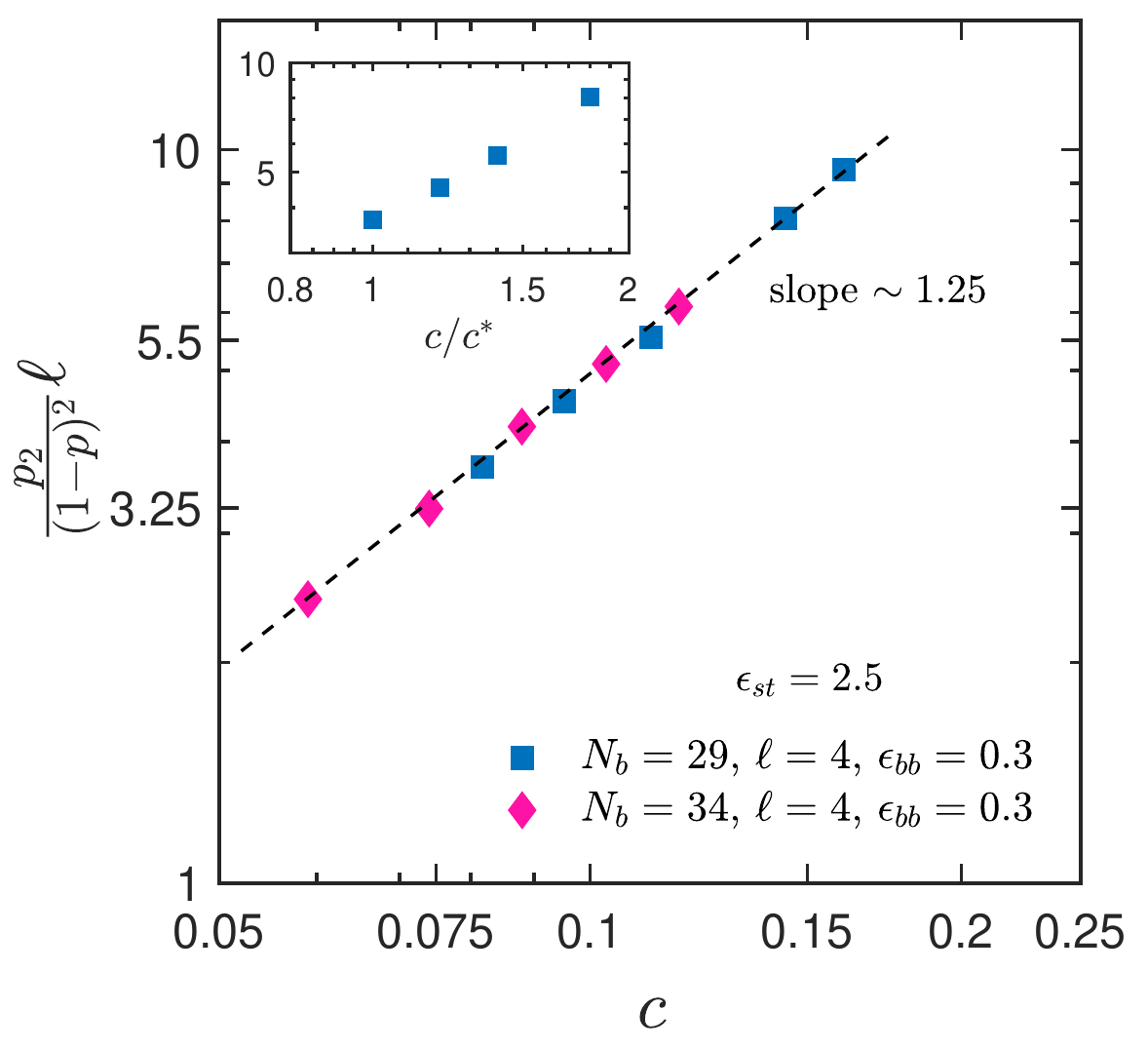} \\
          (a) &  (b) \\
        \multicolumn{2}{c}{\includegraphics[width=65mm]{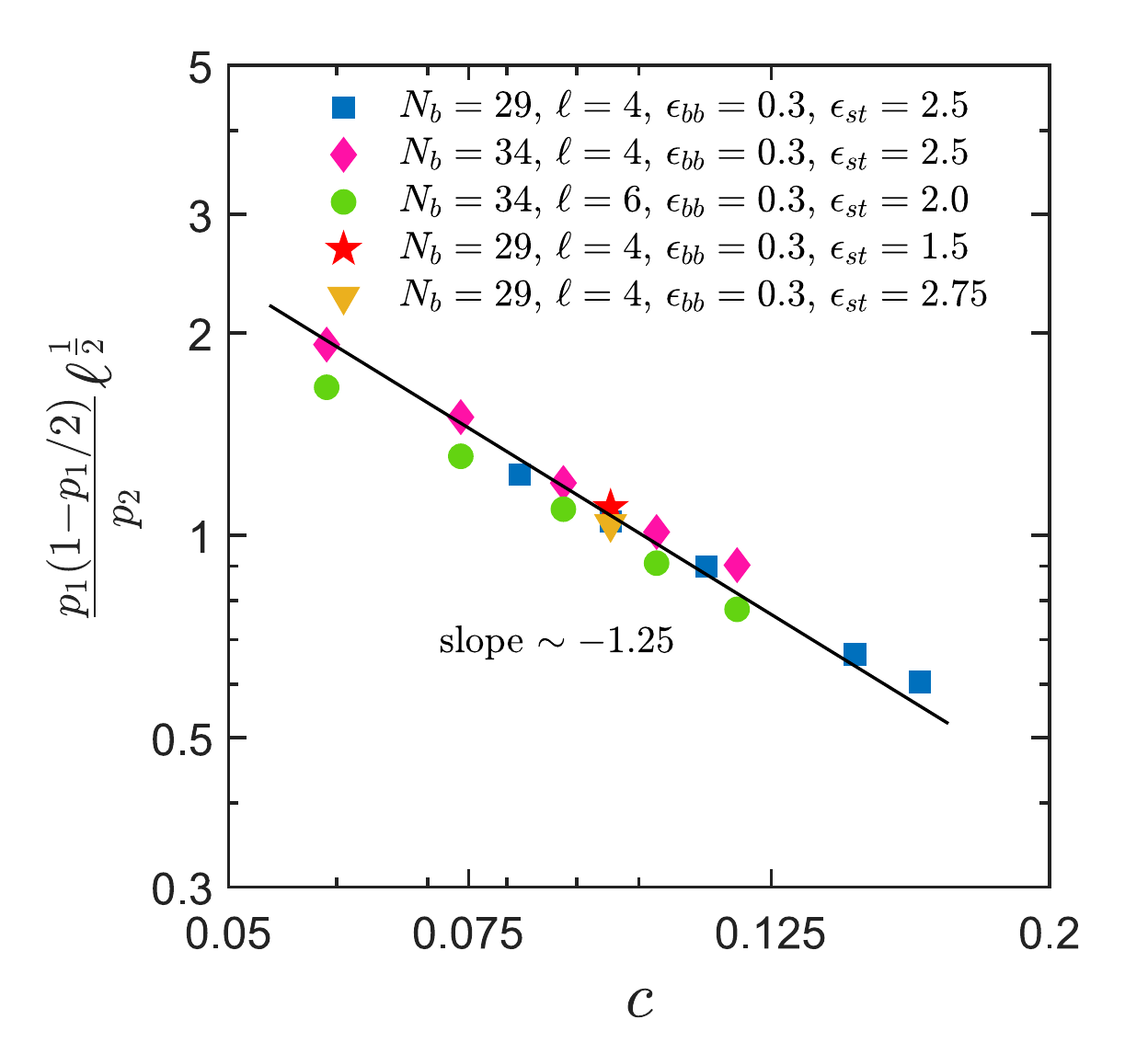}} \\
        \multicolumn{2}{c}{ (c)} \\
    \end{tabular}
    }
\caption{The dependence of ratios involving (a) intra-chain (b) inter-chain degrees of conversion, and (c) the ratio of intra-chain and inter-chain association fractions, on the monomer concentration, $c$, with the chain backbone under Regime~I ($\ell < g_T < g_c$) conditions. The spacer length dependence is absorbed in the $y$-axis. Simulations are carried out at constant solvent quality parameter $\hat \tau$, and a constant sticker strength $\epsilon_{st}$, except in~(c) where several different values of  $\epsilon_{st}$ have been considered. The values $\nu = 3/5$ and $\theta_2 = 1/3$ have been used (see Table~\ref{tab:Scaling_rel}). The dashed and the solid lines are drawn with slopes equal to the prediction by scaling theory, while symbols represent simulation data.}
\label{fig:RI_constStk}
\end{figure*}

It is clear from the values given in Table~\ref{tab:parameters} for the various simulation parameters used in the results reported here, that a more extensive variation of parameters has been carried out in Regime~II compared to the other regimes. {As discussed above}, this is essentially because of the relative ease of simulating Regime~II, both due to the physics of sticky polymer solutions, and due to the constraints of the current computational algorithm. {All the same, as will be clear from the results in this section, the simulations that have been carried out in Regimes~I and~III are still sufficient to establish the validity of the scaling relations in these regimes.} 

The scaling relations summarised in Table~\ref{tab:Scaling_rel} are examined here in two steps. We first consider the dependence of the degrees of intra and inter-chain conversion on the length of the spacer segment between stickers, $\ell$, and the monomer concentration $c$ in section~\ref{sec:ell_and_c}, followed by an examination of their dependence on the solvent quality parameter, $\hat \tau$, and the function of the effective sticker strength, $g_{ss}$ in section~\ref{sec:gss}. {The crossover behaviour from Regime~I to Regime~II, and from Regime~II to Regime~III is examined in section~\ref{sec:cross}}. The difference in the scaling of the radius of gyration with concentration between homopolymers and sticky polymers is discussed in section~\ref{sec:Rg2vc}. Finally, the special situation where the sticky polymer chain as a whole is under $\theta$-solvent conditions is examined in section~\ref{sec:thetasticky}. The data presented in this section for the dependence of $R_g$, $p_1$ and $p_2$ on the various parameters $\{N_b,\ell, f, \epsilon_{bb}, \epsilon_{st}, c, c/c^*\}$, in the form of figures, is also given in tabular form in the Supplementary Information, so that they are readily available for comparison with any model predictions that may be made in the future. 

\subsection{\label{sec:ell_and_c} Dependence on spacer length and monomer concentration}

The dependence of $p_1$ and $p_2$ on $\ell$ and $c$ is considered in this section, while keeping $\hat \tau$ and sticker strength $\epsilon_{st}$ constant, in each of the different scaling regimes. We first consider the case where the backbone monomers are under Regime~I conditions, followed by a consideration of Regimes~II and III. The case of $\theta$-solvent conditions for the backbone is examined simultaneously with Regime~II.

\subsubsection{\label{sec:Regime_I} Regime~I}

The validity of scaling predictions for Regime~I, as given in the  {first} row of Table~\ref{tab:Scaling_rel}~(b) with $\ell < g_T < g_c$, with $\hat \tau$ and sticker strength $\epsilon_{st}$ held constant, are shown in Figs.~\ref{fig:RI_constStk}. It should be noted that in order to express the ratio involving intra-chain and inter-chain associations only as a function of concentration, $c$, the spacer length ($\ell$) dependence has been absorbed into the $y$-axis.  {According to the prediction of scaling theory in Regime~I, the ratio $\left[ p_1 (1-p_1/2)/(1-p^2) \right] \, \ell^{3/2}$ is expected to be independent of monomer concentration $c$. On the other hand, Figure~\ref{fig:RI_constStk}~(a) appears to suggest a weak dependence of this ratio on concentration. It should be noted that while the fraction of intra-chain associated stickers $p_1$ decreases with increasing concentration, the total fraction of associated stickers $p$ increases with increasing concentration due to the dominant role played by the increase in the fraction of inter-chain associated stickers $p_2$ with increasing concentration. This might be responsible for the observed weak dependence, but simulations for larger chain lengths would be needed to confirm that it is not a result of finite size effects.} According to scaling theory, the ratio $ \left[ p_2/(1-p^2) \right]  \, \ell$ increases with concentration in this regime with an exponent of 1.25 (when $\theta_2$ is chosen to be 1/3), which is validated by the simulation results displayed in Fig.~\ref{fig:RI_constStk}~(b). Finally,  {as can be seen in Fig.~\ref{fig:RI_constStk}~(c), the ratio of these conversions, $\left[ p_1 (1-p_1/2)/p_2 \right]  \, \ell^{1/2}$, also largely follows the predicted dependence on concentration. This can be expected given the weak dependence of the ratio involving $p_1$ on concentration, and the dominant role played by the ratio involving $p_2$} . Apart from the overall agreement between the predictions and simulation results, there are a few other observations worth noting.

 {In Regime~I, only the ratio involving $p_2$ depends on the des Cloizeaux exponent $\theta_2$, as reflected in its dependence on the monomer concentration $c$ (see first row in Table~\ref{tab:Scaling_rel}~(a)). It is striking to observe that the choice of value of $\theta_2 = 1/3$, derived by~\citet{Dob}, leads to a collapse of data for all the simulation parameters examined in Figs.~\ref{fig:RI_constStk}~(b). } This observation is more rigorously illustrated in the scaling behaviour for Regime~II which is investigated for a wider range of parameters. 

\begin{figure*}[t]
    \centerline{
    \begin{tabular}{cc}
       \includegraphics[width=66mm]{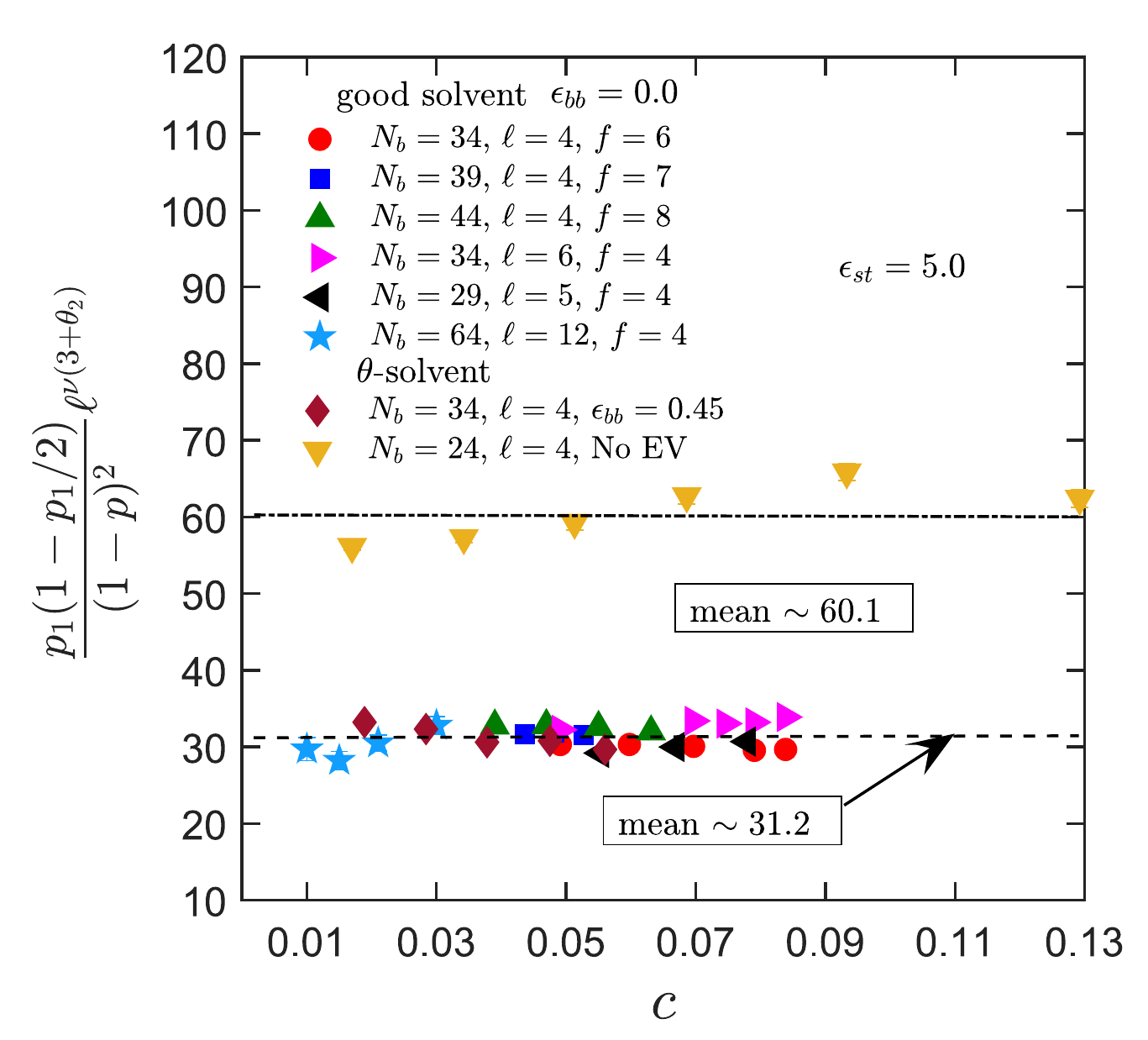} 
       &         \includegraphics[width=63mm]{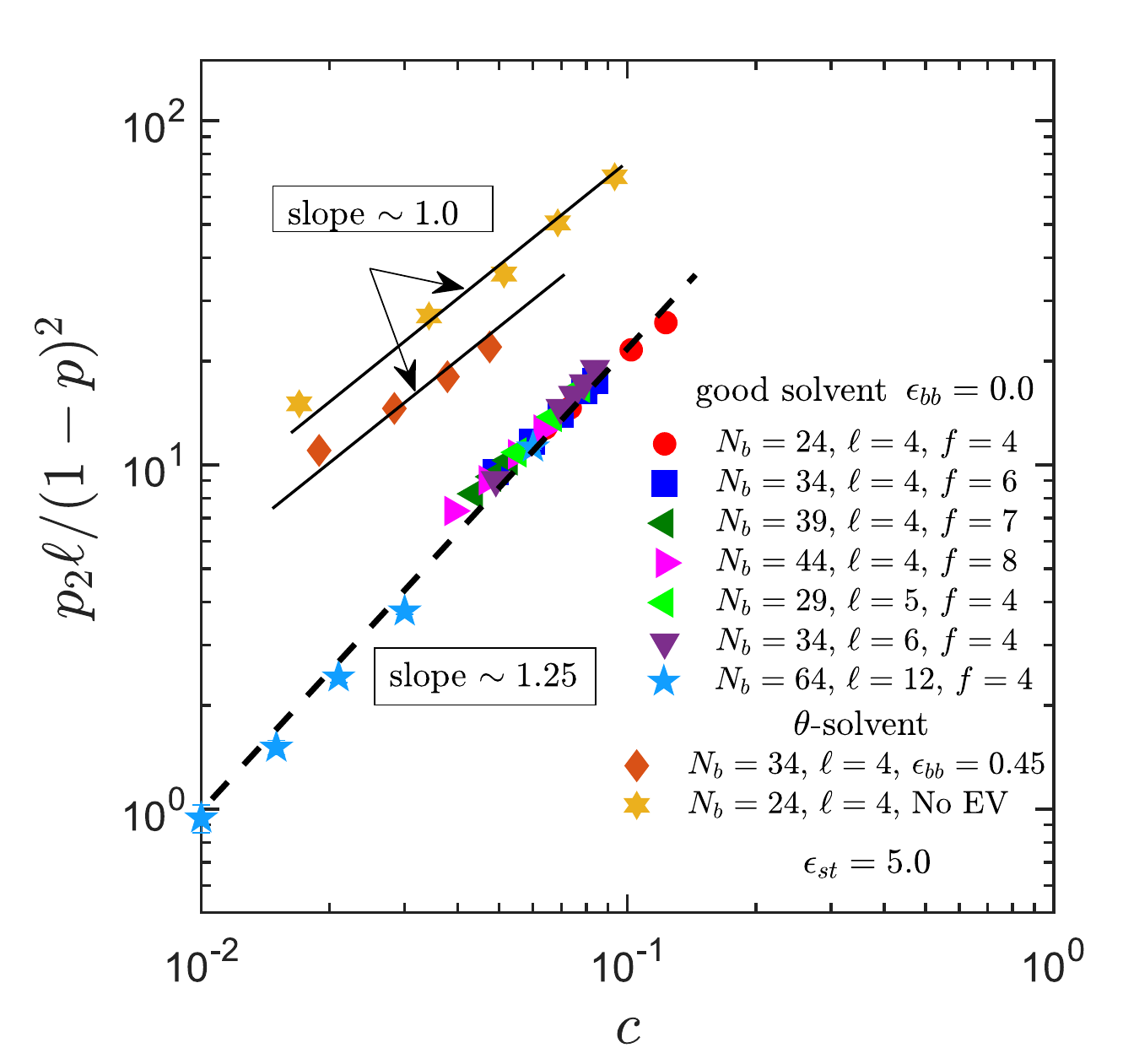} \\
          (a) &  (b) \\
        \multicolumn{2}{c}{\includegraphics[width=65mm]{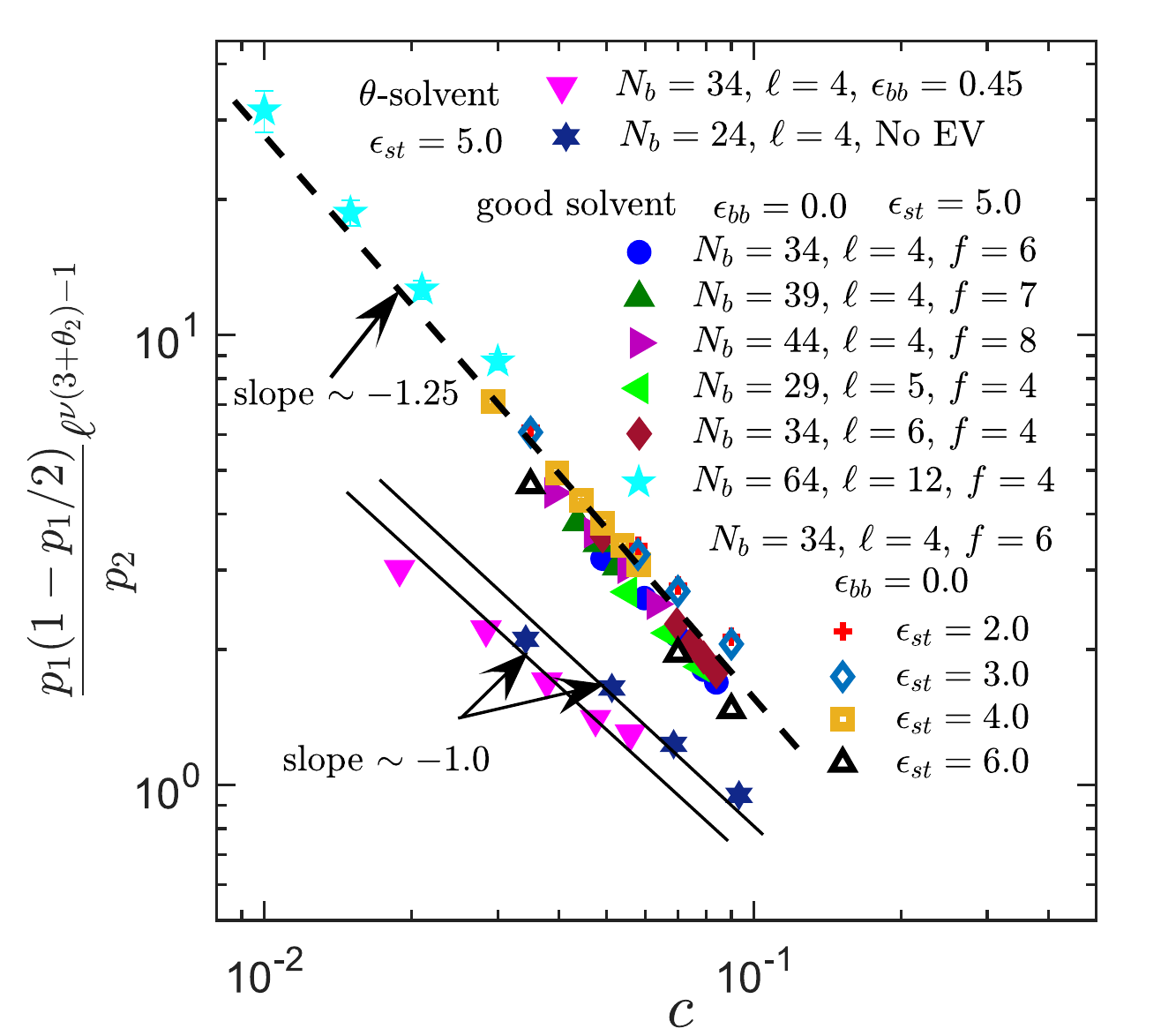}} \\
        \multicolumn{2}{c}{ (c)} \\
    \end{tabular}
    }
  \caption{The dependence of ratios involving (a) intra-chain (b) inter-chain degrees of conversion, and (c) the ratio of intra-chain and inter-chain association fractions, on the monomer concentration, $c$, with the chain backbone under $\theta$-solvent and Regime~II ($g_T < \ell < g_c$) conditions. The spacer length dependence is absorbed in the $y$-axis. Simulations are carried out at constant solvent quality parameter $\hat \tau$ and sticker strength $\epsilon_{st}$, except in~(c) where several different values of  $\epsilon_{st}$ have been considered. The values of $\theta_2$ and $\nu$ are $0$ and $1/2$, respectively for the backbone under $\theta$-solvent conditions, and $1/3$ and $3/5$, respectively, for chains with $\epsilon_{bb}=0$ (see Table~\ref{tab:Scaling_rel}). The dashed and the solid lines are drawn with slopes equal to the prediction by scaling theory, while symbols represent simulation data.}
\label{fig:RII_constStk}
\end{figure*}

Even though the spacer length $\ell$ has been absorbed into the $y$-axis for consistency with the representation in other regimes as shown later, the dependence on $\ell$ has not been examined in Figs.~\ref{fig:RI_constStk}{~(a) and~(b)} since all the simulations have been carried out for a single value of $\ell =4$. This is because, {as discussed earlier in section~\ref{sec:scaling},} changing $\ell$ changes the effective sticker strength $g_{ss}$, even if $\epsilon_{st}$ is held constant. {Interestingly, however, according to scaling theory, the ratio of intra-chain and inter-chain association fractions, $\left[ p_1 (1-p_1/2)/p_2 \right]  \, \ell^{1/2}$, is independent of sticker strength since both intra-chain and inter-chain association fractions have the same dependence on $g_{ss}$. As a result, different values of $\ell$ and $\epsilon_{st}$ should have no influence on the value of this ratio. This is demonstrated in Fig.~\ref{fig:RI_constStk}~(c) for two values of  $\ell$, and a few different values of $\epsilon_{st}$. A similar collapse of data for a wider range of values of $\ell$, and $\epsilon_{st}$ is demonstrated for this ratio in Regime~II below.}

It should be noted that when there are many values of chain length $N_b$ involved in the same plot, it is not possible to plot the dependence of the ratios involving $p_1$ and $p_2$ on $c/c^*$, since $c^*$ depends on $N_b$. The dependence on $c/c^*$ for a single value of $N_b$ is consequently shown in the insets to Figs.~\ref{fig:RI_constStk}~(a) and~(b), to give an idea of the range of values of the scaled concentration that have been examined here. The range of values of $c/c^*$ examined in all the scaling regimes is also indicated in Table~\ref{tab:parameters}. 

\subsubsection{\label{sec:theta_and_II} $\theta$-solvent and Regime~II}

The scaling relations corresponding to $\theta$-solvent and Regime~II conditions for backbone monomers are given in the fourth and second rows of Table~\ref{tab:Scaling_rel}~(b), respectively, and the results of simulations  in these regimes, with $\hat \tau$ and sticker strength $\epsilon_{st}$ constant, are shown in Figs.~\ref{fig:RII_constStk}. In order to display both cases in the same set of plots, the dependence on spacer length $\ell$ is absorbed into the $y$-axis in Figs.~\ref{fig:RII_constStk}~(a) and~(c), noting that $\ell^{\nu(3+\theta_2)}$ reduces to the $\theta$-solvent case for $\nu = 1/2$ and $\theta_2 =0$, and to the Regime~II case when $\nu = 3/5$ and $\theta_2 =1/3$. This substitution is not necessary for the fraction of inter-chain associations (Fig.~\ref{fig:RII_constStk}~(b)), since the dependence on $\ell$ is the same in both cases. {Interestingly, as discussed in greater detail in section~\ref{sec:gss} below, it turns out that $g_{ss}$ does not depend on $\ell$ for the special case when $\epsilon_{bb} = 0$, i.e., when the backbone monomers are in an athermal solvent. This independence enables an examination of the dependence of the degrees of conversion on $\ell$ in Regimes~II and~III, independently of $\epsilon_{st}$, unlike in  the $\theta$-solvent and Regime~I regimes.} We now consider the Regime~II and $\theta$-solvent cases in turn. 

The scaling with monomer concentration of the ratios involving the intra-chain and inter-chain degrees of conversion, under Regime~II conditions, can be seen  in Figs.~\ref{fig:RII_constStk}~(a) and~(b) to obey scaling predictions (given in the  {second} row of Table~\ref{tab:Scaling_rel}~(b)). Similar to the observation in Regime~I, the choice of value $\theta_2 =1/3$, leads to data collapse across the entire range of parameter values considered in Regime~II as well. It is clear from Table~\ref{tab:Scaling_rel}~(a) that $\theta_2$ appears in the scaling exponents for both the variables $\ell$ and $c$ in Regime~II. The impressive collapse of data seen in Figs.~\ref{fig:RII_constStk} consequently provides convincing evidence of the validity of Dobrynin's estimate of the $\theta_2$ exponent in sticky polymer solutions. 

\begin{figure*}[t]
    \centerline{
    \begin{tabular}{cc}
       \includegraphics[width=75mm]{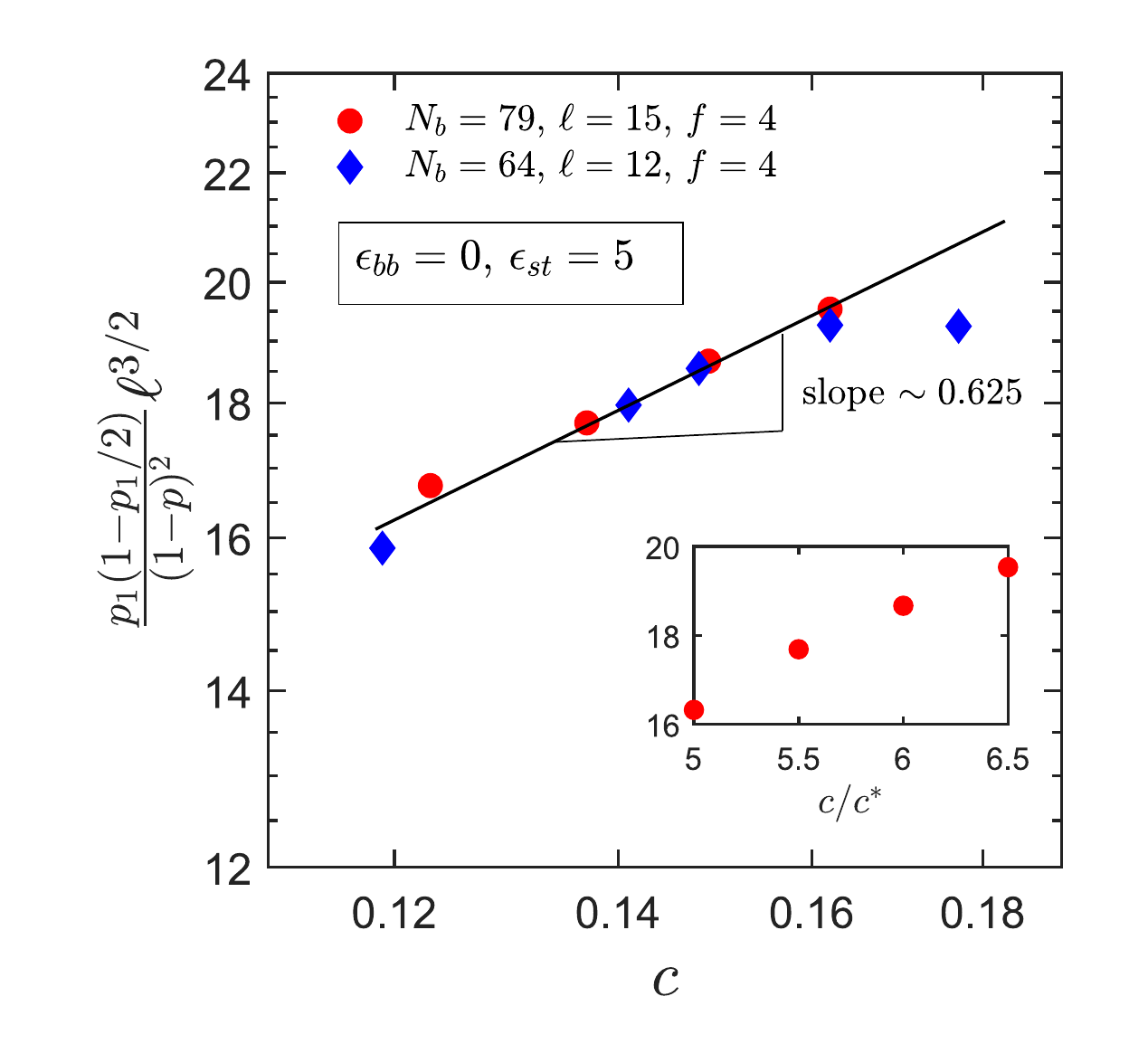} 
       &         \includegraphics[width=71mm]{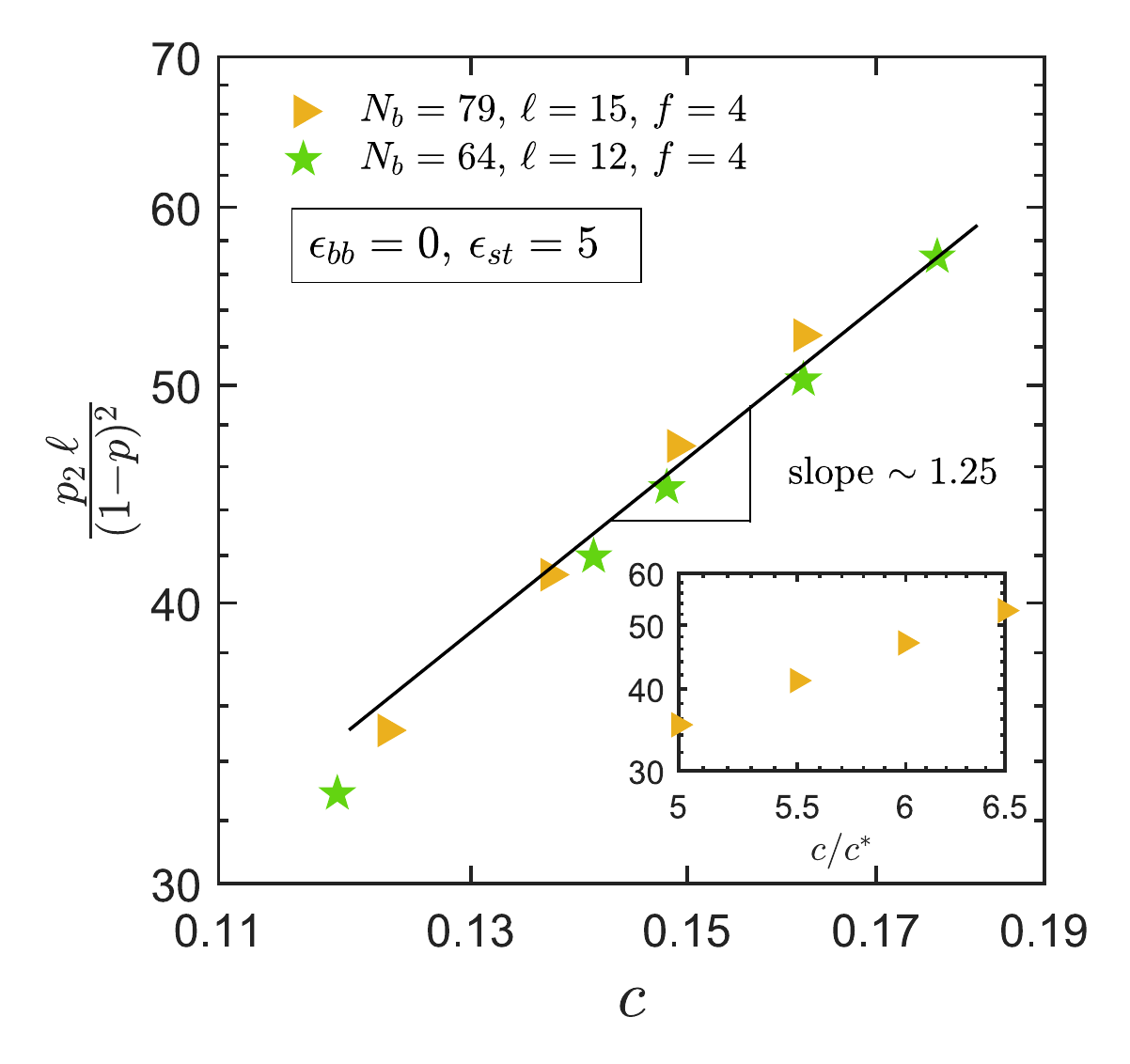} \\
          (a) &  (b) \\
        \multicolumn{2}{c}{\includegraphics[width=71mm]{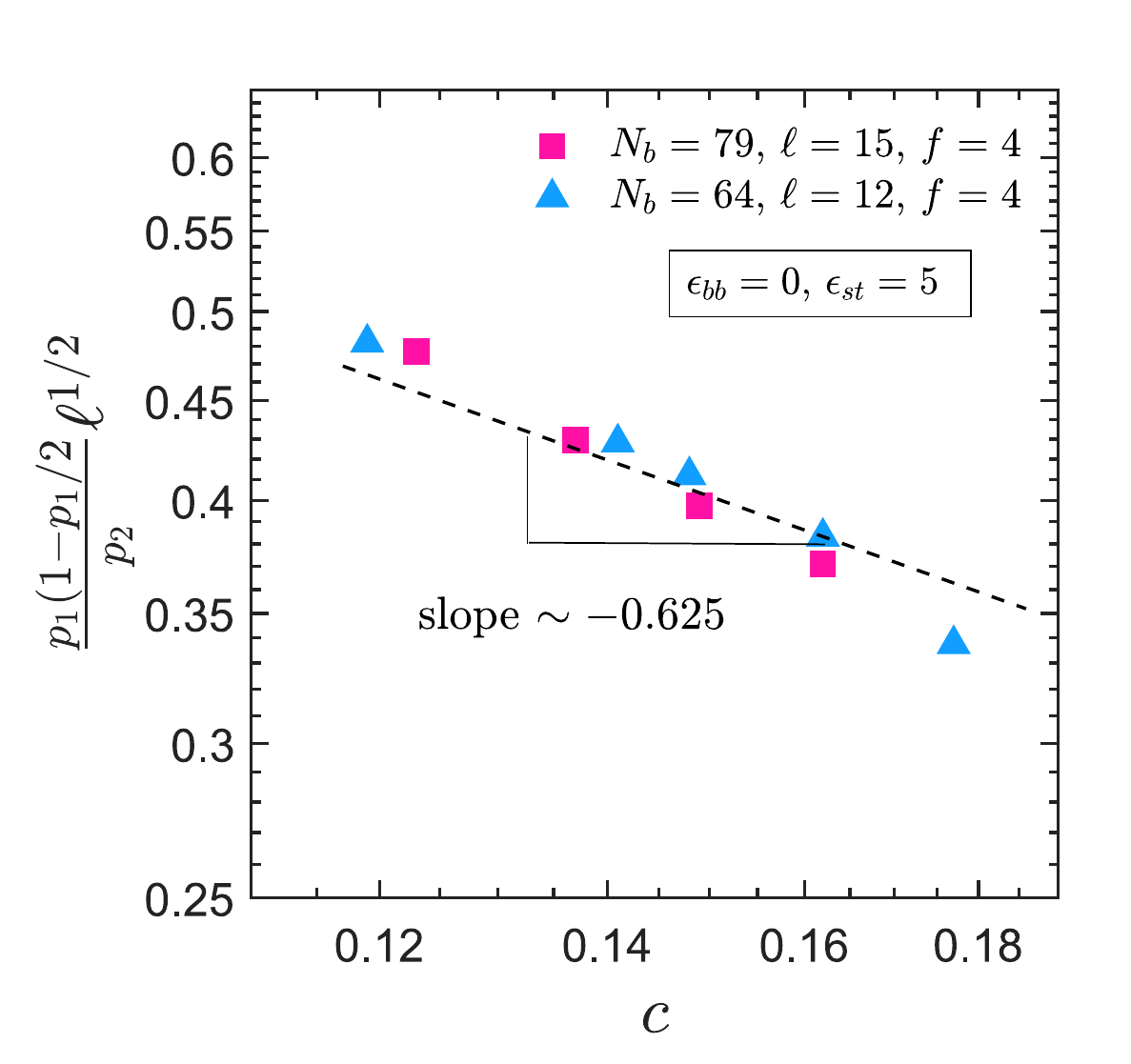}} \\
        \multicolumn{2}{c}{ (c)} \\
    \end{tabular}
    }
  \caption{The dependence of ratios involving (a) intra-chain (b) inter-chain degrees of conversion, and (c) the ratio of intra-chain and inter-chain association fractions, on the monomer concentration, $c$, with the chain backbone under Regime~III conditions. The spacer length dependence is absorbed in the $y$-axis. Simulations are carried out at constant solvent quality parameter $\hat \tau = 1$, and constant sticker strength, $\epsilon_{st} = 5.0$. The values of $\theta_2$ and $\nu$ are $1/3$ and $3/5$, respectively  (see Table~\ref{tab:Scaling_rel}). Insets in (a) and (b) show the range of concentration in terms of $c/c^*$. The dashed and the solid lines are drawn with slopes derived from the prediction and the symbols are simulation data.}
\label{fig:RIII_constStk}
\end{figure*}

Another observation from Figs.~\ref{fig:RII_constStk}, which is common to both the intra and inter-chain association fractions in Regime~II, is that the collapse of data for different values of $\ell$ indicates that the dependence on spacer length  is captured accurately by scaling theory. {Note that the exponent of $\ell$ in the ratio involving $p_1$ is 2 (for $\nu = 3/5$ and $\theta_2 =1/3$), and as a consequence, a variation of $\ell$ between 4 and 12 in Figs.~\ref{fig:RII_constStk}~(a) represents an exploration over  a considerably wide range of the values of $\ell$.} Concurrently, Fig.~\ref{fig:RII_constStk}~(c) shows that the ratio of intra-chain and inter-chain conversions, $\left[ p_1 (1-p_1/2)/p_2 \right]  \, \ell$, also follows the predicted scaling. Moreover this ratio is independent of the effective sticker strength, since both intra and inter-chain degrees of conversion have an identical dependence on $g_{ss}$. As can be seen from Fig.~\ref{fig:RII_constStk}~(c), this prediction is supported by simulations that show data collapse for  {several} different values of $\epsilon_{st}$.

Recall that the case with backbone monomers under $\theta$-solvent conditions has been simulated here with two different approaches. The first is to neglect excluded volume interactions altogether, and to treat,  as is commonly done, the simulation of $\theta$ conditions to be identical to simulating ideal (or ghost) chains that can cross each other. The second approach is to use a value of $\epsilon_{bb} = \epsilon_{\theta} = 0.45$, which has been shown to 
reproduce scaling predictions for homopolymer chains consistent with $\theta$-solvent conditions~\cite{Aritra2019}. As can be seen from Table~\ref{tab:Scaling_rel}~(b), scaling theory predicts that in this case as well, the ratio $\left[ p_1 (1-p_1/2)/(1-p^2) \right]  \, \ell^{3/2}$ is independent of monomer concentration. Fig.~\ref{fig:RII_constStk}~(a), which displays the results of the two approaches, demonstrates the validation of this prediction. When the SDK potential with $\epsilon_{bb} = \epsilon_{\theta}$  is used, the numerical value of the ratio is identical to that for the Regime~II case (with the appropriate scaling with $\ell$ taken into account). On the other hand, the value of the ratio is higher for the case of ideal chains. As will be demonstrated in section~\ref{sec:gss}, this difference arises from a difference in the function $g_{ss}$ in the two cases.

The exponent $\theta_2$ is not relevant for backbone monomers under $\theta$-solvent conditions, and according to Table~\ref{tab:Scaling_rel}~(b), scaling theory predicts that the ratio $ \left[ p_2/(1-p^2) \right]  \, \ell$ increases linearly with concentration in this case. As can be seen from Fig.~\ref{fig:RII_constStk}~(b), this prediction is validated by both the approaches used here to simulate a backbone chain under $\theta$-solvent conditions. It is clear from Fig.~\ref{fig:RII_constStk}~(c) that the ratio of intra and inter-chain degrees of conversion, $\left[ p_1 (1-p_1/2)/p_2 \right]  \, \ell^{1/2}$, also follows the predicted dependence on monomer concentration $c$. A discussion of the dependence on the variables $\ell$ and $\epsilon_{st}$, in this case, is postponed to section~\ref{sec:gss}. As will be seen subsequently, the values of concentrations depicted in Figs.~\ref{fig:RII_constStk}~(a) and~(b) and listed in Table~\ref{tab:parameters} for the good solvent backbone are well into the regime where the sticky polymer solution is in the gel phase. As a consequence, all the simulation results presented here so far, clearly indicate that the scaling relations hold true in both the sol and gel phases, and as pointed out by~\citet{Dob}, do not distinguish between them.

\subsubsection{\label{sec:Regime_III} Regime~III}

 {The constraints associated with carrying out simulations in Regime~III ($g_T < g_c <  \ell$)  have been detailed in section~\ref{subsec:simcon}. }  We have considered two values of chain length, $N_b = 64$ and $79$, respectively,  {in this regime}, with spacer lengths $\ell = 12$ and $15$.   {According to Eq.~(\ref{Eq:gTgc}), with $N_b = 64$, $g_c < 11$, for $c/c^* > 4$, while with $N_b = 79$, $g_c < 14$, for $c/c^* > 4$. Thus, with the number of monomers in a thermal blob $g_T = 1$ (since $\epsilon_{bb} =0$), any value of $c/c^*$ in the range $4 < c/c^* < 28 \, (33)$ (corresponding to $g_c \approx 1$ for $N_b = 64 \, (79)$) would correspond to Regime~III conditions for both these values of chain length.} As indicated in the insets to Figs.~\ref{fig:RIII_constStk}~(a) and~(b) and Table~\ref{tab:parameters}, a range of values of $c/c^*$ from {4} to 6.5 has been simulated here to explore Regime~III, due to limitations of the computational cost for simulating larger values of $c/c^*$. 

Similar to the simulation results in Regime~I and II, Figs.~\ref{fig:RIII_constStk} show that the scaling of the intra-chain and inter-chain conversion ratios with monomer concentration  in Regime~III, at constant $\hat{\tau}$ and $\epsilon_{st}$, are also in good agreement with the theoretical prediction (given in the  {third} row of Table~\ref{tab:Scaling_rel}~(b), assuming $\theta_2=1/3$).  {The absorption of the dependence on spacer length into the $y$-axis and the collapse of data seen for the two different values of $\ell$ simulated here, is inline with the predicted dependence on $\ell$ by scaling theory. However, it would be desirable to carry out simulations for a greater range of values of $\ell$ for a thorough validation of the scaling prediction}. The deviation from scaling theory observed in Fig.~\ref{fig:RIII_constStk}~(a) for the simulated value of the ratio involving $p_1$ at the highest value of $c$, when $N_b = 64$, suggests that there are probably too few monomers in a correlation blob ($g_c \approx 6$) for the scaling ansatz to be valid at this chain length. This is, however, not the case when $N_b = 79$, or for the ratio involving $p_2$ (at both the values of chain length used here), since it can be seen from Figs.~\ref{fig:RIII_constStk}~(a)~and~(b), respectively, that scaling predictions for both the ratios are confirmed by simulation results.

\subsection{\label{sec:gss} Dependence on solvent quality parameter and sticker strength}

\begin{figure}[t]
    \centerline{
    \begin{tabular}{c}
   {\includegraphics[width=78mm]{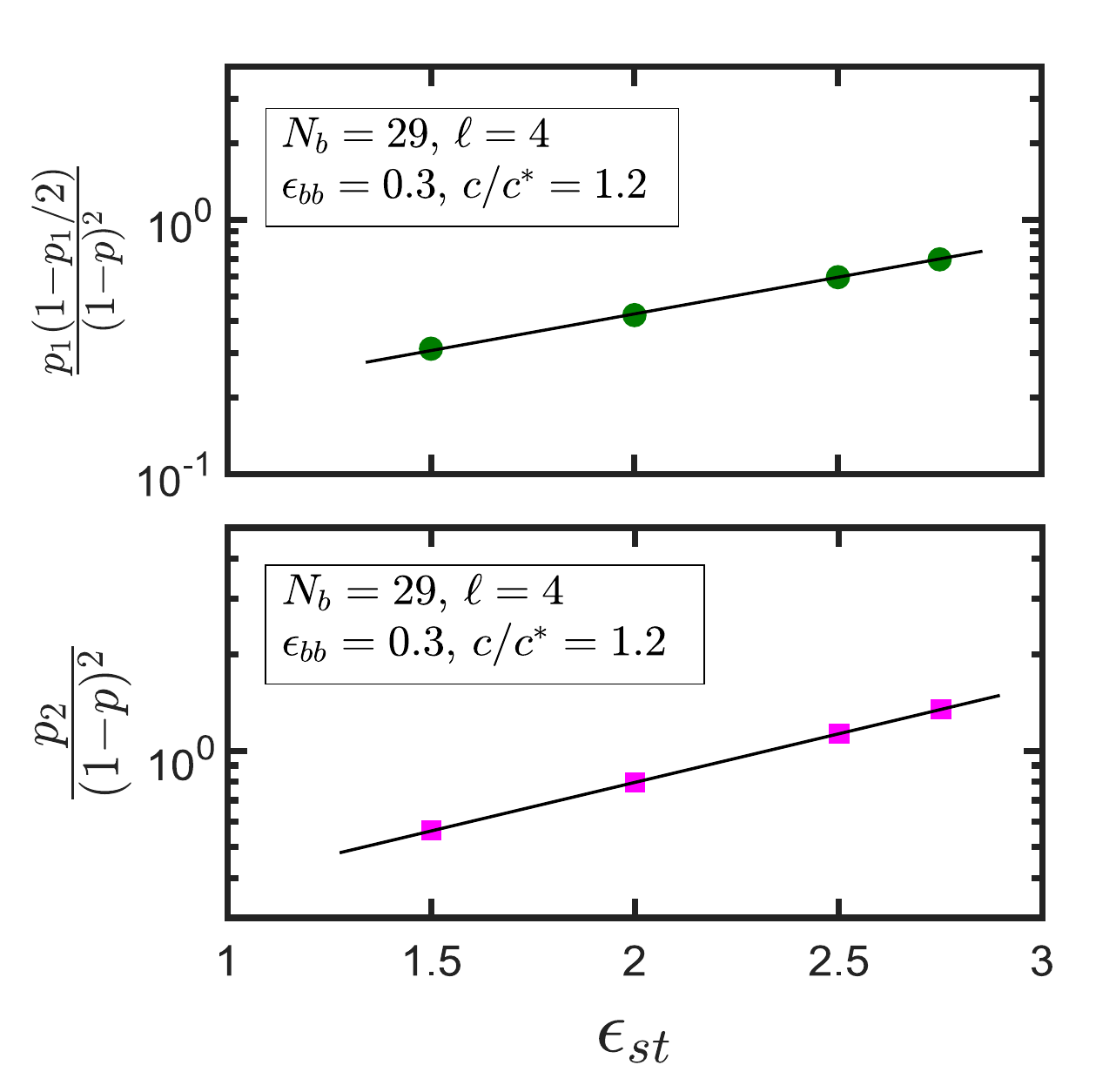}} \\
    { (a) }\\[5pt]
         \includegraphics[width=78mm]{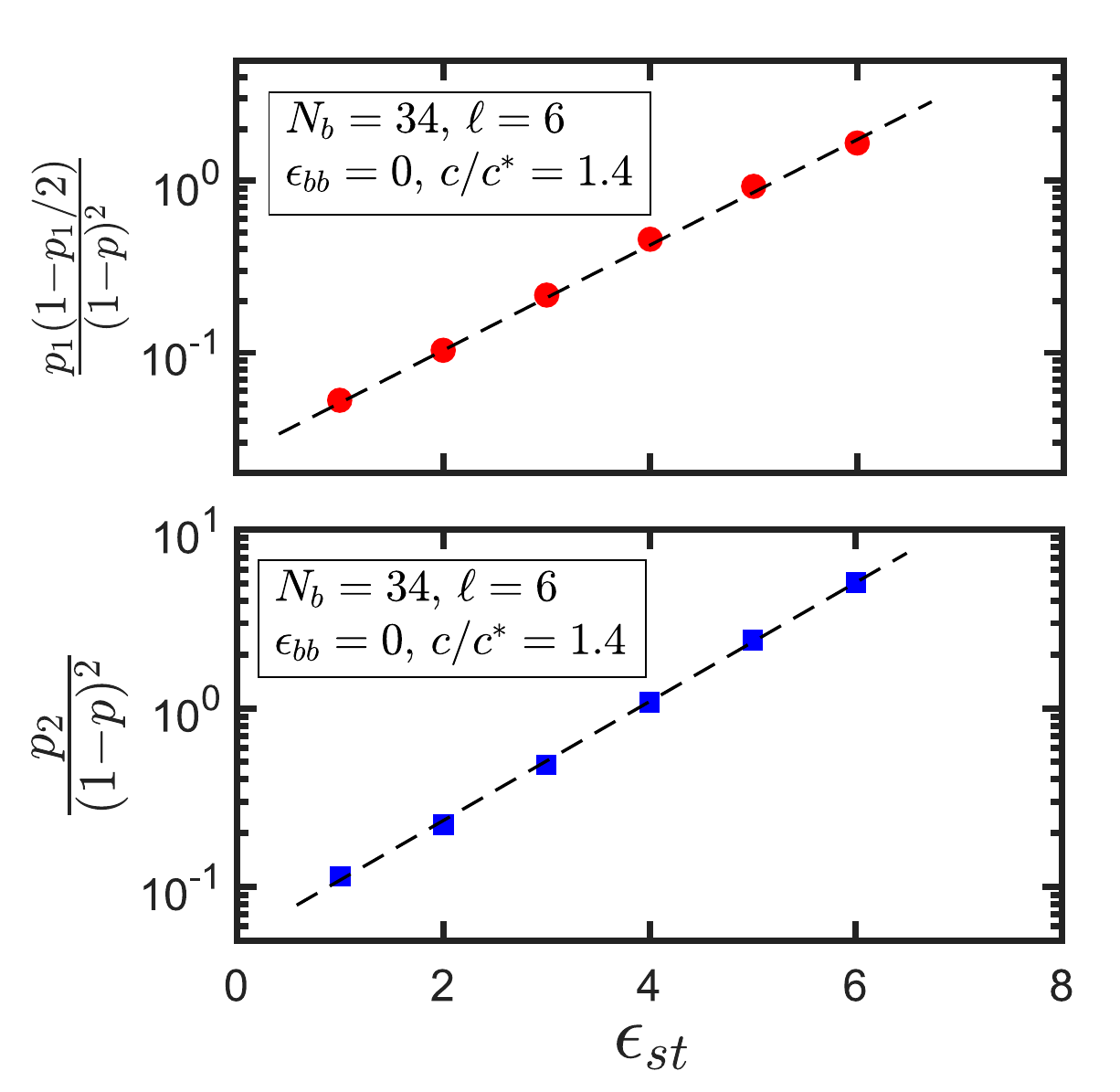}  \\
    { (b) }\\[5pt]
        \end{tabular}
    }
    	\vskip-10pt
	\caption{Scaling of the ratios involving intra-chain and inter-chain degrees of conversion as a function of sticker strength, $\epsilon_{st}$, for systems in (a) Regime~I and (b) Regime~II. Symbols represent results of simulations, while the lines are drawn to guide the eye.}
\label{fig:p1p2Sc_stk}
\end{figure}

Defining the quantities, 
\[ \alpha=\frac{\nu \, \theta_2}{3\nu-1} , \, \quad \text{and} \quad \beta=\nu(3+\theta_2), \]
the scaling relations for the ratios involving intra and inter-chain degrees of conversion given in Table~\ref{tab:Scaling_rel}  can be represented by the following common expressions in both Regimes~I and~II,
\begin{equation}
\label{Eq:common}
\begin{aligned}
\frac{p_1(1-p_1/2)}{(1-p)^2} \, \ell^{\beta_i} \, {\hat \tau}^{2 \beta_i -3} &  \sim g_{ss} \,; \quad i = 1, 2 \\[10pt]
\frac{p_2}{(1-p)^2} \, \ell \, {\hat \tau}^\alpha \, c^{-(1+\alpha)}  & \sim g_{ss}
 \end{aligned}
 \end{equation} 
where $\beta_1 = \beta - (1/2)$ applies in Regime~I, and $\beta_2 = \beta$ applies in Regime~II. 
Setting $\nu = 3/5$ and $\theta_2 = 1/3$, gives $\alpha = 1/4$ and $\beta =2$, and leads to the recovery of the simplified relations displayed in Table~\ref{tab:Scaling_rel}~(b), for these two scaling regimes. The representation of the scaling relations in the forms given in Eqs.~(\ref{Eq:common}), focusses attention on the function of effective sticker strength, $g_{ss}$. According to scaling theory~\cite{Dob}, for fixed values of backbone solvent quality parameter $\hat \tau$ (or equivalently, $\epsilon_{bb}$), and spacer length $\ell$, the ratios  involving intra and inter-chain degrees of conversion should depend exponentially on $\epsilon_{st}$. This expectation is clearly fulfilled in both the scaling regimes, as can be seen in Figs.~\ref{fig:p1p2Sc_stk}, for the particular parameter values that have been examined here.

\begin{table}[tb]
\begin{center}
\setlength{\tabcolsep}{10pt}
\renewcommand{\arraystretch}{1.2}
\begin{tabular}{| c | c | c | c |}
\hline
 $\ell$     & $\epsilon_{bb}$ 
            & $A$ 
            & $B$   
\\                
\hline
\hline
            $6$
            & $0.0$
            & $4.65$
            & $0.76$ 
\\           
            $4$
            & $0.3$
            & $6.17$
            & $0.70$
\\
            $6$
            & $0.35$
            & $5.07$
            & $0.72$
\\
\hline
\end{tabular}
\end{center}
\caption{Values of the functions $A(\epsilon_{bb},\ell)$ and $B(\epsilon_{bb},\ell)$ in Eq.~(\ref{Eq:exp_fit}), for different spacer lengths, $\ell$, and backbone monomer interaction strengths, $\epsilon_{bb}$, determined from fitting data from simulations carried out in scaling regimes~I and~II.}
\label{tab:para_A_B}
\end{table}  

 {As discussed previously, in the present implicit solvent model, the function $g_{ss}$ depends on all three variables $\epsilon_{st}$, $\epsilon_{bb}$, and $\ell$, and cannot, in general, be varied independently of spacer length $\ell$.} Here {we propose} the form, 
\begin{equation}
\label{Eq:gss}
g_{ss} = A(\epsilon_{bb},\ell)\,\exp\left[B(\epsilon_{bb},\ell)\,\epsilon_{st}\right] 
\end{equation} 
which accounts for the expected dependence on all of the three parameters. The functions $A(\epsilon_{bb},\ell)$ and $B(\epsilon_{bb},\ell)$ can be determined by fitting simulation data. Since both the ratios involving intra and inter-chain degrees of conversion have the same dependence on $g_{ss}$, they can be combined to maximise the data available for the purpose of fitting,
\begin{align}
\label{Eq:exp_fit}
 \frac{p_1(1-p_1/2)}{(1-p)^2} \, & \ell^{\beta_i} \, {\hat \tau}^{2 \beta_i -3} 
 + \frac{p_2}{(1-p)^2} \, \ell \, {\hat \tau}^\alpha \, c^{-(1+\alpha)}  \nonumber \\[10pt]
 & = 2\,A(\epsilon_{bb},\ell)\,\exp\left[B(\epsilon_{bb},\ell)\,\epsilon_{st}\right]
\end{align}
Values of the functions $A$ and $B$ obtained in this manner, for the various choices of $\epsilon_{bb}$ and $\ell$ used here, are displayed in Table~\ref{tab:para_A_B}. 

\begin{table}[tb]
\begin{center}
\setlength{\tabcolsep}{10pt}
\renewcommand{\arraystretch}{1.5}
\begin{tabular}{| c | c | c | c | c | c |}
\hline
 $\ell$     & $\epsilon_{bb}$  & $A_1$  & $B_1$ & $A_2$ & $B_2$   \\                
\hline
\hline
            $4$  & $0.45$ & $0.673$ & $0.638$ & $5.25$ & $0.781$
\\           
            $4$ & no EV & $1.614$ & $0.611$ & $9.672$ & $0.693$
\\
            $6$ & $0.45$ & $0.369$ & $0.797$ & $3.477$ & $0.879$
\\
            $6$ & no EV & $1.06$ & $0.73$ & $5.44$ & $0.87$
\\
\hline
\end{tabular}
\end{center}
\caption{Values of the functions $A_1(\epsilon_{bb},\ell)$  and $A_2(\epsilon_{bb},\ell)$, and $B_1(\epsilon_{bb},\ell)$ and $B_2\epsilon_{bb},\ell)$, for different spacer lengths, $\ell$, determined from fitting data from simulations carried out for backbone monomers under $\theta$-solvent conditions. The two approaches correspond to using the SDK potential with $\epsilon_{bb} = \epsilon_{\theta} =0.45$, and ghost chains with no excluded volume interactions.}
\label{tab:para_A1_B1}
\end{table}

\begin{figure}[b]
    \centerline{
   {\includegraphics[width=80mm]{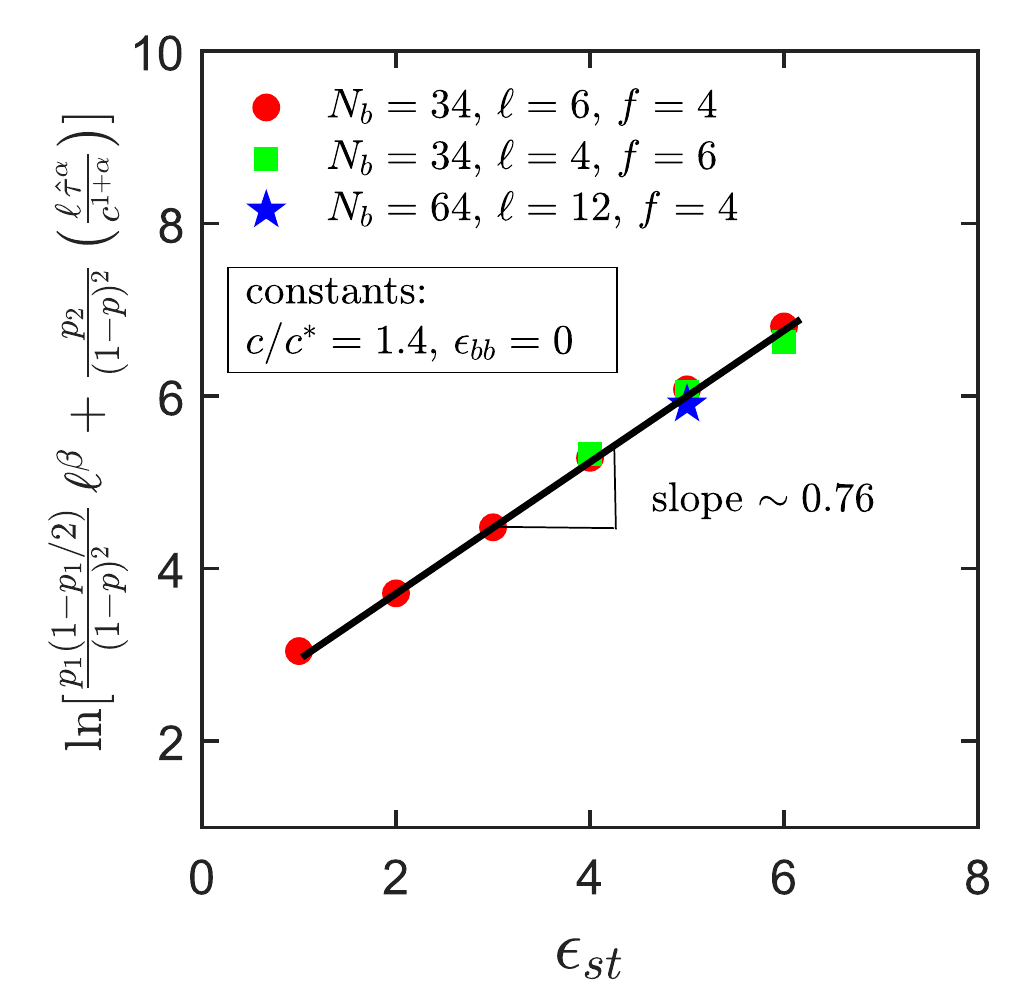}}     
    }
 \vskip-15pt
 \caption{Scaling of the sum of the ratios involving intra-chain and inter-chain association fractions (see Eq.~(\ref{Eq:exp_fit})) as a function of sticker strength, $\epsilon_{st}$, {for three systems with $N_b =34$, $\ell=6$, $f=4$; $N_b =34$, $\ell=4$, $f=6$, and $N_b =64$, $\ell=12$, $f=4$, with} $\epsilon_{bb}=0$, at $c/c^*=1.4$. The symbols are the simulation data and the solid line is an exponential fit to the data. Note that for Regime~II depicted here, $\alpha=1/4$ and $\beta=2$.}
\label{fig:p1p2_stk_l4l6}
\end{figure}

\begin{figure*}[t]
    \centerline{
    \begin{tabular}{c c}
        \includegraphics[width=77mm]{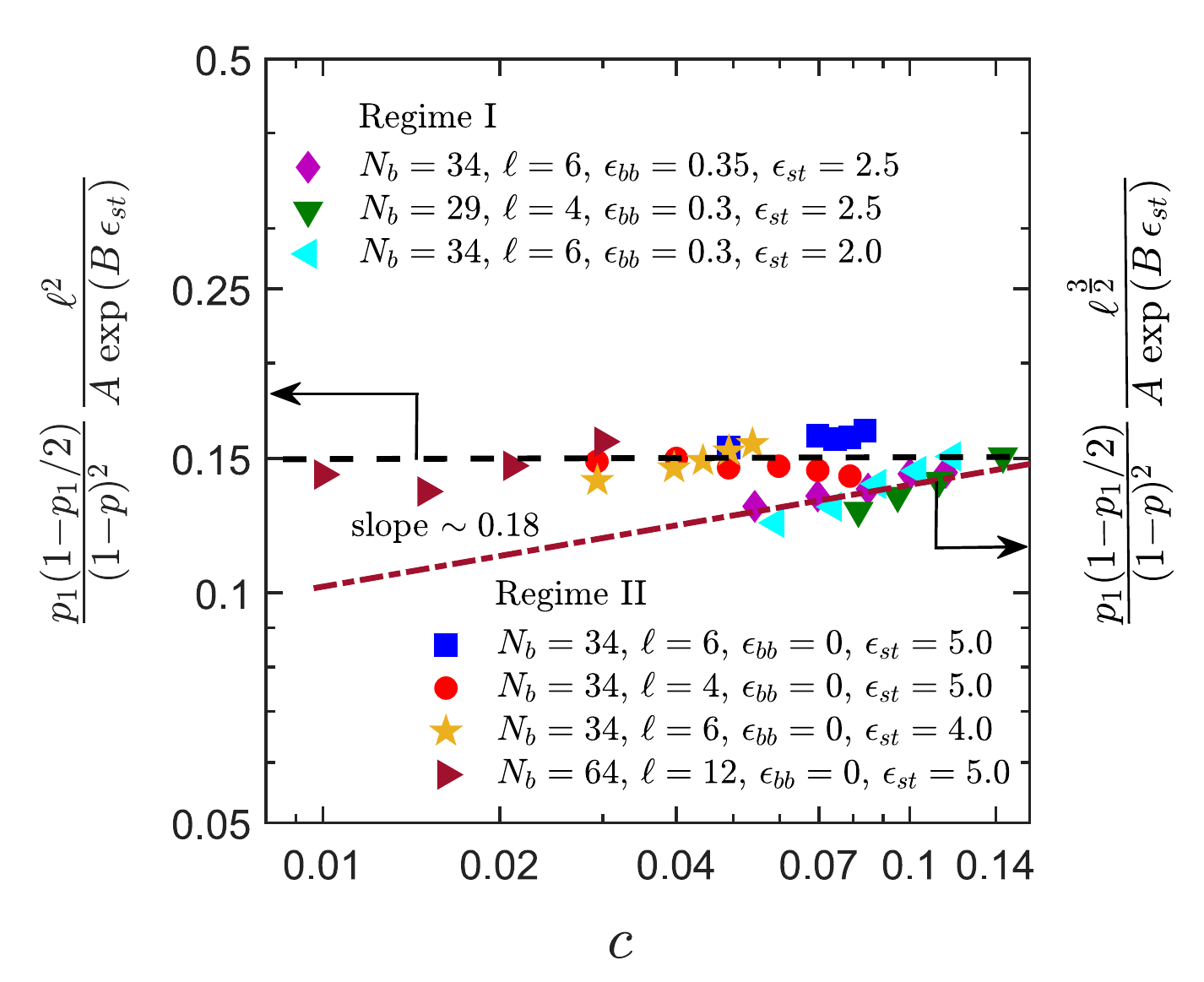} 
        & \includegraphics[width=73mm]{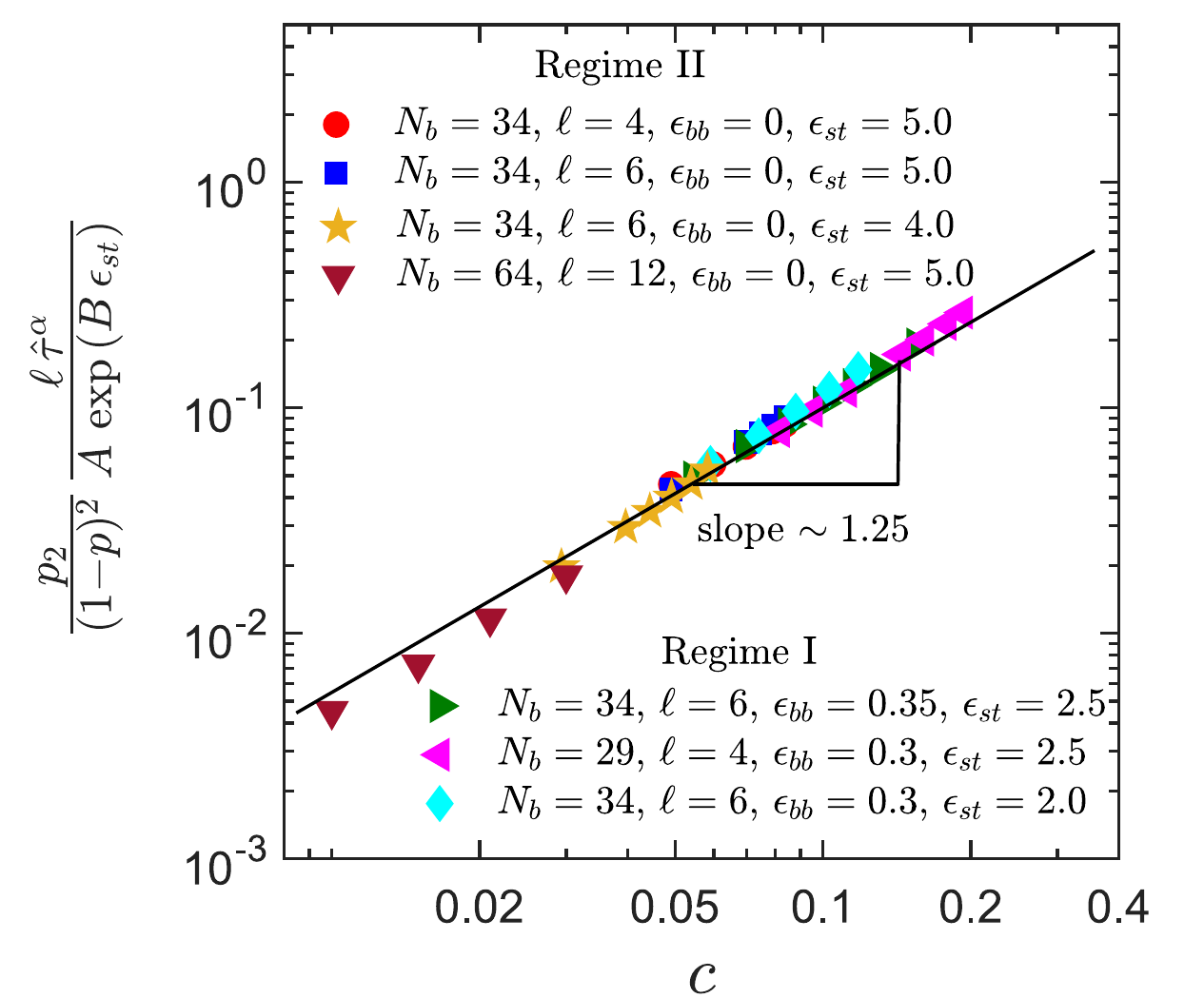} \\
        (a) & (b) \\
        \includegraphics[width=77mm]{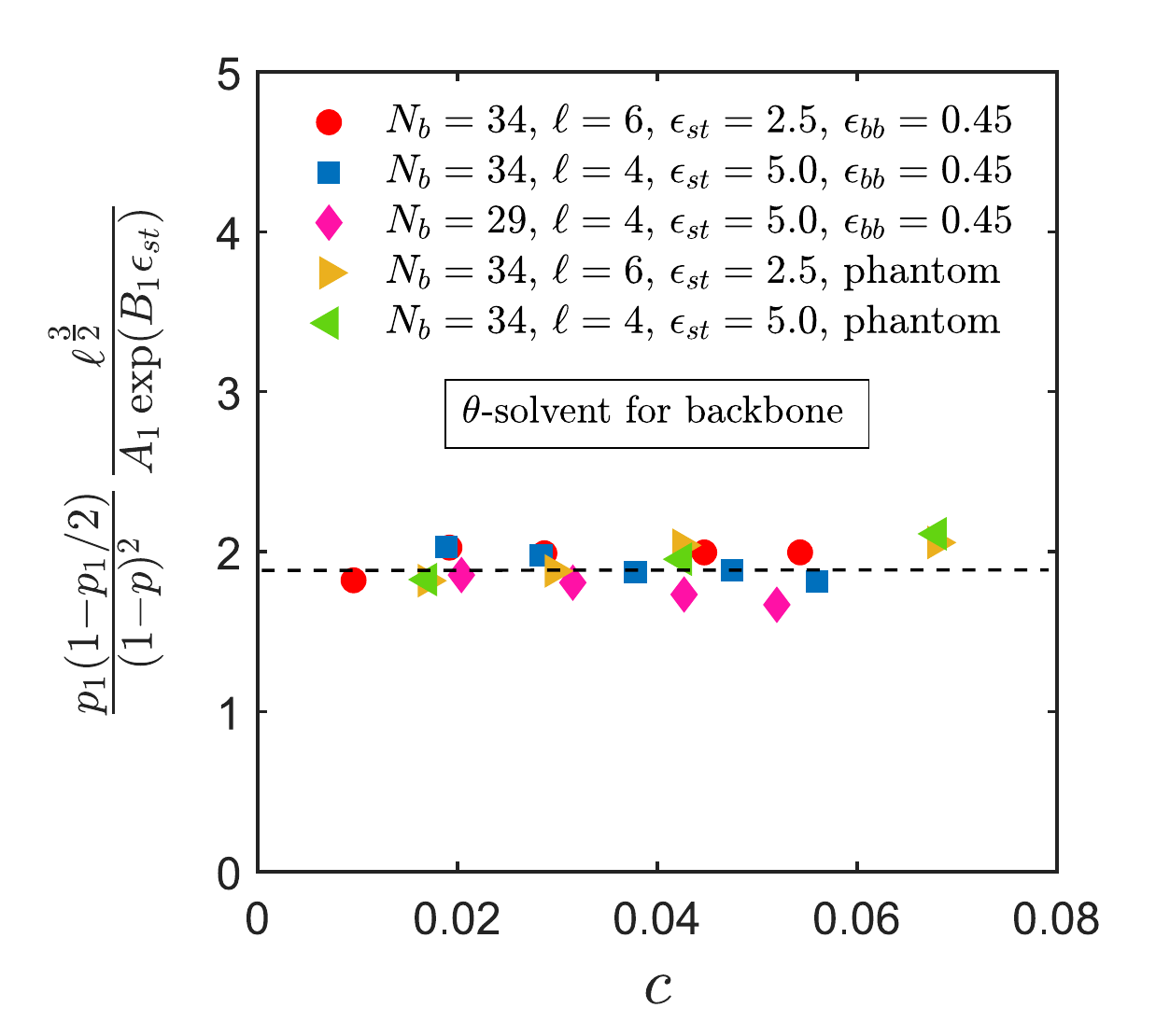} 
        & \includegraphics[width=73mm]{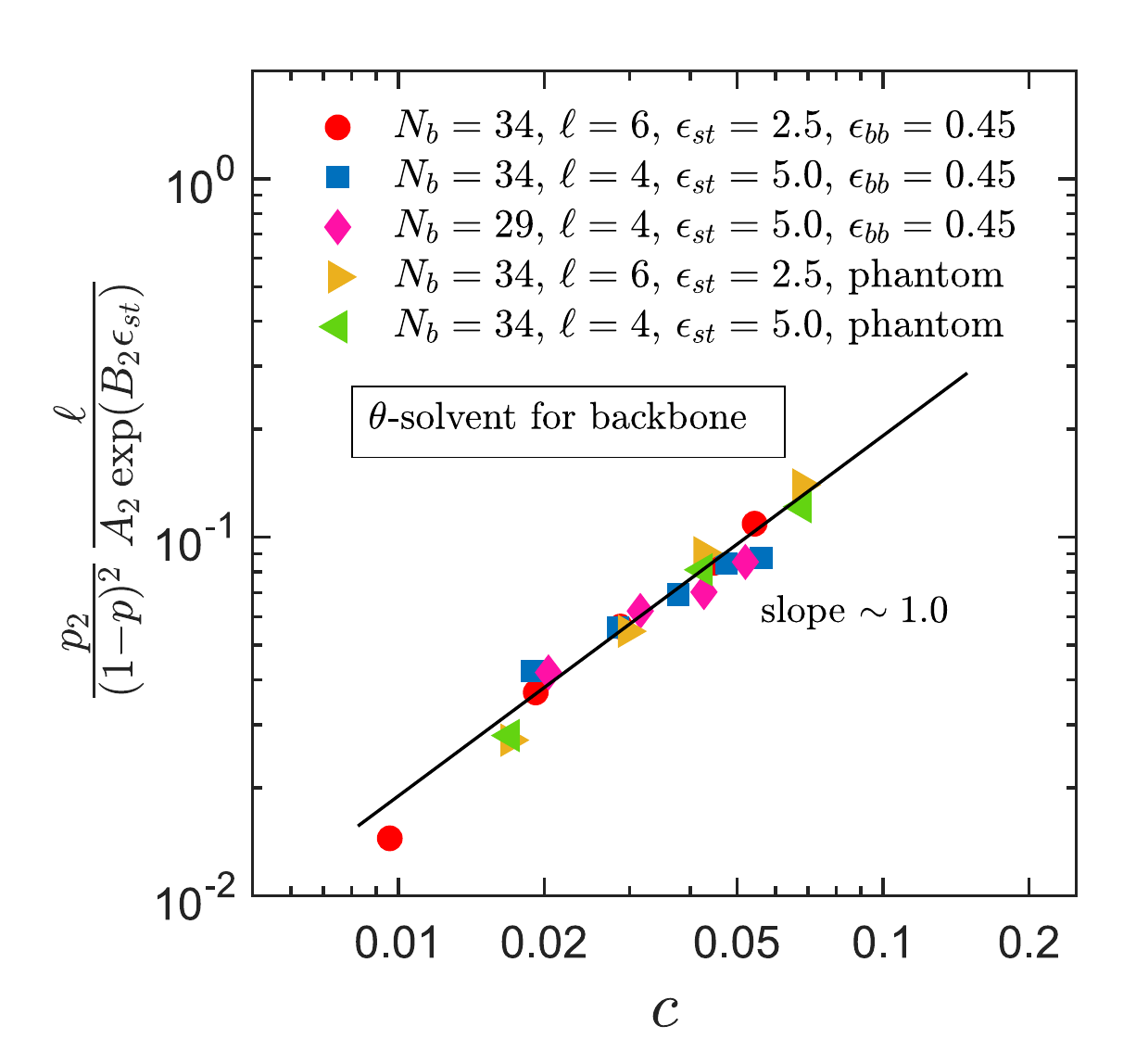} \\
        (c) & (d) \\
    \end{tabular}
    }
	\caption{Master plots demonstrating validation of scaling relations for, (i) the ratio involving the intra-chain degree of conversion $p_1$, in (a) Regimes~I and~II, and (c) $\theta$-solvent conditions for backbone monomers, and, (ii)  the ratio involving the inter-chain degree of conversion $p_2$, in (b) Regimes~I and~II, and (d) $\theta$-solvent conditions for backbone monomers, plotted as a function of monomer concentration, $c$, for different spacer segment lengths $\ell$, sticker strengths, $\epsilon_{st}$, and solution temperatures, $\hat{\tau}$. The exponent $\alpha = 1/4$. The dashed and the solid lines are drawn with slopes equal to the prediction by scaling theory, while symbols represent simulation data.}
\label{fig:p1p2Sc_R1_R2}
\end{figure*}

Simulations carried out for the case where backbone monomers are under $\theta$-solvent conditions indicate that the function $g_{ss}$ is not the same for the ratios involving intra and inter-chain degrees of conversion, and that they cannot be combined together, as was done in Eq.~(\ref{Eq:exp_fit}) for backbone monomers under good solvent conditions. The scaling relations for the two ratios in the $\theta$-solvent case, displayed in the {last} row of Table~\ref{tab:Scaling_rel}~(b), can be recovered from Eqs.~(\ref{Eq:common}) by setting $\beta_i = 3/2$ and $\alpha =0$. Using $A_1(\epsilon_{bb},\ell)$ and $B_1(\epsilon_{bb},\ell)$ to denote the functions occurring in the fit to the function $g_{ss}$ for the ratio involving $p_1$, and similarly, $A_2(\epsilon_{bb},\ell)$ and $B_2(\epsilon_{bb},\ell)$ for the ratio involving $p_2$, their estimated values are given in Table~\ref{tab:para_A1_B1}.

Interestingly, as mentioned earlier, for the case when $\epsilon_{bb} =0$ (which is the value used here to simulate the good solvent conditions corresponding to Regimes~II and~III), it can be seen from Fig.~\ref{fig:p1p2_stk_l4l6} that the function $g_{ss}$ appears to be independent of spacer length $\ell$. This lack of dependence is responsible for the collapse of data for different values of $\ell$ shown in Figs.~\ref{fig:RII_constStk}~(a) and~(b) for Regime~II, and Figs.~\ref{fig:RIII_constStk}~(a) and~(b) for Regime~III, while  the dependence of $g_{ss}$ on $\ell$ in Regime~I, and for backbone monomers under $\theta$-solvent conditions, implies that a similar collapse cannot be considered for these cases. 

\begin{figure*}[t]
    \centerline{
    \begin{tabular}{c c}
        \includegraphics[width=79mm]{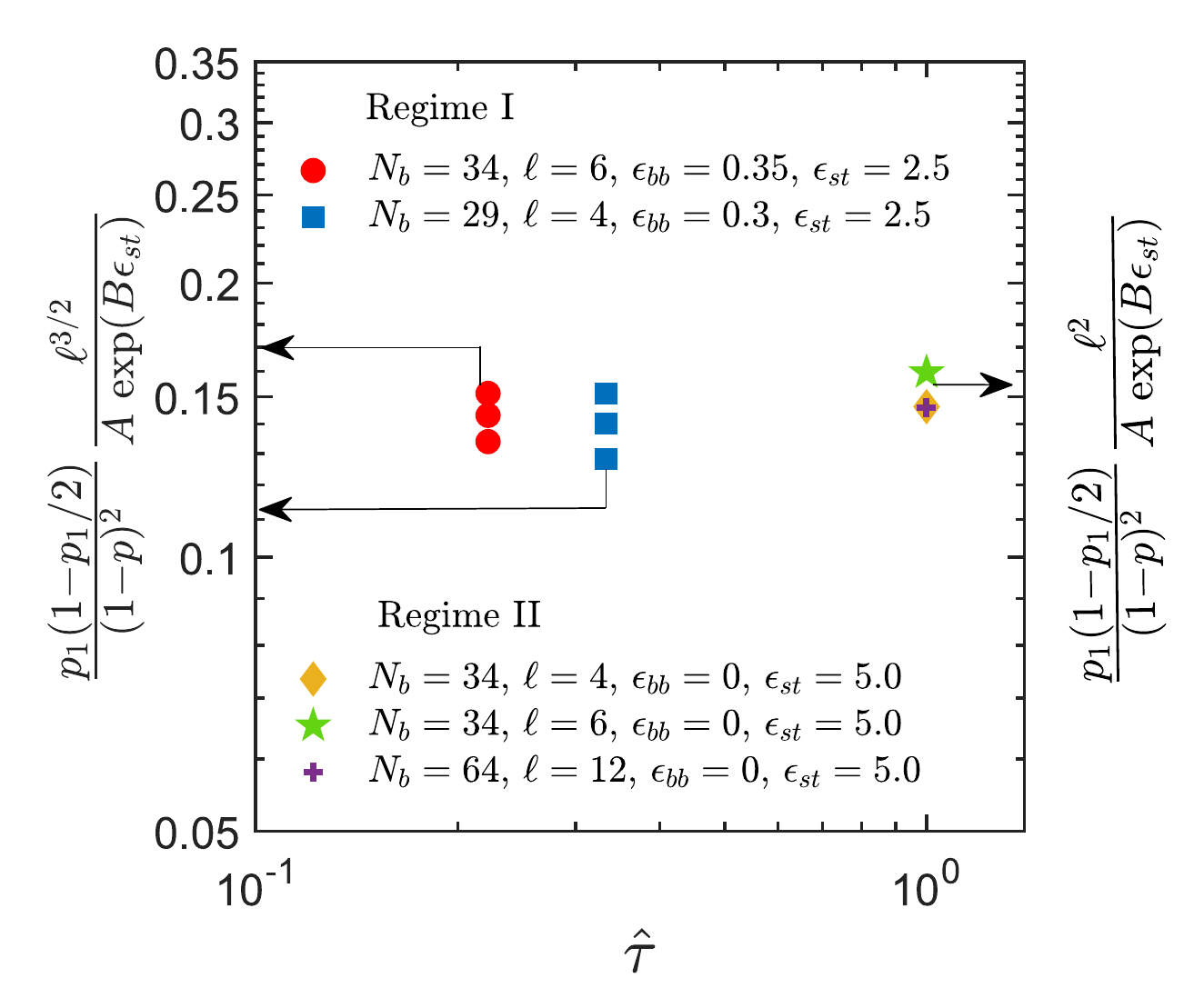} 
        & \includegraphics[width=72mm]{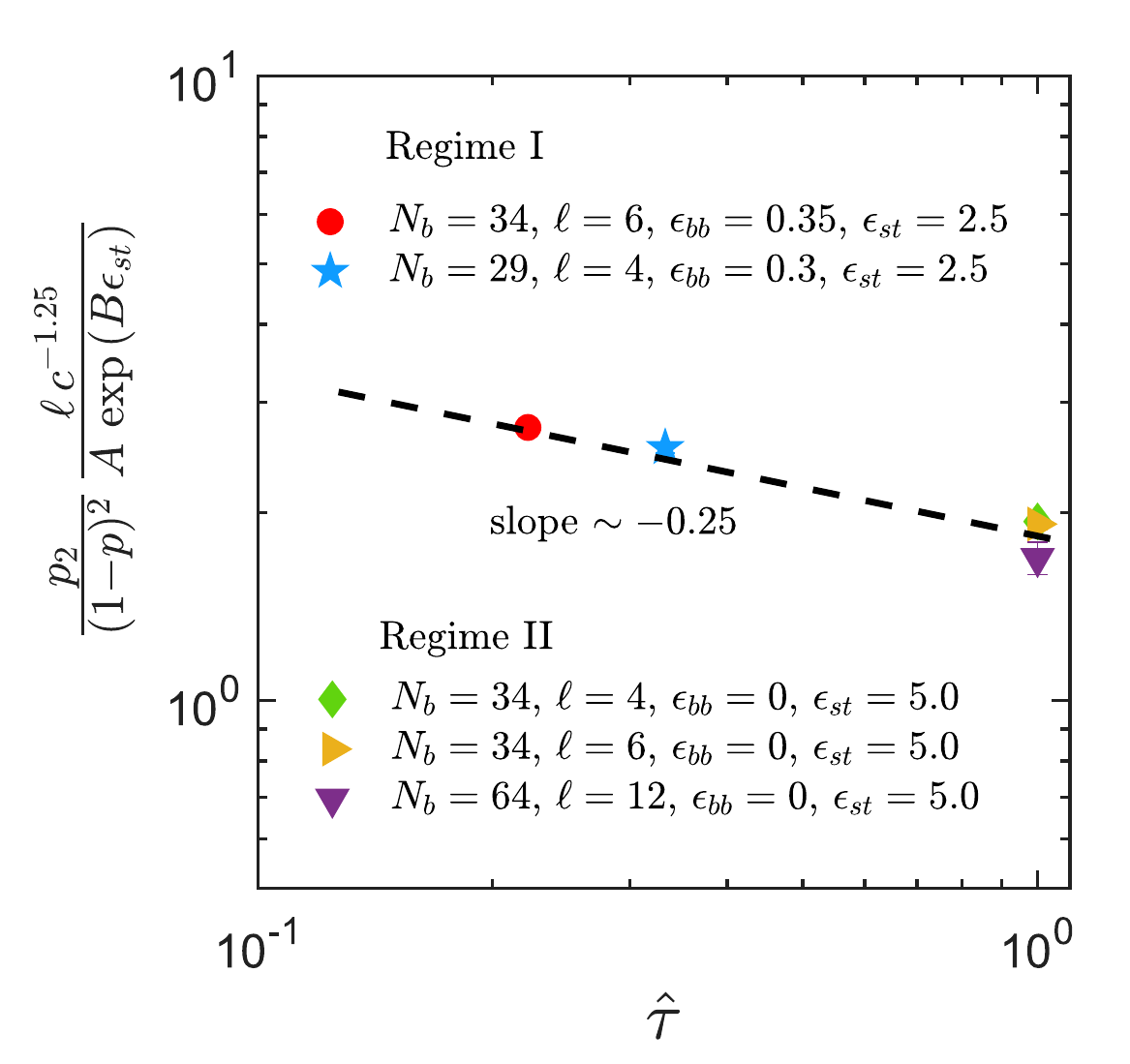} \\
        (a) & (b) \\
        \includegraphics[width=75mm]{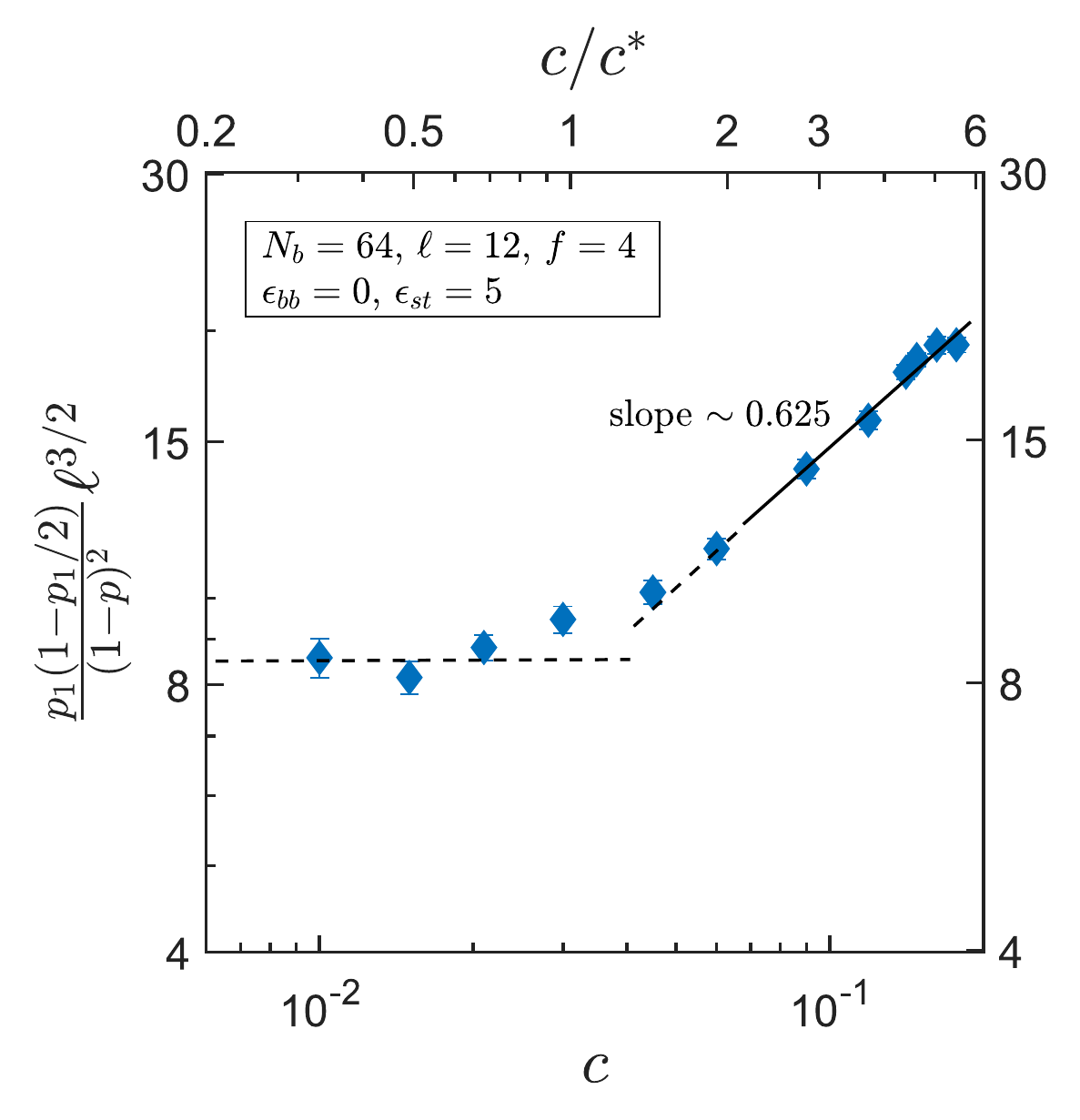} 
        & \includegraphics[width=79mm]{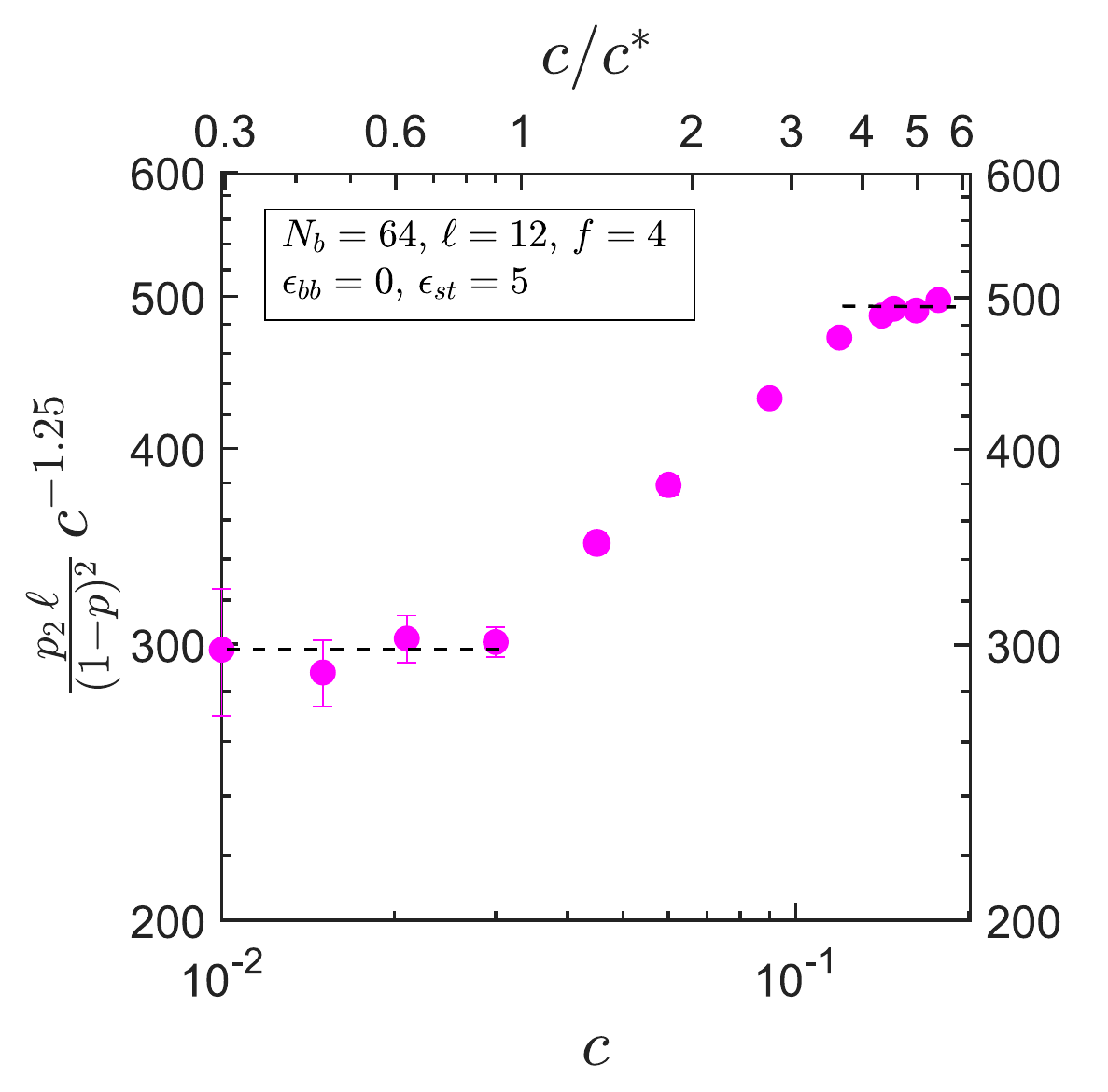} \\
        (c) & (d) \\
    \end{tabular}
    }
	\caption{{Crossover behaviour between the different scaling regimes. (a) Change in the ratio involving the intra-chain degree of conversion $p_1$ from Regime~I, where $\ell < g_T < g_c$ to Regime~II, where $g_T < \ell < g_c$, due to a change in the solvent quality parameter $\hat{\tau}$. (b) Change in the ratio involving the inter-chain degree of conversion $p_2$ from Regime~I to Regime~II due to a change in $\hat{\tau}$.  (c) Change in the ratio involving the intra-chain degree of conversion $p_1$ from Regime~II, where $g_T < \ell < g_c$, to Regime~III, where $g_T < g_c < \ell$, due to a change in the concentration $c$. (d)  Change in the ratio involving the inter-chain degree of conversion $p_2$ from Regime~II to Regime~III, due to a change in $c$.} }
\label{fig:cross}
\end{figure*}

The validation of the scaling relations displayed in Table~\ref{tab:Scaling_rel}~(b) for the ratios involving $p_1$ and $p_2$, in terms of all the relevant scaling variables, for (i) backbone monomers under good solvent conditions corresponding to Regimes~I and~II, and (ii) backbone monomers under $\theta$-solvent conditions, is demonstrated in the respective subfigures of Fig.~\ref{fig:p1p2Sc_R1_R2}. It is clear that when the dependence of the effective sticker strength on the spacer length $\ell$ is taken into account, then all the simulation data can be collapsed onto master plots for the dependence of the ratios involving $p_1$ and $p_2$ on monomer concentration $c$. {Note that the weak dependence on concentration observed for the ratio involving $p_1$ in Regime~I (contradictory to the prediction of scaling theory) appears to persist at other values of $\epsilon_{st}$ as well. The values of the ratio in the Regimes~I and~II approach each other with increasing concentration. Whether the weak power law dependence on $c$ in Regime~I persists at higher concentrations and higher chain lengths needs further examination}. Even though Eq.~(\ref{Eq:common}) indicates that the ratio ${p_1(1-p_1/2)}/{(1-p)^2}$ depends on $\hat \tau$ in Regime~II, the choice $\epsilon_{bb} = 0$ implies that $\hat \tau = 1$, and consequently it has not been included in the $y$-axis label in Fig.~\ref{fig:p1p2Sc_R1_R2}~(a). {Nevertheless, it is worth noting that the pre-factor to the solvent quality parameter $\hat \tau$ affects the value of the ratio displayed in Fig.~\ref{fig:p1p2Sc_R1_R2}~(a) for  Regime~II}.

In the case of $\theta$-solvent conditions for the backbone, we have seen in Figs.~\ref{fig:RII_constStk} that although the scaling of the ratios involving $p_1$ and $p_2$ with concentration are identical for the two models used to simulate $\theta$ conditions for the backbone, the pre-factors are different, which results in a difference in the numerical values. Nevertheless, from  Fig.~\ref{fig:p1p2Sc_R1_R2}~({c}) and Fig.~\ref{fig:p1p2Sc_R1_R2}~(d) we can conclude that the difference in the pre-factors arises from the factor, $g_{ss}$, which is found to be different for the two models for the $\theta$-solvent condition. By absorbing the dependence of $g_{ss}$ in the $y$-axis we observe the expected data collapse, as can be seen in Figs.~\ref{fig:p1p2Sc_R1_R2}~({c}) and (d).

{\subsection{\label{sec:cross} Crossover behaviour between the scaling regimes}}

{The crossover between Regimes~I and~II is driven by the solvent quality parameter $\hat{\tau}$, while that between Regimes~II and~III is driven by the concentration $c$, as discussed earlier in section~\ref{subsec:simcon}. Within the constraints of the relatively narrow range of parameters that have been explored here (due to the computational intensity of the Brownian dynamics simulations), a preliminary examination of the crossover behaviour between Regimes~I and~II is shown in Figs.~\ref{fig:cross}~(a) and~(b), and that between Regimes~II and~III is displayed in Figs.~\ref{fig:cross}~(c) and~(d).}

{The dependence of the ratio involving $p_1$ on the concentration $c$, in scaling regimes~I and~II, has been plotted together in Fig.~\ref{fig:p1p2Sc_R1_R2}~(a), and discussed in that context. With regard to the dependence on the solvent quality parameter $\hat\tau$,  it is clear from first and second rows of Table~\ref{tab:Scaling_rel}~(b), that the ratio is independent of $\hat\tau$ in Regime~I, and it scales with an exponent $-1$ in Regime~II. As a consequence, when plotted as a function of $\hat\tau$, we expect to see a constant value in Regime~I, and then a crossover into Regime~II with an asymptotic slope of $-1$. With the current set of simulations, however, this behaviour cannot be observed due to the paucity of values of $\hat\tau$ at which the simulations have been carried out. Essentially, two value of $\hat\tau = 0.22$ and $0.33$ have been used in Regime~I, and its value has been set equal to one in Regime~II. The values of the ratio at these values of $\hat\tau$, with the dependence on $\ell$ and $g_{ss}$ absorbed into the $y$-axis, are displayed in Fig.~\ref{fig:cross}~(a). While it would be possible to obtain data at other values of $\hat\tau$ in Regime~II, the marginal difference in the magnitude of the ratio in the two regimes would make it difficult to observe the $-1$ exponent in this regime. Further, the weak dependence on concentration leads to the scatter of the data seen in Regime~I. Clearly, simulations of much longer chains, and over a wider range of values of $\hat\tau$, would be required to adequately describe the crossover of the ratio involving $p_1$ between Regimes~I and~II}.

{ The dependence on concentration $c$ of the ratio involving $p_2$ is predicted to be the same in both Regimes~I and~II according to scaling theory (see first and second rows of Table~\ref{tab:Scaling_rel}~(b)), which has been verified by the simulation results displayed in Fig.~\ref{fig:p1p2Sc_R1_R2}~(b). In order to examine the dependence of the ratio on just $\hat\tau$, the concentration has been absorbed into the $y$-axis in Fig.~\ref{fig:cross}~(b). To avoid the overlaying of data at different concentrations on top of each other, their average value has been reported in the figure. Similarly, the dependence on $\ell$ and $g_{ss}$ has also been absorbed into the $y$-axis, as was done previously in Fig.~\ref{fig:p1p2Sc_R1_R2}~(b). Scaling theory predicts that the ratio depends on $\hat\tau$ with an exponent $\alpha = -1/4$ in both Regimes~I and~II. This expectation is verified in Fig.~\ref{fig:cross}~(b), where the values of the ratio at $\hat\tau = 0.22$ and $0.33$ correspond to Regime~I, and the values at $\hat\tau = 1$ correspond to Regime~II. The crossover between the two regimes appears to be smooth, though it is desirable to confirm this with additional data points in both regimes.}

{The crossover between Regimes~II and~III has been examined for a single system with $N_b = 64$ and $\ell =12$, since amongst the many data sets used in the current simulations, it is one that spans both regimes. Recall that according to scaling theory, the ratio involving $p_1$ is independent of concentration in Regime~II and scales as $c^{5/8}$ in Regime~III, while the ratio involving $p_2$ scales as $c^{5/4}$ in both Regimes~II and~III. These asymptotic scaling regimes and the crossover between them is displayed in Figs.~\ref{fig:cross}~(c) and~(d), where, in the latter, the dependence on concentration has been absorbed into the $y$-axis to highlight the crossover behaviour. As displayed in Table~\ref{tab:parameters}, in the simulations carried out here, the upper bound of the scaled concentration $c/c^*$ in Regime~II is 1.9, while the lower bound in Regime~III is 4.0. With these bounds in mind, it is clear that the ratio involving $p_1$ appears to leave Regime~II around $c/c^* = 1$, but already exhibits Regime~III scaling by  $c/c^* = 2$. On the other hand, while the ratio involving $p_2$ also appears to leave Regime~II around $c/c^* = 1$, it displays the asymptotic scaling of Regime~III only by  $c/c^* = 4$. As discussed earlier in section~\ref{subsec:simcon}, the actual lower bound to Regime~III is  $c/c^* = 1$. As a result, it is expected that for longer chain lengths, the ratio will exhibit asymptotic Regime~III scaling at smaller values of $c/c^*$. While within scaling theory, the pre-factor for the dependence of the ratio involving $p_2$ on concentration $c$ is the same in both regimes, simulations seem to indicate that it may be different  in the two regimes since the asymptotic constants displayed in Figs.~\ref{fig:cross}~(d) are not the same. Establishing the true nature of the crossover, and the values of the pre-factors with greater certainty would require simulations with longer chains. }

\begin{figure}[tb]
    \centerline{
   {\includegraphics[width=85mm]{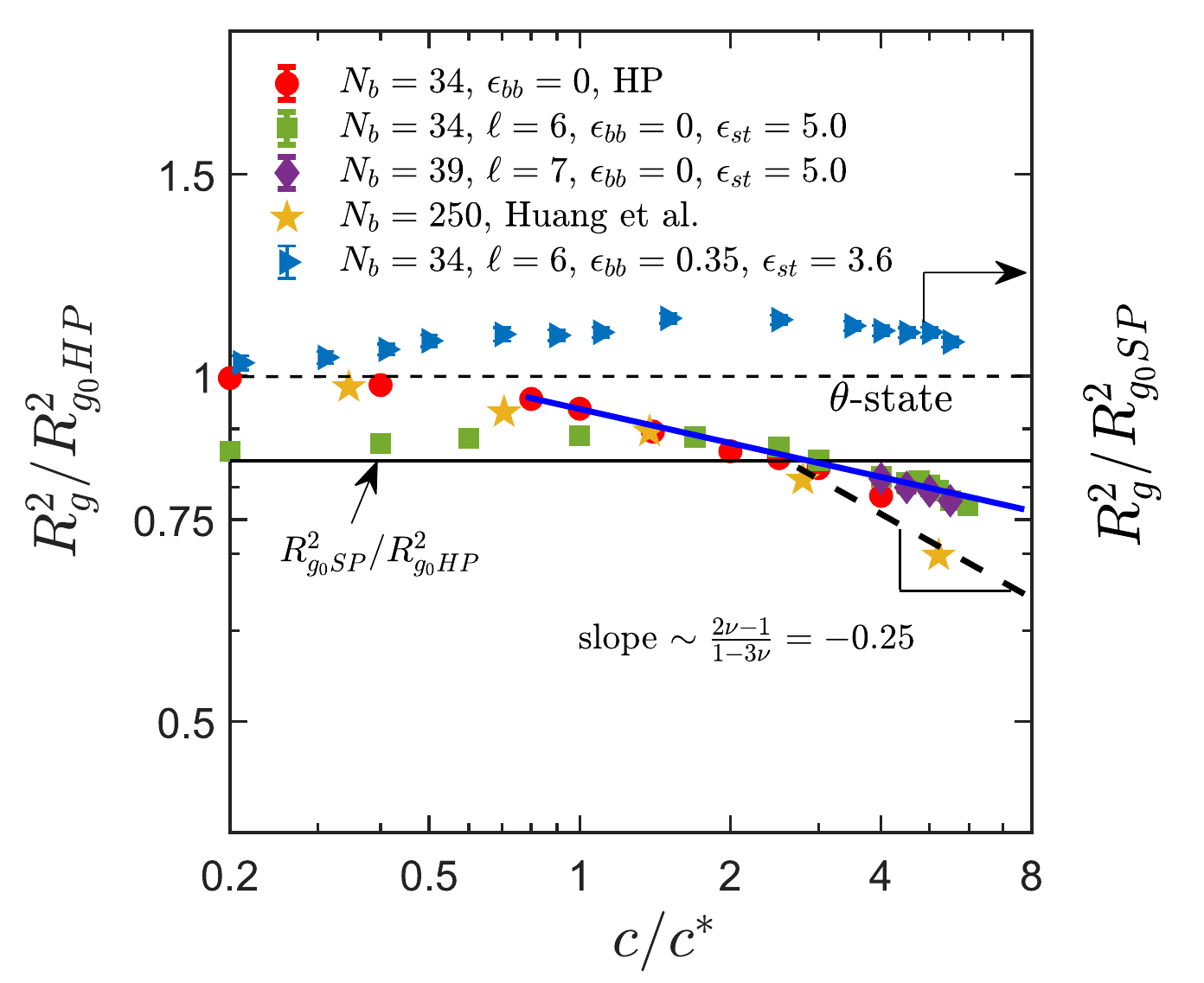}}     
    }
 \vskip-5pt
 \caption{{Ratios of the} radius of gyration as a function of scaled concentration, $c/c^*$. $R_{g0H \!P}^2$  {and $R_{g0S \!P}^2$ are} the mean-squared radius of gyration of the homopolymer (HP) {and the sticky polymer (SP), respectively}, in the dilute limit. {The backbone monomers in both polymers are in an athermal solvent}. The filled red symbols represent data for homopolymer  solutions obtained from current simulations, while the yellow stars are from MPCD simulations by~\citet{Huang2010}\  The filled green squares and purple diamonds represent sticky polymer solutions. The filled blue right triangles represent a situation in which the sticky polymer chain as a whole is under $\theta$-solvent conditions. These different scenarios are achieved with different backbone monomer interaction strengths, $\epsilon_{bb}$, sticker strengths, $\epsilon_{st}$, and spacer lengths $\ell$. The dashed line is the theoretical scaling prediction (Eq.~(\ref{Eq:Rg2vscbycstr})) for unentangled semidilute homopolymer solutions. The filled {blue} line is drawn to guide the eye.}
\label{fig:Rg2_vs_c}
\end{figure}

\vskip10pt

\subsection{\label{sec:Rg2vc} Scaling of the radius of gyration}

It is interesting to observe the variation with $c/c^*$ of the radius of gyration ratio, $R_g^2/R_{g0H \!P}^2$, displayed in Fig.~\ref{fig:Rg2_vs_c}, of an individual chain in a sticky polymer solution,  and compare it with the behaviour of a chain in a homopolymer solution.  Here, $R_{g0H\!P}$ is the radius of gyration of the homopolymer chain in the dilute limit. The asymptotic scaling law for the radius of gyration ratio as a function of the scaled concentration $c/c^*$, in semidilute unentangled homopolymer solutions, is well known~\cite{DoiEdwards,Daoud1975,Pelissetto2008,Huang2010},
\begin{equation}\label{Eq:Rg2vscbycstr}
\left. \frac{R_{g}^2}{R_{g0}^2} \right\vert_{H\!P} = \left(\frac{c}{c^{\ast}}\right)^{(2\nu-1)/(1-3\nu)}
\end{equation} 
Equation~(\ref{Eq:Rg2vscbycstr}) describes the shrinking of individual chains with increasing concentration due to the presence of Flory screening. It is clear from the filled red symbols, which are the results of current simulations, and the yellow stars, which are the results of MPCD simulations by~\citet{Huang2010}, that the radius of gyration ratio for homopolymer solutions is constant at low concentrations (as expected for dilute solutions), and then decreases in a broad crossover region between $c/c^* = 0.4$ to $c/c^* \approx 3$ {{(indicated by the filled blue line in Fig.~\ref{fig:Rg2_vs_c})}, as it changes from the dilute to the asymptotic semidilute scaling regime, where it {finally} decreases with a power law. The dashed black line in Fig.~\ref{fig:Rg2_vs_c} is drawn with slope equal to the asymptotic scaling exponent $-0.25$. On the other hand, the filled green and purple symbols, representing sticky polymer solutions with backbone under athermal solvent ($\epsilon_{bb}=0$), reflect a very different behaviour. 

The {ratio of the} radius of gyration for a chain in a sticky polymer solution { to that for an equivalent homopolymer chain in the dilute limit, $R_{g0S \!P}^2/R_{g0H \!P}^2$}, is less than {one} {because of the presence of stickers and the concomitant existence of intra-chain associations}. {Its value, determined  from single chain simulations, {is indicated by the filled black line in Fig.~\ref{fig:Rg2_vs_c}}. {At low concentrations, the magnitude of the ratio {(filled green symbols)} remains close to the dilute limit value, and} appears to increase gradually with increasing concentration. {The} gradual increase can be ascribed to the increase in inter-chain associations with increasing concentration, that occur at the expense of  intra-chain associations. {It could also be a finite size effect, which could be confirmed with simulations for longer chains}.  The start of the crossover into the semidilute regime due to Flory screening seems to be delayed until $c/c^* \approx 2$, and it is clear from  Fig.~\ref{fig:Rg2_vs_c} that the crossover seems to persist beyond $c/c^* = 6$, with the asymptotic scaling regime not yet reached at this concentration, {{as indicated by the filled blue line}. It should be noted the onset of this crossover is expected to depend on the parameters $\epsilon_{st}$, $\epsilon_{bb}$ and $\ell$. These dependencies have not been studied in the present work and are worthy of investigation in the future. The behaviour displayed in Fig.~\ref{fig:Rg2_vs_c} indicates that polymer conformations in solutions of sticky polymers are significantly different from those of homopolymer chains in good solvent conditions, upon which the scaling theory is based. This aspect will be considered further in section~\ref{sec:thetasticky} below, when sticky polymer solutions in which chains as a whole are under $\theta$-solvent conditions ($\epsilon_{bb}=0.35$, $\epsilon_{st}=3.6$) (filled blue triangles in Fig.~\ref{fig:Rg2_vs_c}) are discussed.

\subsection{\label{sec:thetasticky} $\theta$-solvent conditions for sticky polymer chains}

\begin{figure*}[t]
    \centerline{
    \begin{tabular}{c c}
        \includegraphics[width=92mm]{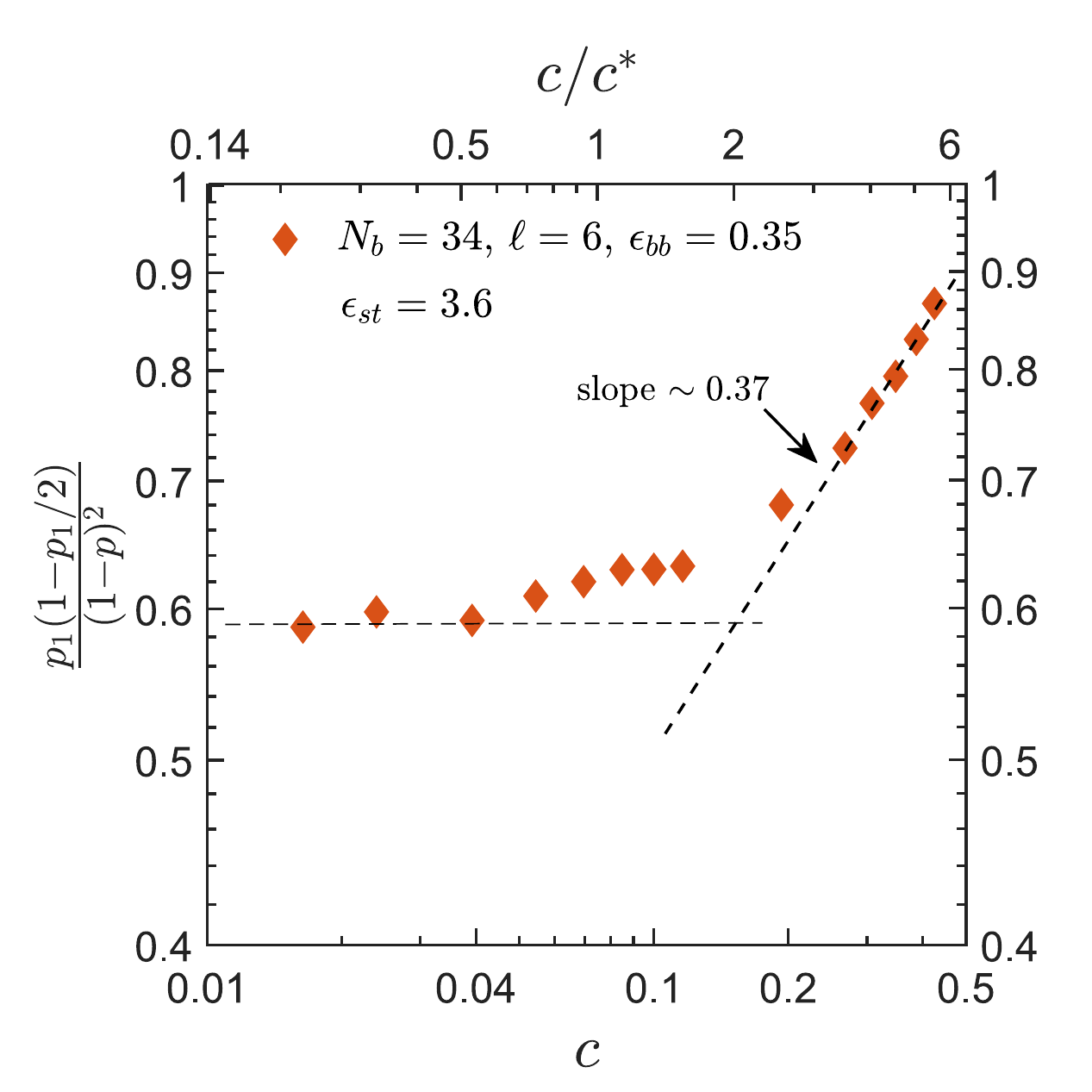} 
        & \includegraphics[width=90mm]{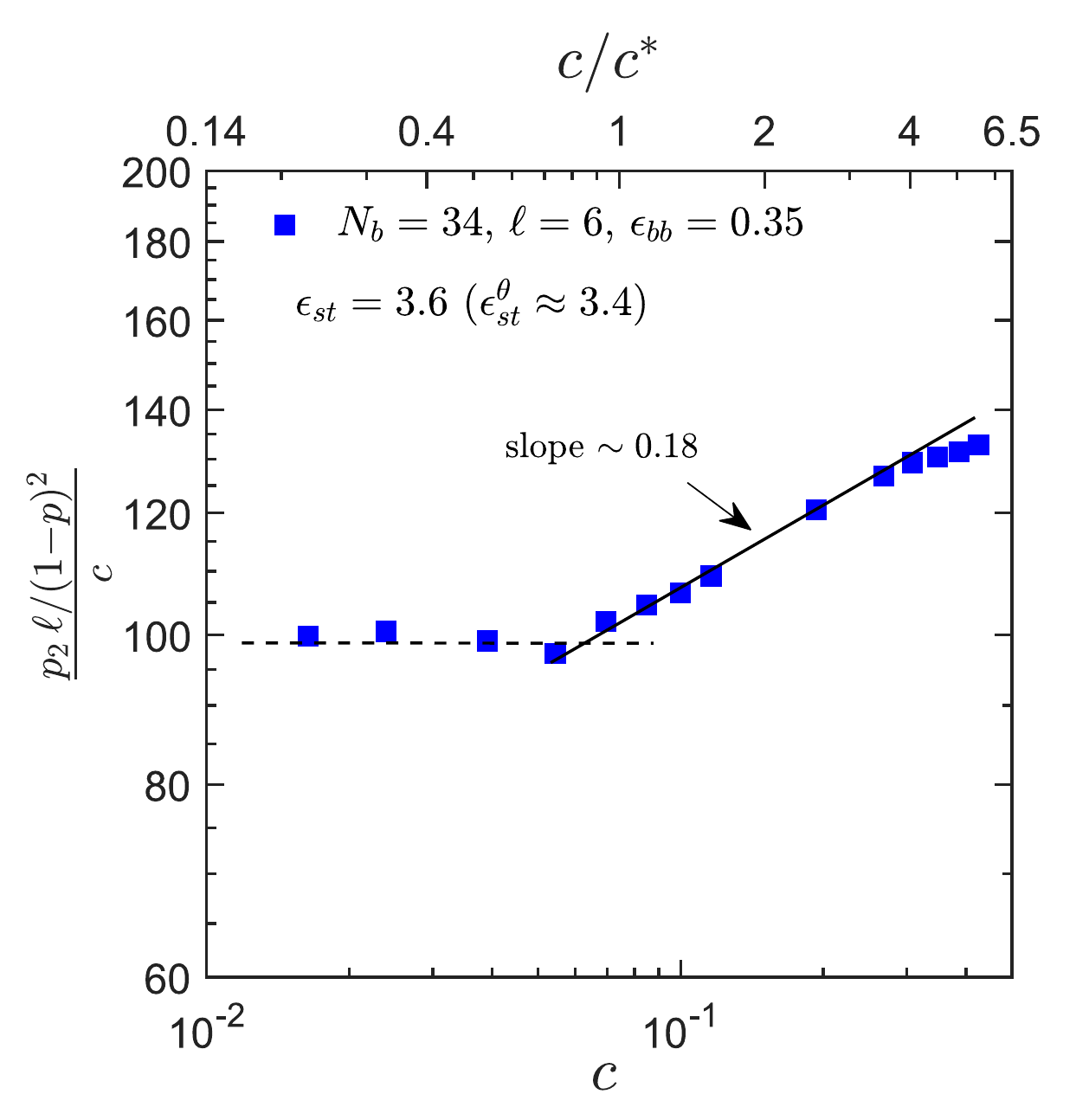} \\
        (a) & (b) \\
    \end{tabular}
    }
	\vskip-5pt
	\caption{The dependence of ratios involving (a) intra-chain, and (b) inter-chain degrees of conversion, on the monomer concentration, $c$, with sticky polymer chains under nearly $\theta$-solvent conditions for sticky chains as a whole. This is achieved, for a chain with $N_b=34$, $\ell=6$, $f=4$, and $\epsilon_{bb}=0.35$, by setting $\epsilon_{st}=3.6$, which is close to the value $\epsilon_{st}^\theta = 3.4$ computed for these parameter values.  The dashed and the solid lines are drawn with slopes that are a good fit to the symbols, which represent simulation data.
\label{fig:p2_overaltheta}}
\end{figure*} 

The scaling relations listed in Table~\ref{tab:Scaling_rel} have all been derived by considering the quality of the solvent relative to backbone monomers on the sticky polymer chain~\cite{Dob}, with the spacer length between stickers, {the solvent quality parameter} and the monomer concentration determining the particular scaling regime that is relevant. In this section, we briefly consider a situation that has not been not treated so far within the framework of scaling theory, namely, one in which the sticky polymer chain as a whole is under $\theta$-solvent conditions. As discussed in section~\ref{sec:gov_eqs}, $\theta$-solvent conditions for a sticky polymer chain can be realised by setting $\epsilon_{st}$ equal to the corresponding value of  $\epsilon_{st}^\theta$, for the given values of $\epsilon_{bb}$ and $\ell$.

\begin{figure}[b]
    \centerline{
   {\includegraphics[width=80mm]{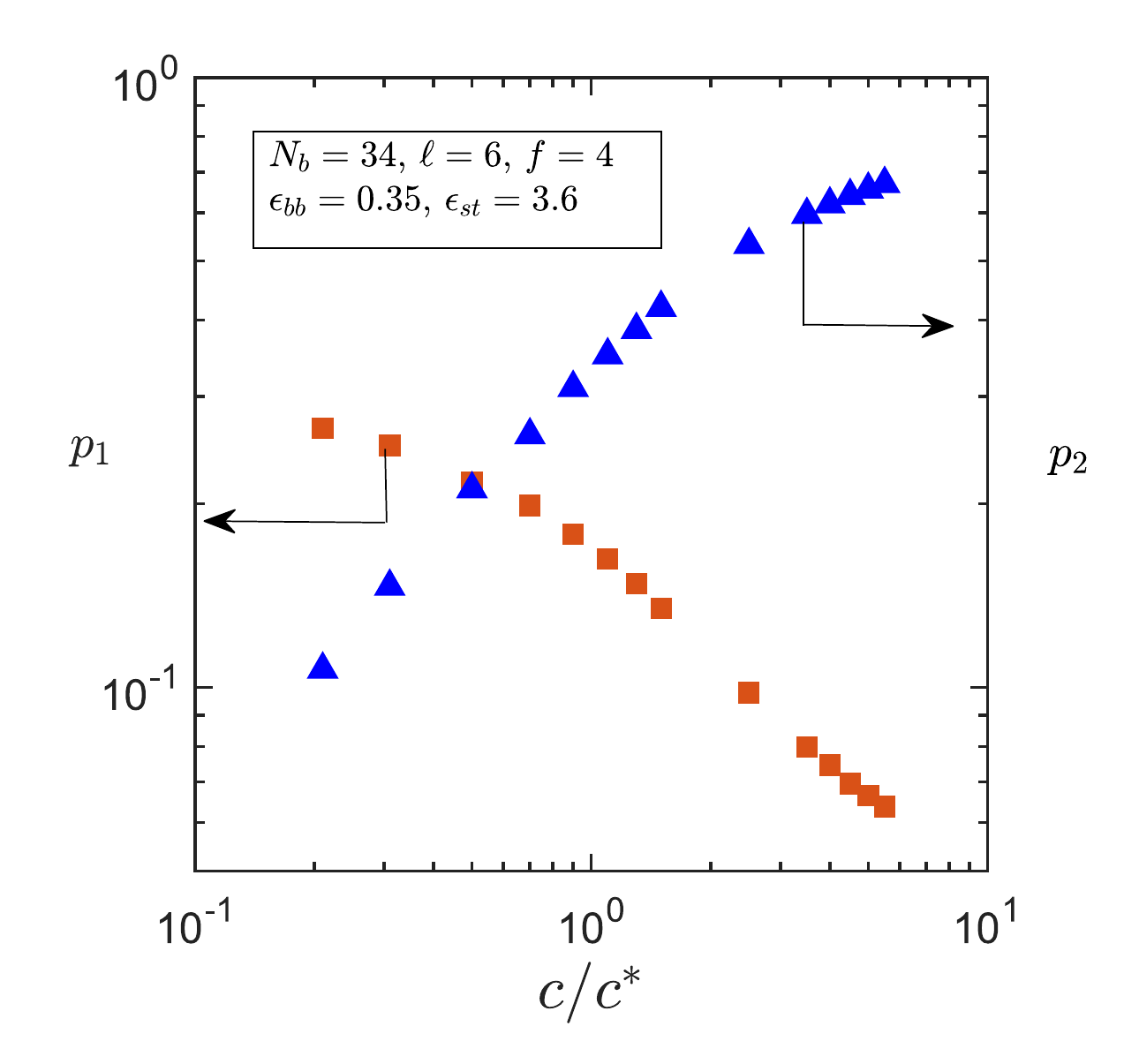}}     
    }
 \vskip-15pt
 \caption{Variation of intra-chain ($p_1$) and inter-chain ($p_2$) association fractions as a function of scaled monomer concentration, $c/c^*$, with sticky polymer chains under nearly $\theta$-solvent conditions for sticky chains as a whole.}
\label{fig:p1p2_thetaSurf}
\end{figure}

The dependence of the ratios involving $p_1$ and $p_2$ on $c$, for a system in which the sticky polymer chains as a whole are under $\theta$-solvent conditions, is displayed in Figs.~\ref{fig:p2_overaltheta}~(a) and (b), respectively, for a chain with $N_b=34$, $\ell=6$, $f=4$, and $\epsilon_{bb}=0.35$. For these parameter values, {using the method discussed previously in}~\citet{Aritra2019} it can be shown that $\epsilon_{st}^\theta \approx 3.4 \,{ \pm \, 0.4}$. The simulation results reported in Figs.~\ref{fig:p2_overaltheta} were carried out with $\epsilon_{st}=3.6$, which is  {in the range of} values required to achieve $\theta$-solvent conditions. It is clear from the figures that the observed dependence on monomer concentration of the intra and inter-chain association fractions is unlike that seen in any of the scaling regimes studied previously. 

\begin{figure*}[t]
    \centerline{
    \begin{tabular}{c c}
        \includegraphics[width=82mm]{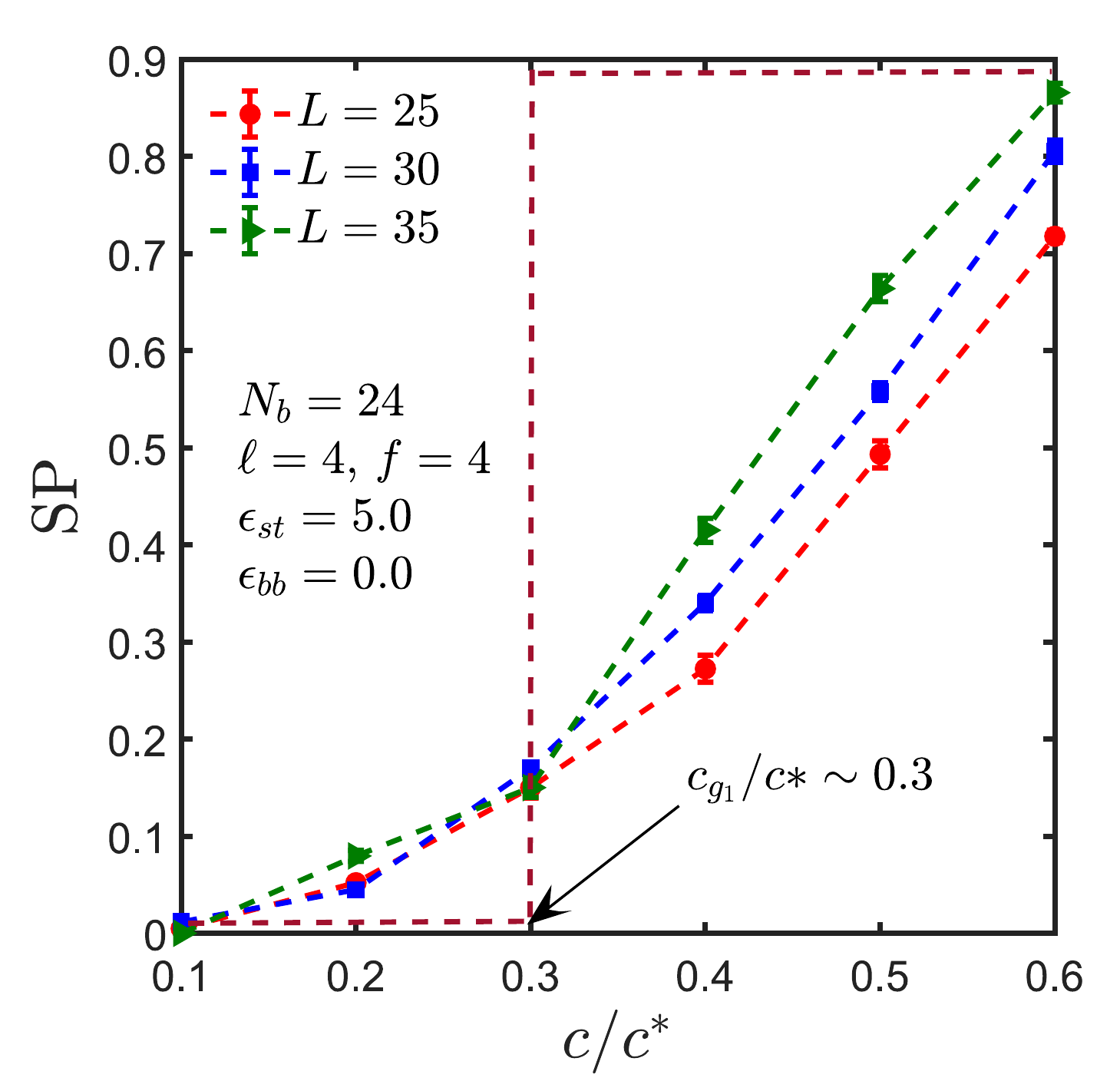} 
        & \includegraphics[width=80mm]{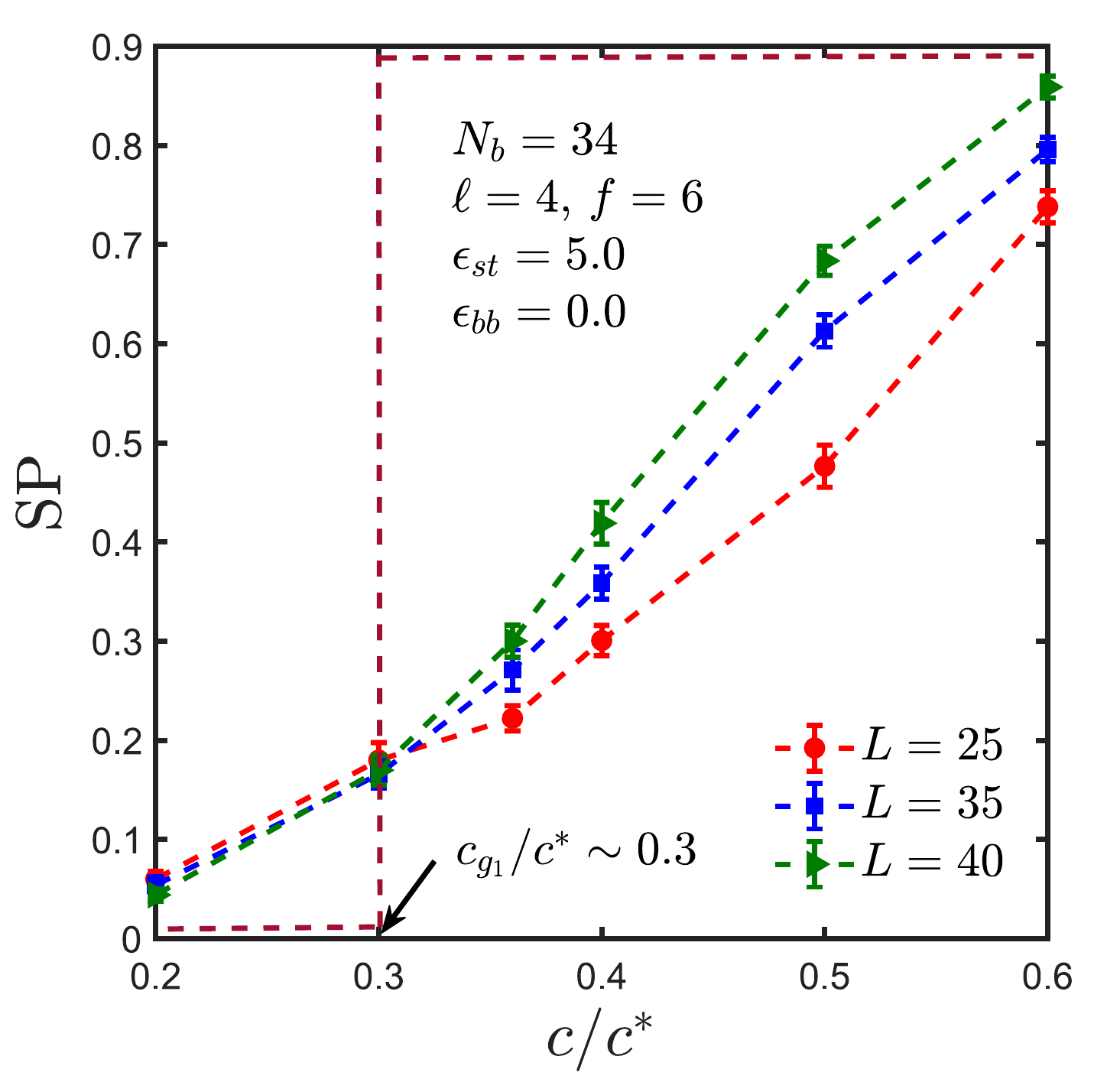} \\
        (a) & (b) \\
    \end{tabular}
    }
	\vskip-10pt
	\caption{Spanning probability (SP) as a function of monomer concentration in Regime~II for (a) $N_b=24$, $\ell=4$, $f=4$ and (b) $N_b=34$, $\ell=4$, $f=6$, with sticker strength, $\epsilon_{st}=5.0$. The point of divergence of the curves at different box sizes is assumed to represent the concentration at the gelation threshold, and is indicated as occuring at $c_{g_1}/c^{\ast} \approx 0.3$. 
	\label{fig:gelpt}}
\end{figure*} 

{Before discussing the results in this section, it is worth making a few remarks about the system that has been studied here ($N_b = 34, \ell = 6, f = 4, \epsilon_{st} = 3.6$)  in the context of the scaling theory. Firstly, in terms of just the backbone monomers, the chain is under good solvent conditions, since $\epsilon_{bb} = 0.35 < \epsilon_\theta = 0.45$. It is the presence of stickers that makes the conditions $\theta$-like for the overall chain. Since $\epsilon_{bb}$ is held fixed at a value of 0.35, the number of monomers in a thermal blob is fixed at $g_T = 20$, independent of the concentrations that have been studied. Since the overall chain under $\theta$-solvent conditions has not been considered in the scaling theory, a wide range of concentrations has been explored here ($0.2 \le c/c^* \le 6)$, independent of the bounds imposed by the different scaling regimes. For the chain of backbone monomers (with $N_b$ fixed at $N_b = 34$), this variation of the scaled concentration implies that the number of monomers in a correlation blob varies in the range $254 \ge g_c \ge 4$, as $c/c^*$ increases from $0.2$ to $6$. It is straightforward to show from Eq.~(\ref{Eq:gTgc}) that $g_c < g_T$ for $c/c^* > 1.5$, and $g_c <  \ell = 6$, for $c/c^* > 4.0$. Both these situations have not been examined within the framework of scaling theory, even for sticky polymer chains under good solvent conditions.}

With this discussion in mind, it can be observed from Figs.~\ref{fig:p2_overaltheta}~(a) and (b) that both association fractions exhibit a change in scaling behaviour at around $c/c^* \approx 0.7$. The ratio involving $p_1$ is fairly independent of concentration until this value, at which point the dependence grows and reaches an asymptotic slope of about $0.37$. In order to more clearly examine the dependence on concentration of  the ratio involving $p_2$, its value has been divided by $c$ on the $y$-axis. Clearly, the scaling with concentration is linear at low concentrations, similar to that observed for chains whose backbone monomers are under $\theta$-solvent conditions. Beyond the value of $c/c^* \approx 0.7$, the slope assumes a value $\approx 1.18$, which is less than the slope of 1.25  observed for chains with backbones under good solvent conditions. However, for $c/c^* \gtrsim 4$, the ratio deviates from this scaling presumably due to the number of monomers in a correlation blob becoming {smaller than the spacer length $\ell$} as a result of the relatively short chain length, $N_b=34$, used in the current simulations. Note that, while in general the number of monomers in a correlation blob is independent of chain length $N_b$, here $g_c$ decreases with increasing $c/c^*$ because $N_b$ is held fixed at $N_b = 34$, as the concentration $c$ is increased. 

It is instructive to study the dependence on concentration of both the degrees of conversion in conjunction with the variation with $c/c^*$ of the radius of gyration ratio,  {$R_g^2/R_{g_{0}S \!P}^2$, displayed in Fig.~\ref{fig:Rg2_vs_c} (filled blue right triangles, $y$-axis on the right of figure)}. For a homopolymer solution under $\theta$-solvent conditions, this ratio is constant, independent of concentration, since there is no Flory screening. In the case of the sticky polymer chains under $\theta$-solvent conditions considered here, the radius of gyration ratio appears to be a weak function of concentration. The ratio increases gradually, followed by a slow decrease beyond the threshold value of $c/c^* \approx 1$, which coincides with the value at which the change in concentration dependence is observed for the ratios involving $p_1$ and $p_2$ in Figs.~\ref{fig:p2_overaltheta}. The initial increase in size can be correlated with the reduction in the intra-chain association fraction $p_1$ displayed in Fig.~\ref{fig:p1p2_thetaSurf}, in which it can be seen that $p_1$  rapidly decreases beyond $c/c^* \approx 1$, with the inter-chain association fraction $p_2$ then becoming the dominant mode of association. 

{The scaling of the radius of gyration behaviour observed here for a sticky polymer chain under overall $\theta$-conditions may well be due to finite size effects. In principle, one expects that for sufficiently long chains, the renormalization of the solvent quality that occurs due to setting $\epsilon_{st} = \epsilon_{st}^\theta$ would lead to a true $\theta$-state, with a radius of gyration that is independent of concentration}. The special case considered in this section has not been investigated further here. Nevertheless, the preliminary results clearly indicate that the scaling behaviour of the intra and inter-chain association fractions is intimately connected to the underlying conformations of the sticky polymer chains.

\begin{figure*}[t]
    \centerline{
    \begin{tabular}{c c}
        \includegraphics[width=80mm]{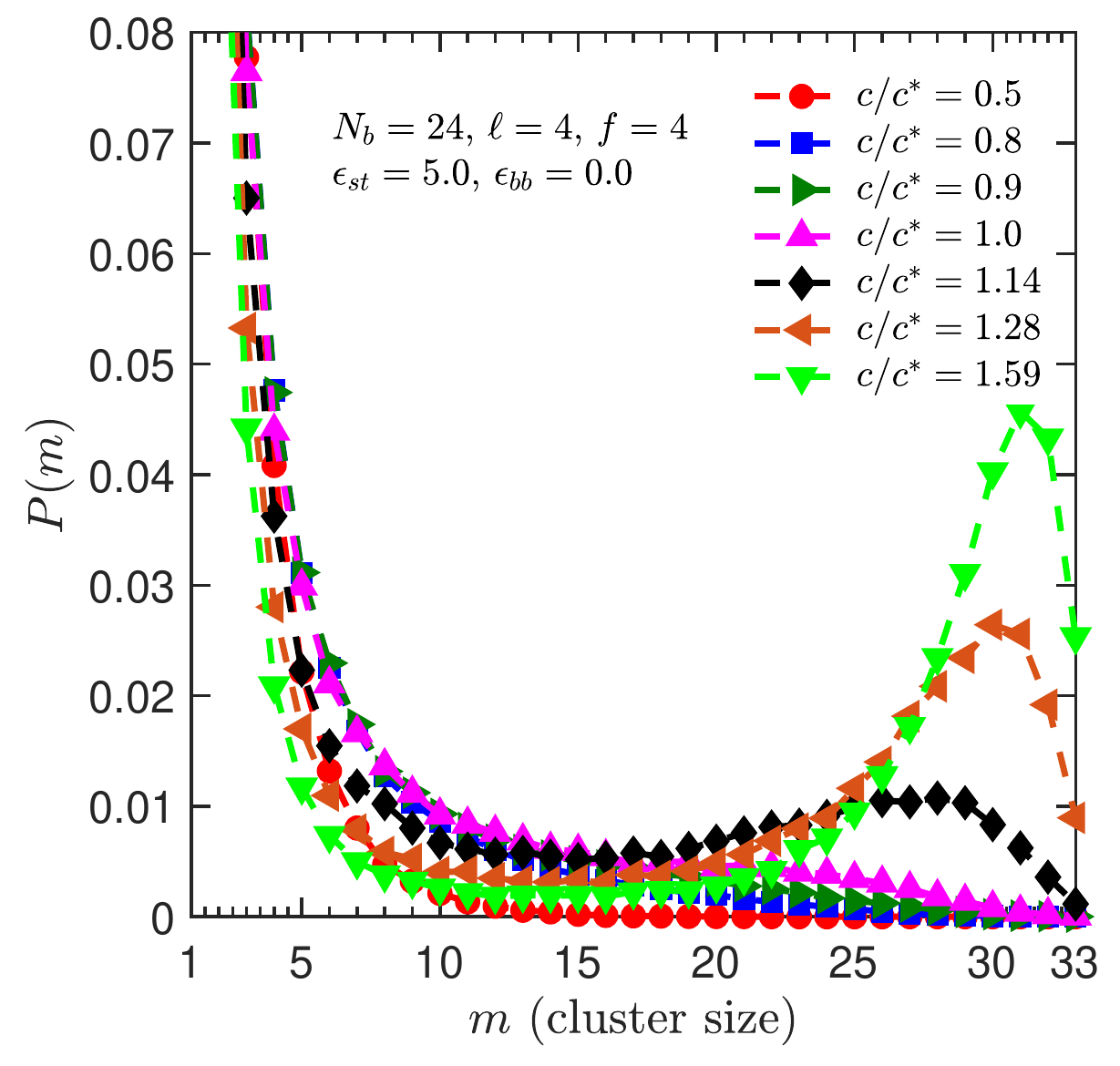} 
        & \includegraphics[width=82mm]{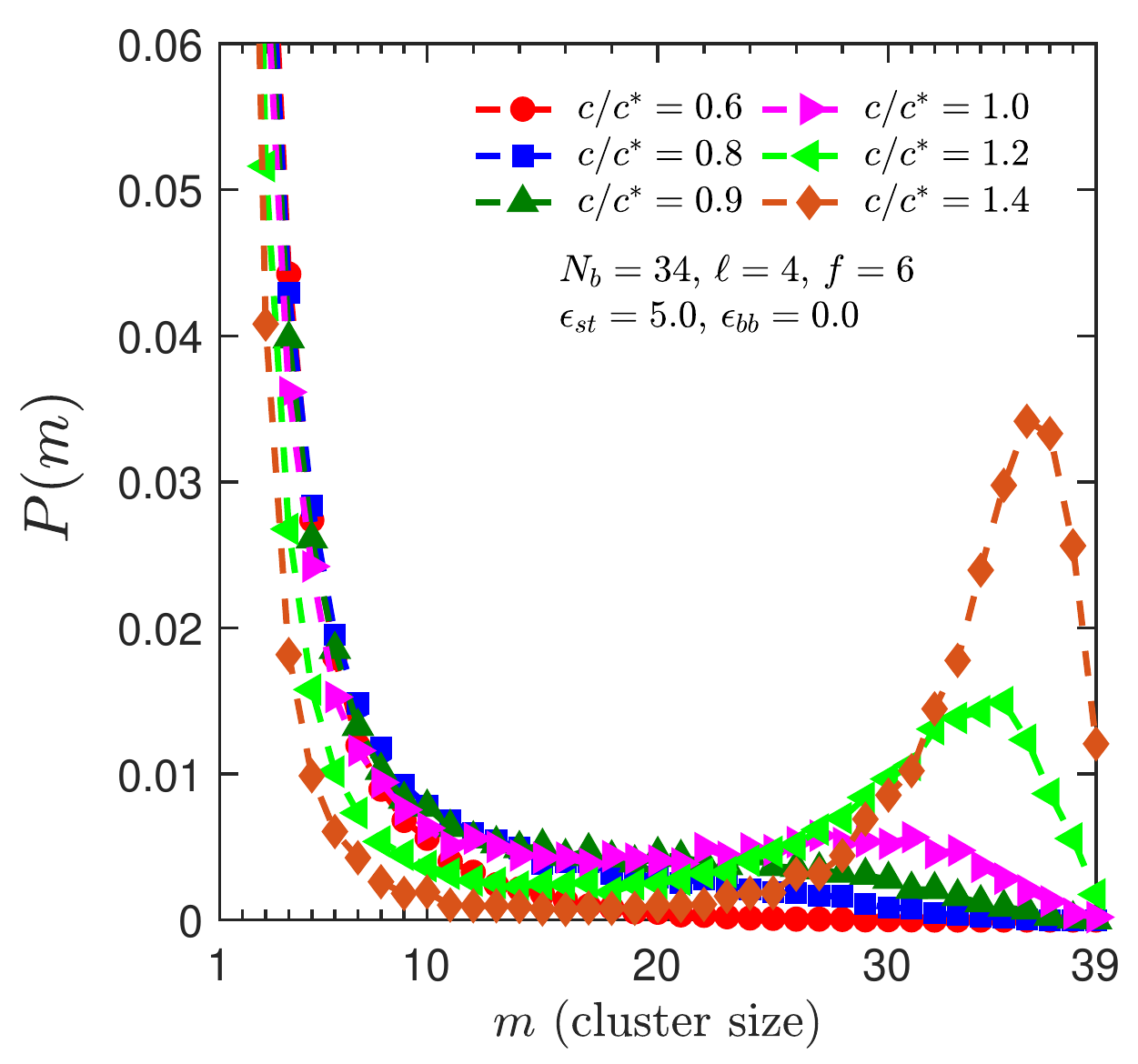} \\
        (a) & (b) \\
    \end{tabular}
    }
	\vskip-10pt
	\caption{Chain cluster-size distribution as a function of monomer concentration in Regime~II for (a) $N_b=24$, $\ell=4$, $f=4$, and (b) $N_b=34$, $\ell=4$, $f=6$, with sticker strength, $\epsilon_{st}=5.0$. The onset of bimodality, which is assumed to represent a signature of gelation, occurs at $c_{g_2}/c^* \approx 1.0$, for each of the three chain lengths.
	\label{fig:clustdist}}
\end{figure*}

\section{\label{sec:gel} Characterization of gelation and the gelation line}

The mean-field theoretical framework has been used by \citet{RnSstatics} and \citet{Dob} to map out the phase diagram of associative polymer solutions in the temperature-concentration plane, and within the phase diagram, to identify different domains in the single phase sol and gel states. Essential to the demarcation of the different phase boundaries in these theories, is the use of the original Flory-Stockmayer expression, Eq.~(\ref{FS}), that relates the fraction of inter-chain associations $p_2$  at the sol-gel transition to the number of stickers $f$ on a chain. Typically, Eq.~(\ref{FS}), rather than the Dobrynin modified~\cite{Dob} Flory-Stockmayer expression, Eq.~(\ref{FS-D}), is used since the fraction of intra-chain associations $p_1$ is considered to be negligibly small. While this is a reasonable assumption at sufficiently high concentrations, $p_1$ and $p_2$ are of comparable magnitudes for most of the concentrations examined here. Consequently, substituting Eq.~(\ref{FS-D}) (rather than Eq.~(\ref{FS})) into the expression for $p_2$ in the second of Eqs.(\ref{Eq:common}), leads to the following expression for the dependence of the monomer concentration, $c_g$,  along the \textit{gelation line} that separates the sol and gel states, on all the system parameters,  
\begin{equation}
\label{Eq:gel_line}
c_g \sim \left(\frac{\hat{\tau}^{\tfrac{\nu\theta_2}{(3\nu-1)}}\,\ell}{[(1-p_1^g)f-1](1-p_g)^2g_{ss}}\right)^{\tfrac{3\nu-1}{\nu(3+\theta_2)-1}}
\end{equation}
Here, $p_1^g$ is the fraction of intra-chain associated stickers and $p_g$ is the total fraction of associated stickers at the sol-gel transition. Clearly, both Eq.~(\ref{FS-D})  and Eq.~(\ref{Eq:gel_line}) are testable elements of the scaling theory, which have not  been examined so far by molecular simulations. In this work, we examine the validity of Eqs.~(\ref{FS-D}) and~(\ref{Eq:gel_line}) in a limited way, i.e., we confine our attention to determining the dependence of $p_2$ on $p_1$ and $f$, and the variation of $c_g$  along the gelation line, for fixed values of the solvent quality parameter $\hat \tau$ and sticker strength $\epsilon_{st}$, in the special case where the backbone monomers are in good solvent conditions corresponding to scaling regime~II. Additionally, we examine the dependence of $p_1^g$ and $p_g$ on $\ell$, and on $f$, in order to eliminate them from Eq.~(\ref{Eq:gel_line}), and as a consequence, obtain the dependence of $c_g$ on just the sticky chain properties, $\ell$ and $f$. 

\begin{figure}[t]
    \centerline{
   {\includegraphics[width=80mm]{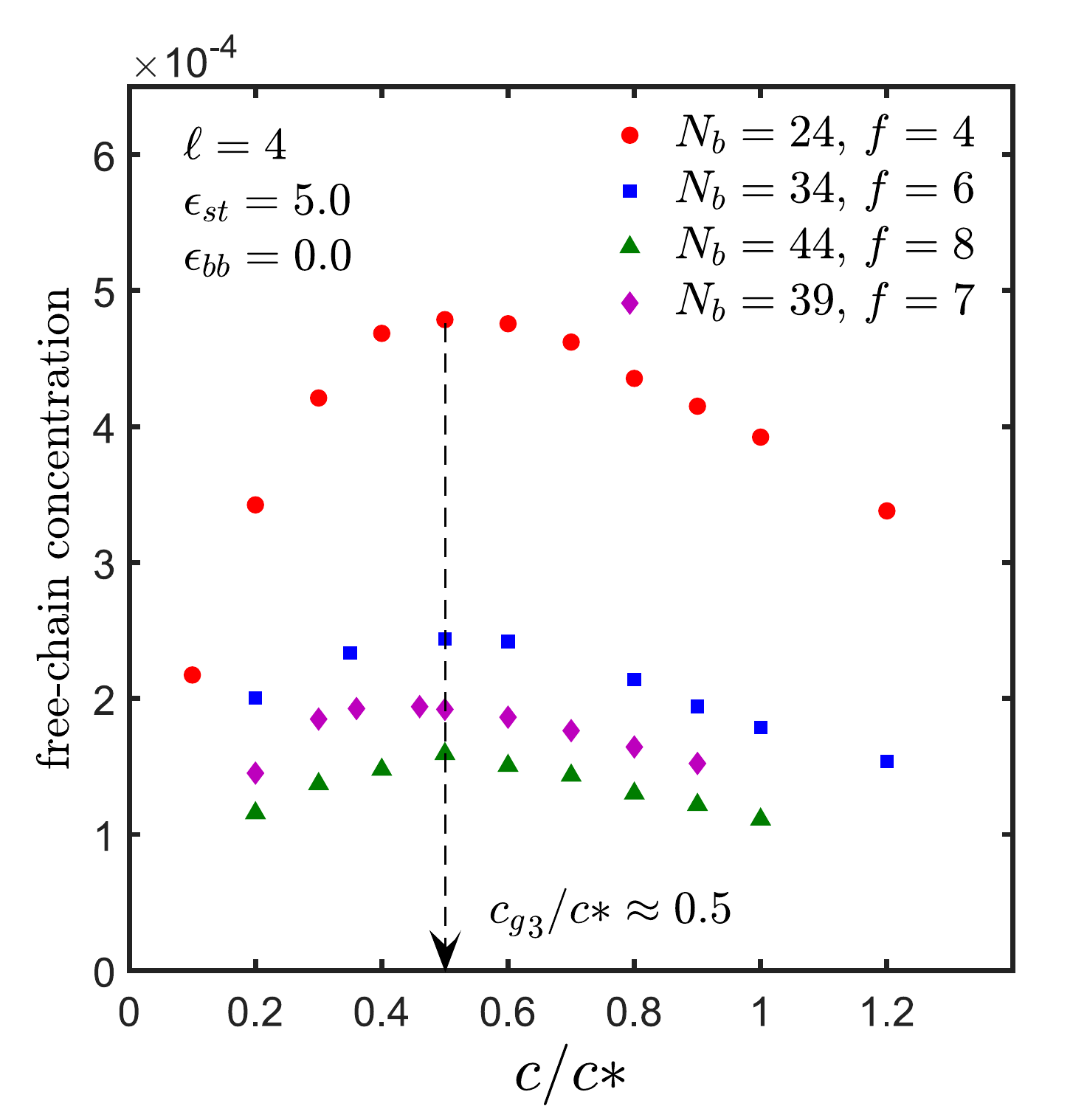}}     
    }
     \vskip-15pt
 \caption{Free chain concentration as a function of monomer concentration in Regime~II for different chain lengths with spacer length, $\ell = 4$, sticker strength, $\epsilon_{st}=5.0$ and $\epsilon_{bb}=0$. The maxima in the free-chain concentration is observed at $c_{g_3}/c^*  \approx 0.5$.
\label{fig:freechain} }
\vskip-15pt
\end{figure}

In order to verify if the prediction of the gelation line by scaling theory is accurate, it is first necessary to locate the concentration at which the sol-gel transition occurs. As mentioned in section~\ref{sec:intro}, there are at least three different approaches in the literature with regards to this question, and here we examine each of them in turn. 

\begin{figure*}[t]
    \centerline{
    \begin{tabular}{c c}
        \includegraphics[width=79mm]{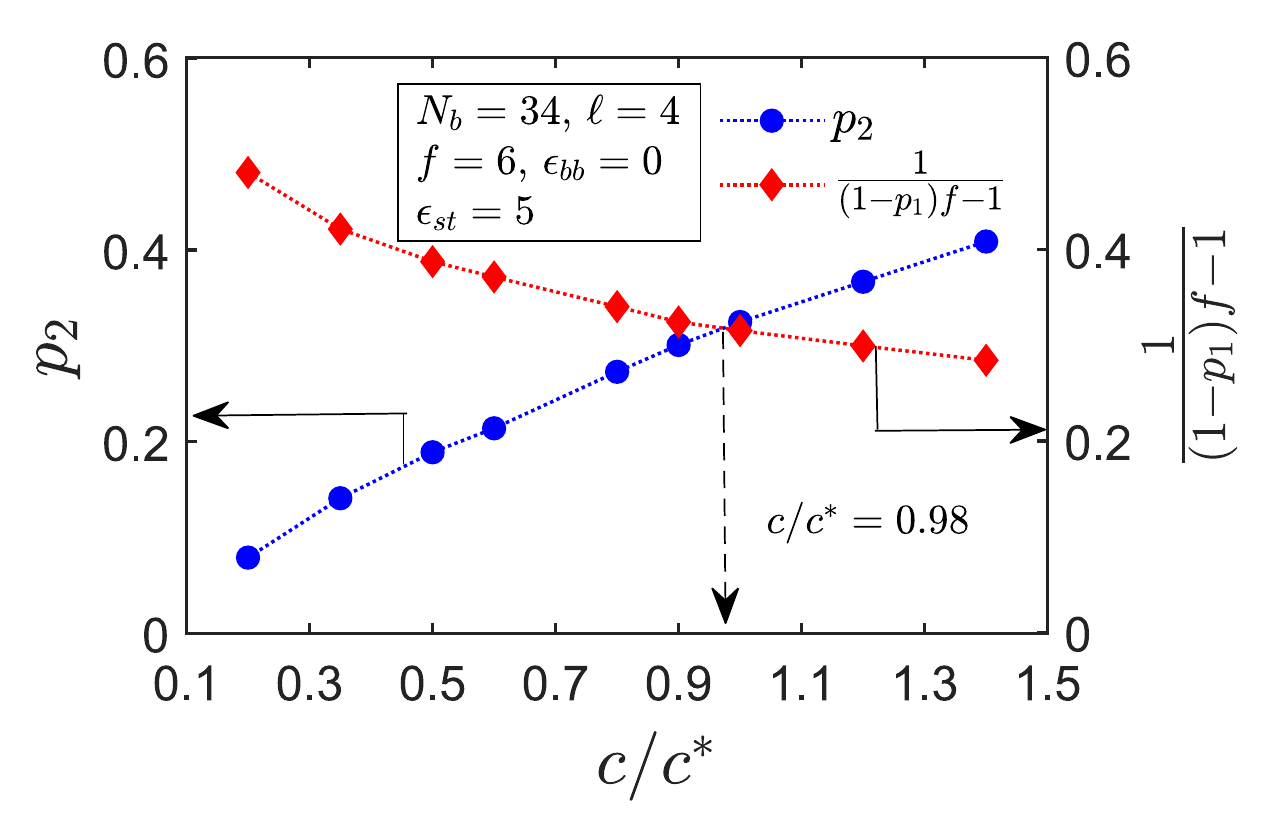} 
        & \includegraphics[width=80mm]{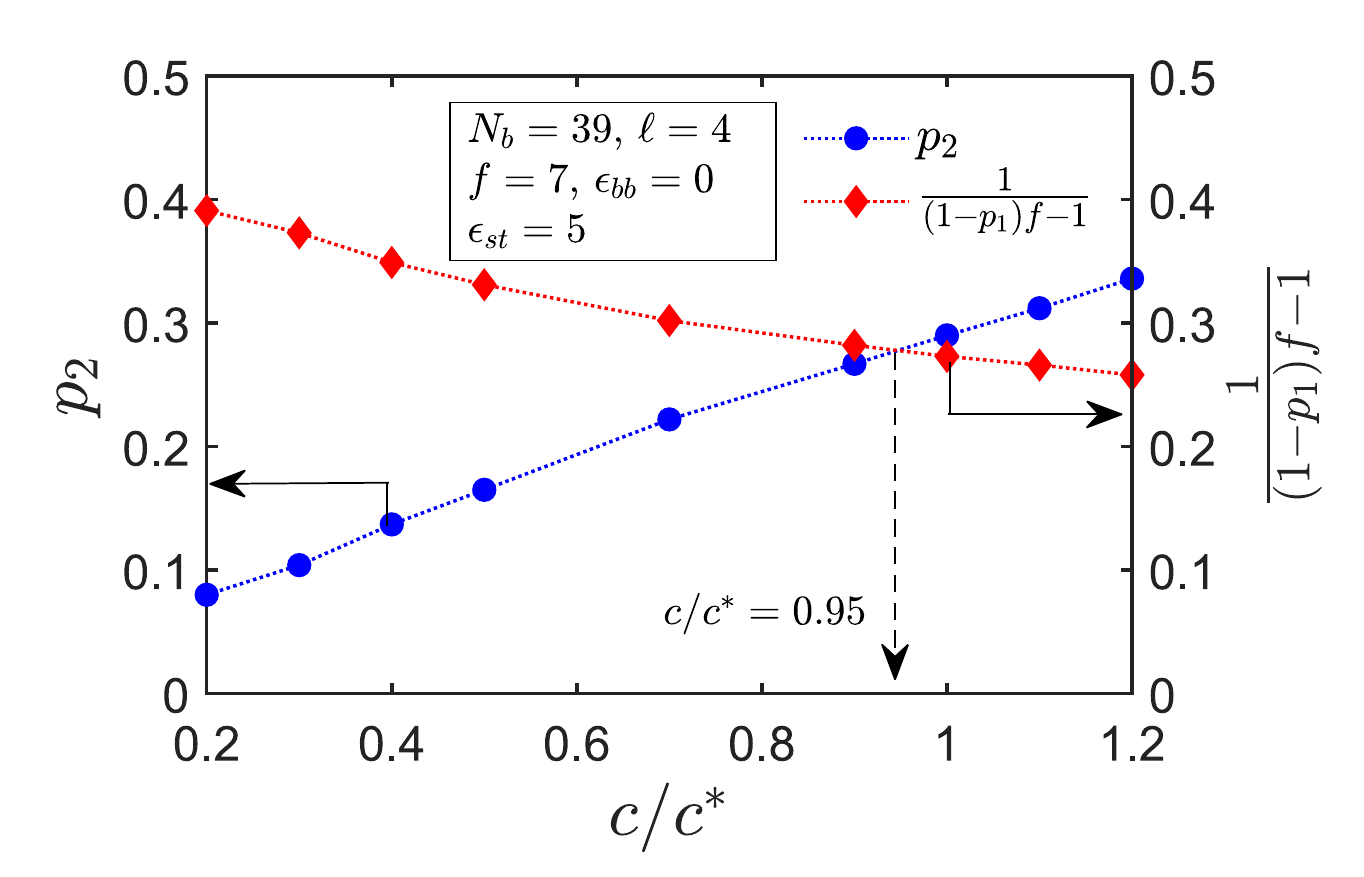} \\
        (a) & (b) \\
        \includegraphics[width=80mm]{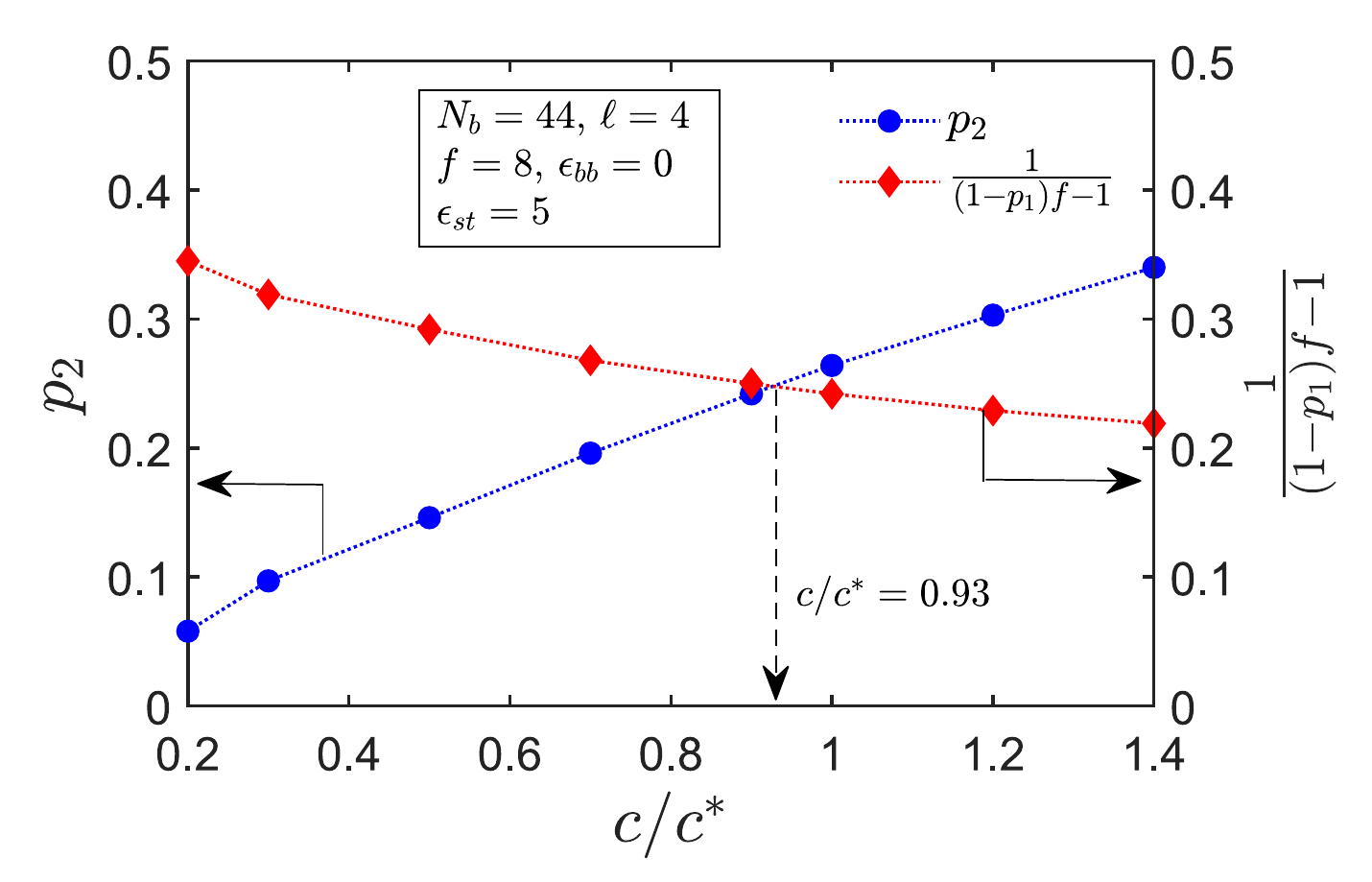} 
        & \includegraphics[width=80mm]{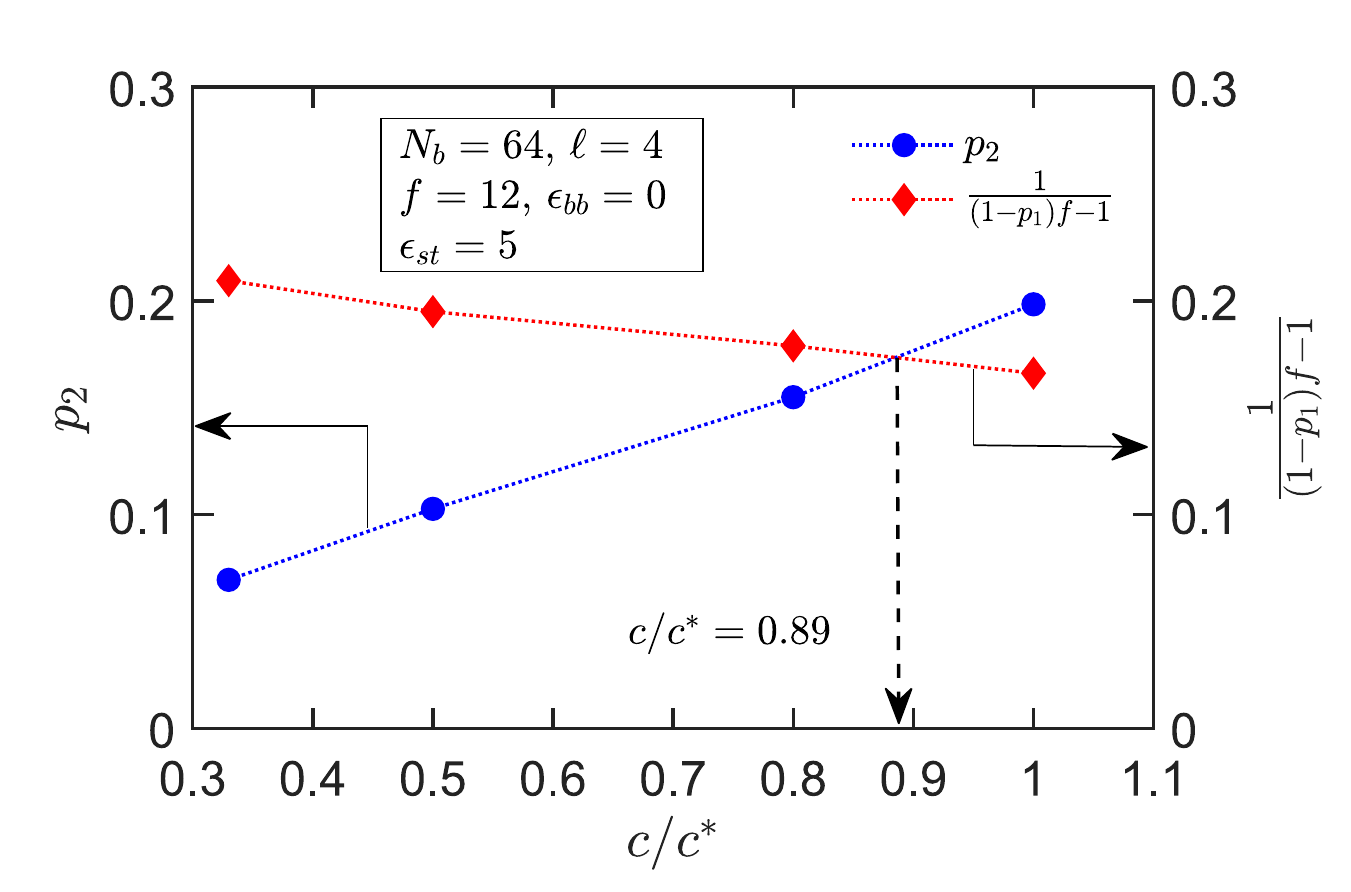} \\
        (c) & (d) \\
    \end{tabular}
    }	
	\vskip-5pt
	\caption{Variation of inter-chain conversion, $p_2$, and the ratio $1/[(1-p_1)f-1]$ with scaled concentration, $c/c^*$, for systems with chain lengths (a) $N_b=34$, (b) $N_b=39$, (c) $N_b=44$ and (d) $N_b=64$. For all the chain lengths, the values of spacer length, $\ell=4$, sticker strength, $\epsilon_{st}=5$, and backbone monomer interaction strength, $\epsilon_{bb}=0$, are held constant. The values of $c/c^*$ at the point of intersection of the two curves for various chain lengths are (a) $0.98$, (b) $0.95$, (c) $0.93$ and (d) $0.89$.}
\label{fig:classical_gel}
	\vskip-15pt
\end{figure*} 

From a geometric perspective, the inception of gelation can be defined as the monomer concentration at which a system spanning network occurs~\cite{Stauffer,TanakaPRL1989,Tanaka1998}. The concentration at which such a percolation transition occurs, denoted here by $c_{g_1}$, can be determined by calculating the probability of finding a cluster of chains that spans the simulation box, and estimating how this probability changes with changing concentration. The so-called spanning probability  is computed here by identifying the chains that belong to a cluster from the chain connectivity matrix, and comparing the maximum span of the cluster with the box size, $L$. If the span of a cluster of chains along any direction is greater than or equal to the box size, the cluster is identified as system spanning. The spanning probability is computed over an ensemble of 64 to 128 independent trajectories,  where each trajectory consists of a set of data collected at an interval of 1000 to 5000 non-dimensional time steps over the entire production run. For an infinitely large simulation box, the probability of finding a cluster that spans the entire box, at a low monomer concentration below the gelation threshold, is essentially zero. With increasing concentration, the spanning probability is expected to undergo a sharp transition at the monomer concentration that corresponds to the percolation transition, and instantly attain a value of one. For a finite box size, however, the variation of spanning probability with concentration is expected to be more gradual, since even at low concentrations, there is a finite probability of finding a system spanning cluster. In this case, the gelation threshold can be determined by computing the spanning probability for a number of systematically increasing box sizes. It is expected that if the studied systems are large enough, their spanning probability curves will intersect at a common point, which represents an accurate estimate of the percolation threshold~\cite{Stauffer,Christensen}. Here, simulations have been carried out for three different box sizes, and in each case, the spanning probability (SP) has been computed as a function of monomer concentration, as displayed in Figs.~\ref{fig:gelpt}. Rather than each box size leading to a distinctive spanning probability curve, which intersect at a unique point, it is observed that at low concentrations, the curves for different box sizes overlap within error bars, probably as a result of insufficiently long chains and the box sizes not being large enough. Beyond a certain scaled concentration, however, the curves are seen to separate and diverge. The location of this change in behaviour has been identified here as the concentration at which percolation transition occurs. The value of the scaled concentration, $c_{g_1}/c^{*} \approx 0.3$, is found to be independent of chain length, for systems with a fixed spacer length $\ell$, sticker strength $\epsilon_{st}$, and backbone monomer solvent quality $\epsilon_{bb}$, as can be seen in Figs.~\ref{fig:gelpt}.

\begin{figure}[ptbh]
    \centerline{
    \begin{tabular}{c}
   {\includegraphics[width=68mm]{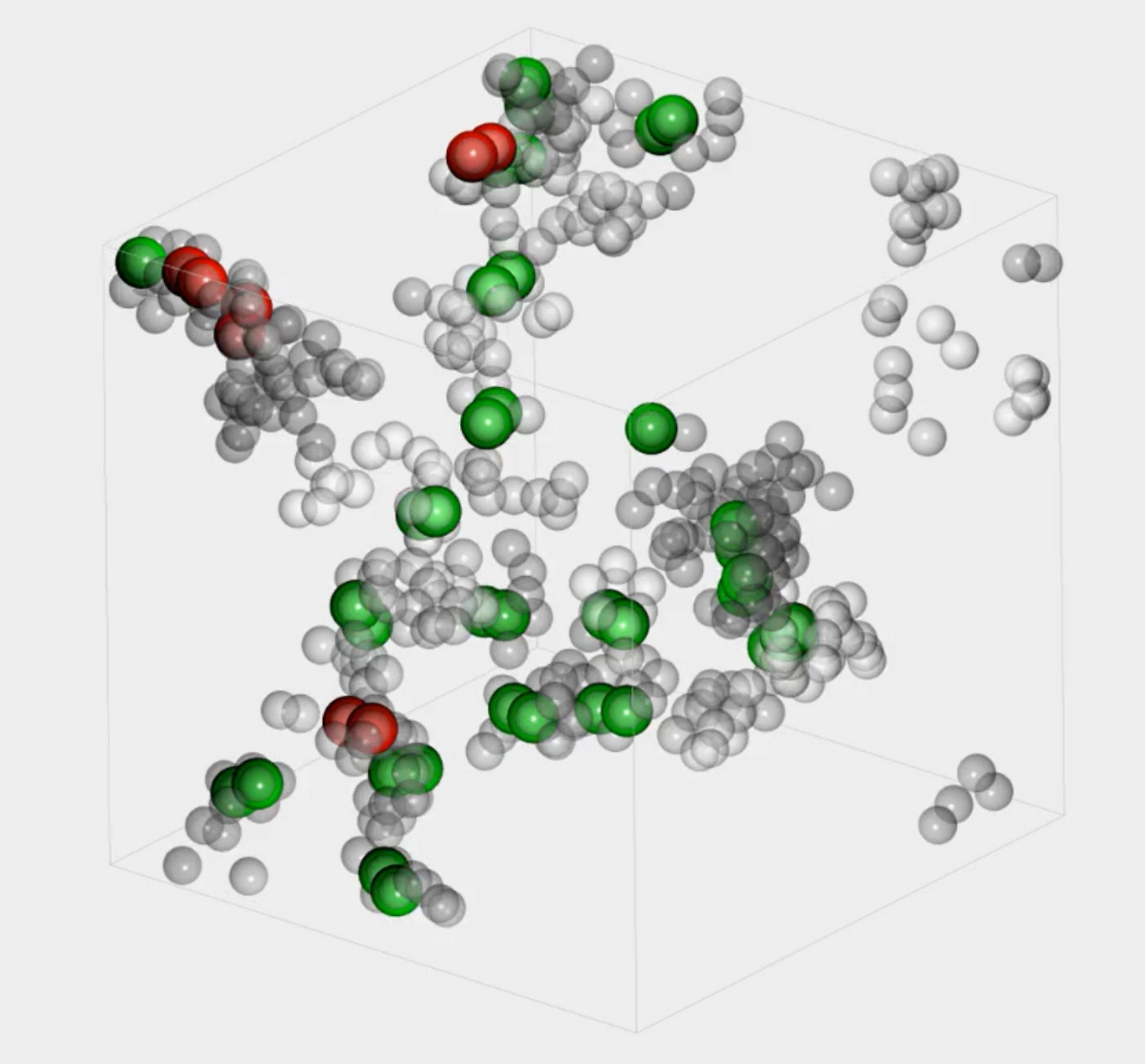}} \\
    { (a) }\\[5pt]
         \includegraphics[width=68mm]{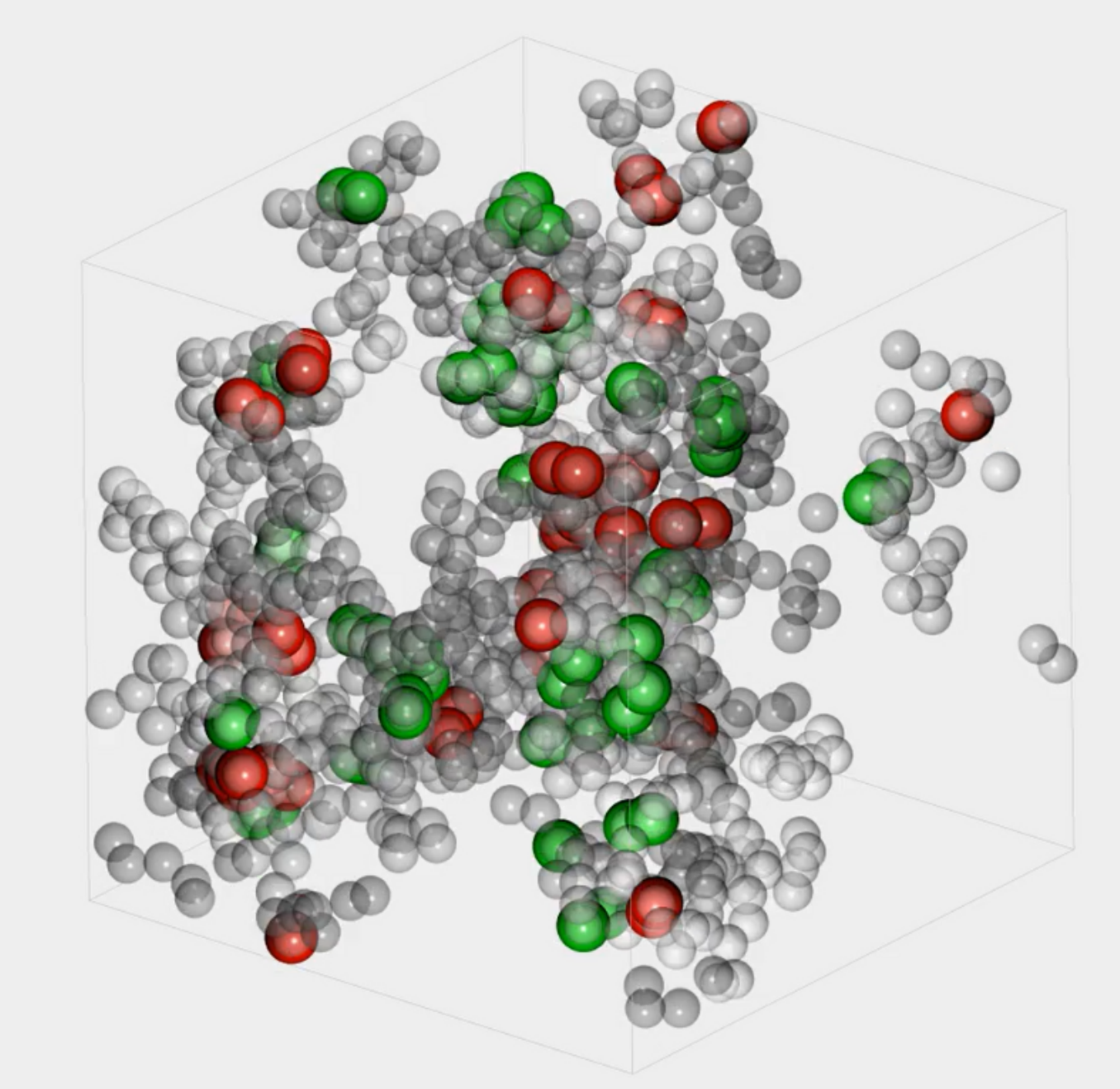}  \\
    { (b) }\\[5pt]
             \includegraphics[width=68mm]{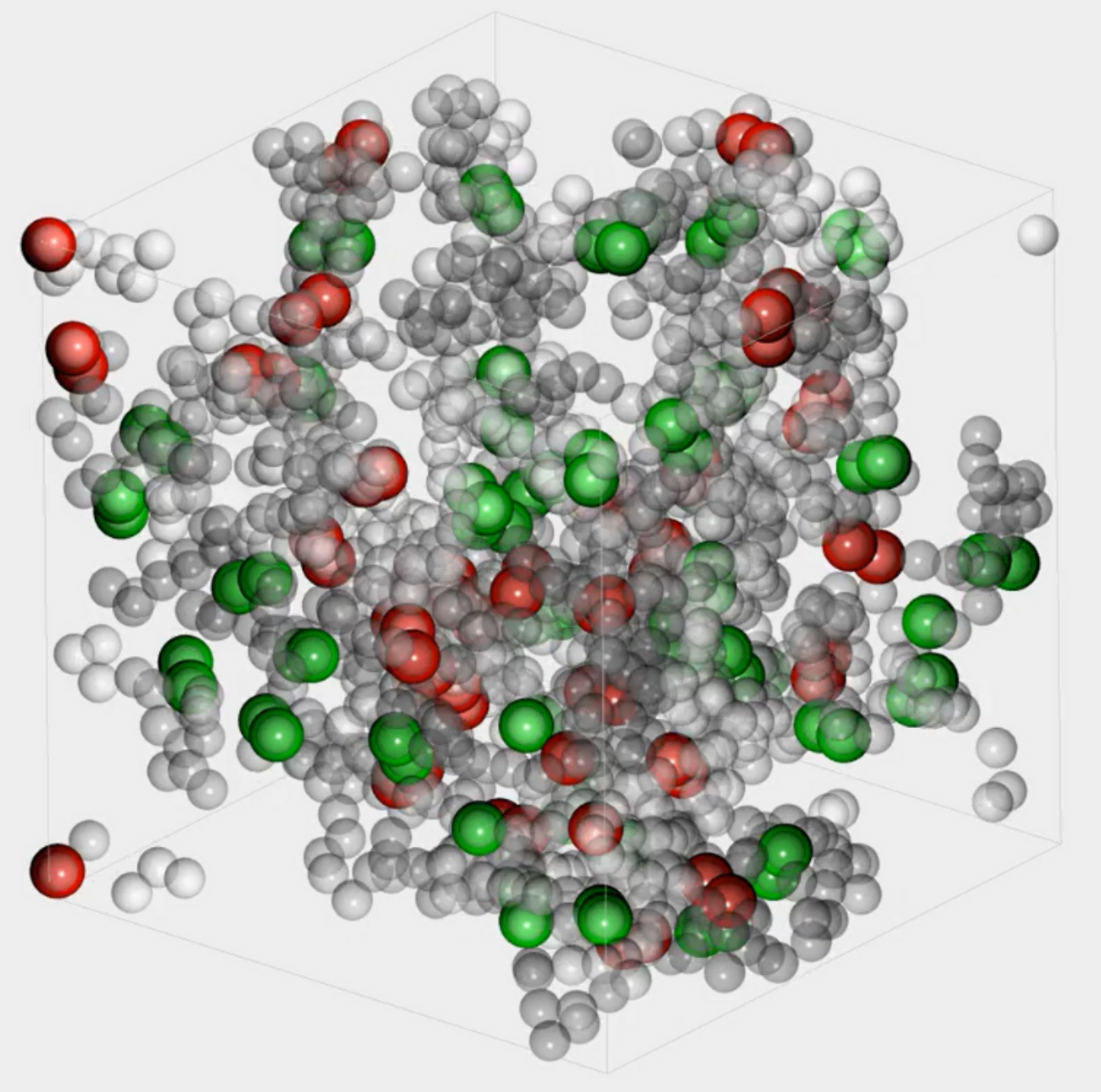}  \\
    { (c) }\\[5pt]
        \end{tabular}
    }
 \caption{{Snapshots of the simulation box for a system with parameters $\{N_b=34, \ell=4, f=6, \epsilon_{bb} = 0, \epsilon_{st}=5 \}$ at (a) $c/c^*=0.2$, (b) $c/c^*=0.5$ and (c) $c/c^*=1.0$. The red beads indicate inter-chain associations, while the green beads represent intra-chain associations}. \label{fig:SnapShot}}
\end{figure} 

In associative polymer solutions, the existence of geometric percolation does not necessarily imply the existence of a persistent network since the bonds between stickers are weak and reversible~\cite{SanatK,SanatDoug}. As mentioned earlier, an alternative approach~\cite{SanatK,SanatDoug} identifies the occurrence of an incipient gel in sticky polymer solutions with the onset of bimodality in the chain-cluster size distribution, $P(m)$, where $m$ is the number of chains in a cluster. Fig.~\ref{fig:clustdist} displays $P(m)$ computed at different monomer concentrations, for two different values of chain length $N_b$, at the specified values of $\ell$, $\epsilon_{st}$ and  $\epsilon_{bb}$. The plots suggest that the distribution function decreases monotonically with increasing $m$ at low monomer concentrations, but becomes bimodal with increasing concentration. The occurrence of a peak at a large cluster size is considered to be correlated with the existence of percolating chain-clusters. Here, the concentration at which the slope of $P(m)$ versus $m$ first becomes positive, at some value of $m$, is considered to be the location of the sol-gel transition, and is denoted by $c_{g_2}$. For the given parameter values, the onset of bimodality is found to occur at $c_{g_2}/c^{*} \approx 1.0$, which is significantly higher than $c_{g_1}/c^*$, the location of the percolation transition. The value of the gelation concentration, $c_{g_2}/c^{*}$, is found to be independent of chain length, as in the case of $c_{g_1}/c^*$. It is apparent from Figs.~\ref{fig:gelpt} that as the monomer concentration approaches $c_{g_2}/c^*$, the spanning probability tends to unity, suggesting that, at this concentration, there is a significant increase in the probability of finding a cluster with size sufficiently large to span the entire system.  

The third and final signature of gelation considered here is the proposal by \citet{RnSstatics} that the maxima in the free chain concentration coincides with the sol-gel transition. Recall that this assumption is the basis for their derivation of Eq.~(\ref{FS}). Figure~\ref{fig:freechain} is a plot of the free chain concentration versus monomer concentration, for various values of chain length $N_b$, at the specified values of $\ell$, $\epsilon_{st}$ and  $\epsilon_{bb}$. Free chains, i.e., those with no inter-chain associations, are essentially chain clusters with only one chain in them, and can consequently be identified with the help of the same cluster computation algorithm used here for determining the other two signatures of gelation. The value of the concentration corresponding to the maximum  for each symbol set in Fig.~\ref{fig:freechain}, denoted here by $c_{g_3}$, is established by fitting a parabola to the data close to the maxima, and finding the location at which the slope is zero. We find that $c_{g_3}/c^* \approx 0.5$, which lies between the two scaled concentrations, $c_{g_1}/c^*$ and $c_{g_2}/c^*$, the locations of the sol-gel transition from the two approaches discussed previously. As is clear from Fig.~\ref{fig:freechain}, consistent with the observations for $c_{g_1}/c^*$ and $c_{g_2}/c^*$, the scaled concentration $c_{g_3}/c^{\ast}$, is also independent of chain length. The fact that $c_{g_3}/c^* > c_{g_1}/c^*$ implies that, even after a system spanning network is formed, new chains added to the system join the sol-phase for a range of concentrations, before joining the gel-phase.

\begin{figure}[tb]
    \centerline{
    \begin{tabular}{c}
   {\includegraphics[width=76mm]{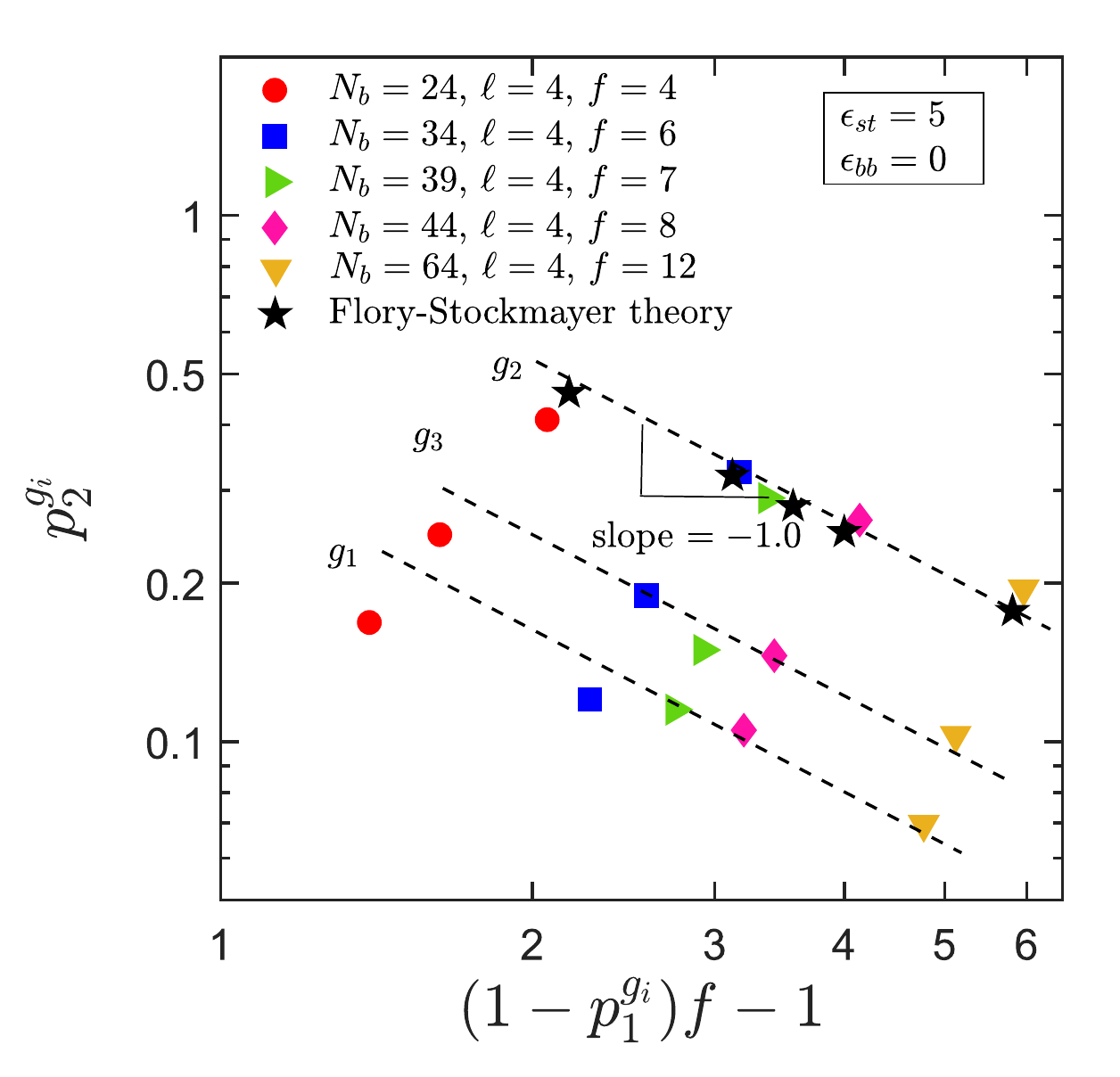}} \\
    { (a) }\\[5pt]
         \includegraphics[width=73mm]{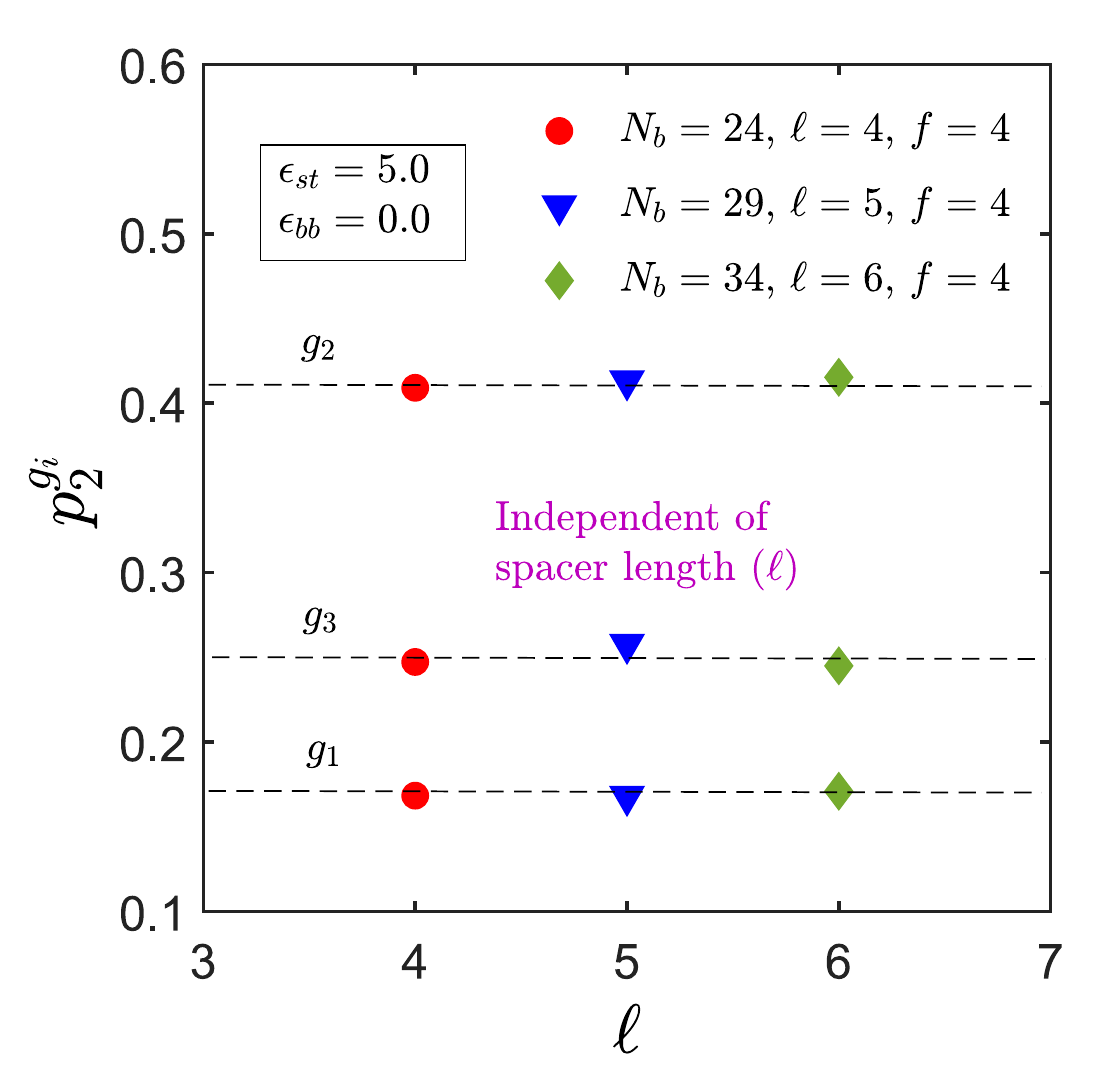}  \\
    { (b) }\\[5pt]
        \end{tabular}
    }
	\vskip-10pt
	\caption{Scaling of inter-chain degree of conversion, $p_2^{g_i}$, at the gel-point  in Regime~II, predicted by the three different signatures of gelation, with (a) $(1-p_1)f-1$, and (b) the number of spacer monomers $\ell$, for systems with constant number of stickers $f$, and different chain lengths $N_b$. The sticker strength and backbone solvent quality are kept constant at $\epsilon_{st}=5.0$ and $\epsilon_{bb}=0$, respectively. Each symbol shape represents a system with a particular chain length $N_b$ and the dashed lines are the Flory-Stockmayer theory predictions at each of the gelation signatures.
\label{fig:p2_vs_f}}
	\vskip-15pt
\end{figure} 

According to Flory-Stockmayer theory~\cite{FloryBook,Stock2} (appropriately modified by Dobrynin~\cite{Dob}), the gel-point coincides with the value of $c/c^*$ at which $p_2 = 1/ [(1-p_1)f-1]$. The variation of inter-chain conversion, $p_2$, and the ratio $1/[(1-p_1)f-1]$ with scaled concentration, $c/c^*$, is presented in Figs.~\ref{fig:classical_gel} for systems with different chain lengths, $N_b$, at constant $\ell$, $\epsilon_{st}$ and $\epsilon_{bb}$. Clearly, the points of intersection between the two curves in the different subfigures of Figs.~\ref{fig:classical_gel} are the Flory-Stockmayer theory estimates of $c/c^*$ {at the gel point,} in all these cases. The estimate of the gel-point  {appears to be} independent of chain length, and close to the value evaluated from the onset of bimodality in the chain-cluster size distribution, {i.e., $c/c^* \approx 1$}.

{It is intriguing that the two gelation concentrations $c_{g_1}/c^* \approx 0.3$ and $c_{g_3}/c^* \approx 0.5$, corresponding to the inception of a system spanning network, and to the free-chain concentration maximum, respectively, occur below the overlap concentration, while that corresponding to the onset of bimodality $c_{g_2}/c^* \approx 1.0$, is more in accord with the intuitive expectation of gelation occurring at $c_g \approx c^*$. This is perhaps related to the fact that geometrical percolation can occur even though the solution is not solid-like, which is the common understanding of a gel. As shown previously~\cite{SanatDoug}, the volume fraction at the percolation threshold is a function of the sticker-sticker interaction strength $\epsilon_{st}$, and it approaches the onset of solid-like behaviour with increasing $\epsilon_{st}$. For relatively low values of $\epsilon_{st}$, while a system spanning network might occur, the frequent pairing and unpairing of stickers leads to a gel that is not rigid~\cite{SanatDoug}. Indeed, as shown by the simulation snapshots in Figs.~\ref{fig:SnapShot}, there is no discernible change in the distribution of chains across the simulation cell when the geometrical (and free-chain maximum concentration) is crossed.} It would be interesting to {study the dependence of the three gelation signatures on the sticker strength,} and to examine if any of the estimates of the sol-gel transition concentration determined here coincides with that determined through rheological experiments~\cite{Winter:2000gw,Li:1997cu, Li:1997gi}, {which would identify the transition to solid-like behaviour}. Addressing this question satisfactorily would require the incorporation of hydrodynamic interactions, in order for the dynamics of sticky polymer solutions to be captured accurately.

\begin{figure*}[t]
    \centerline{
    \begin{tabular}{c c}
        \includegraphics[width=72mm]{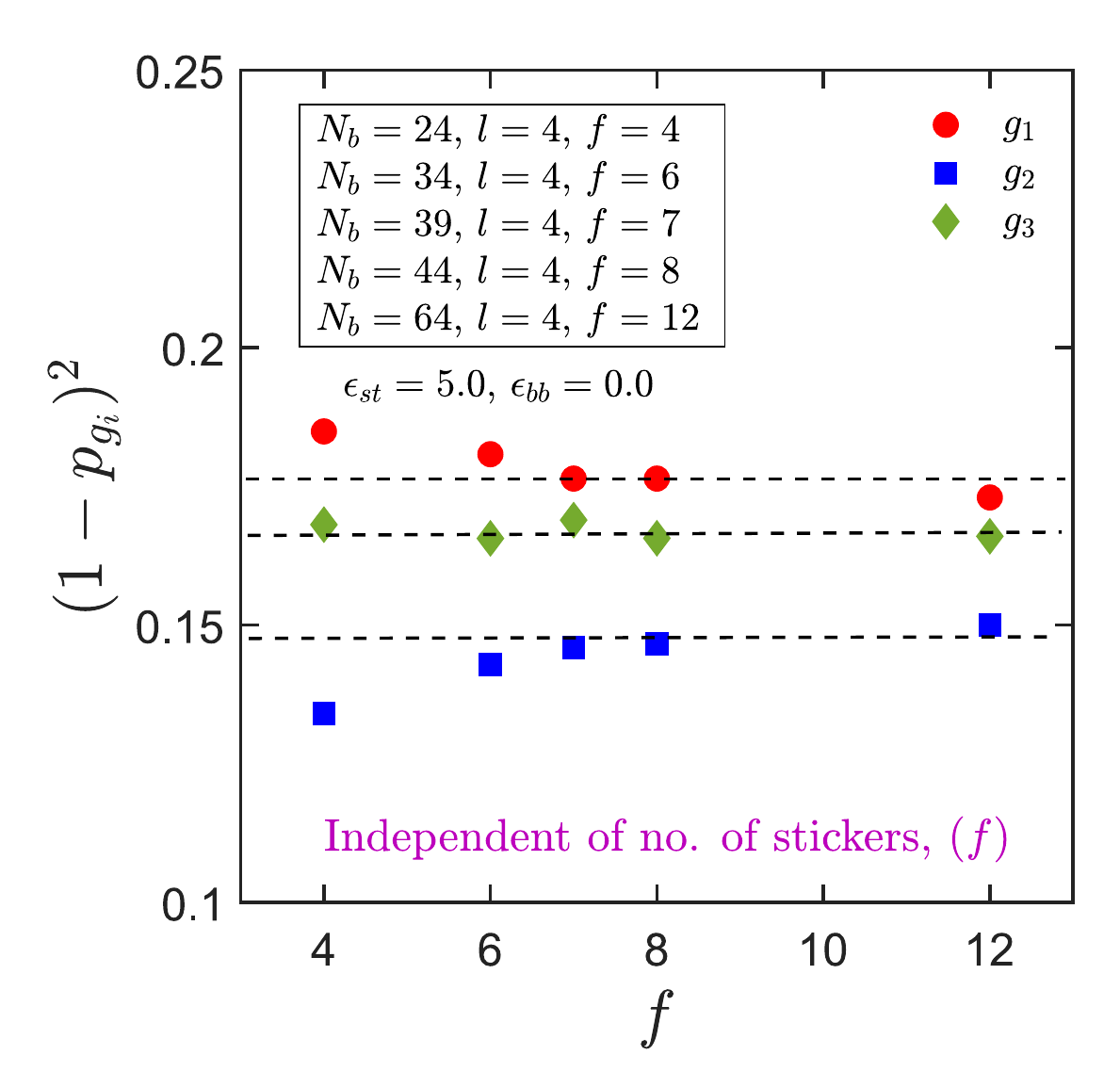} 
        & \includegraphics[width=70mm]{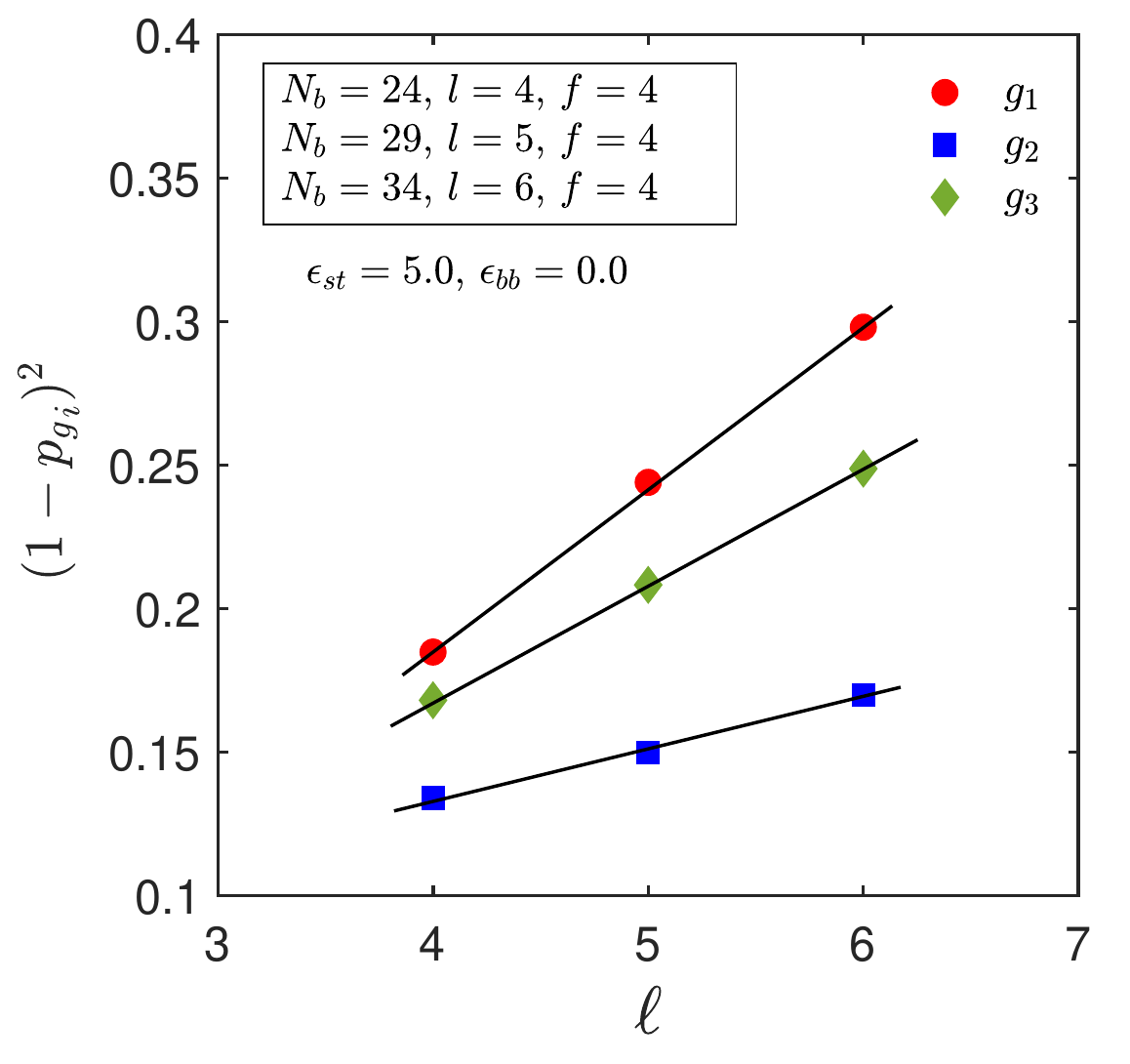} \\
        (a) & (b) \\
        \includegraphics[width=71mm]{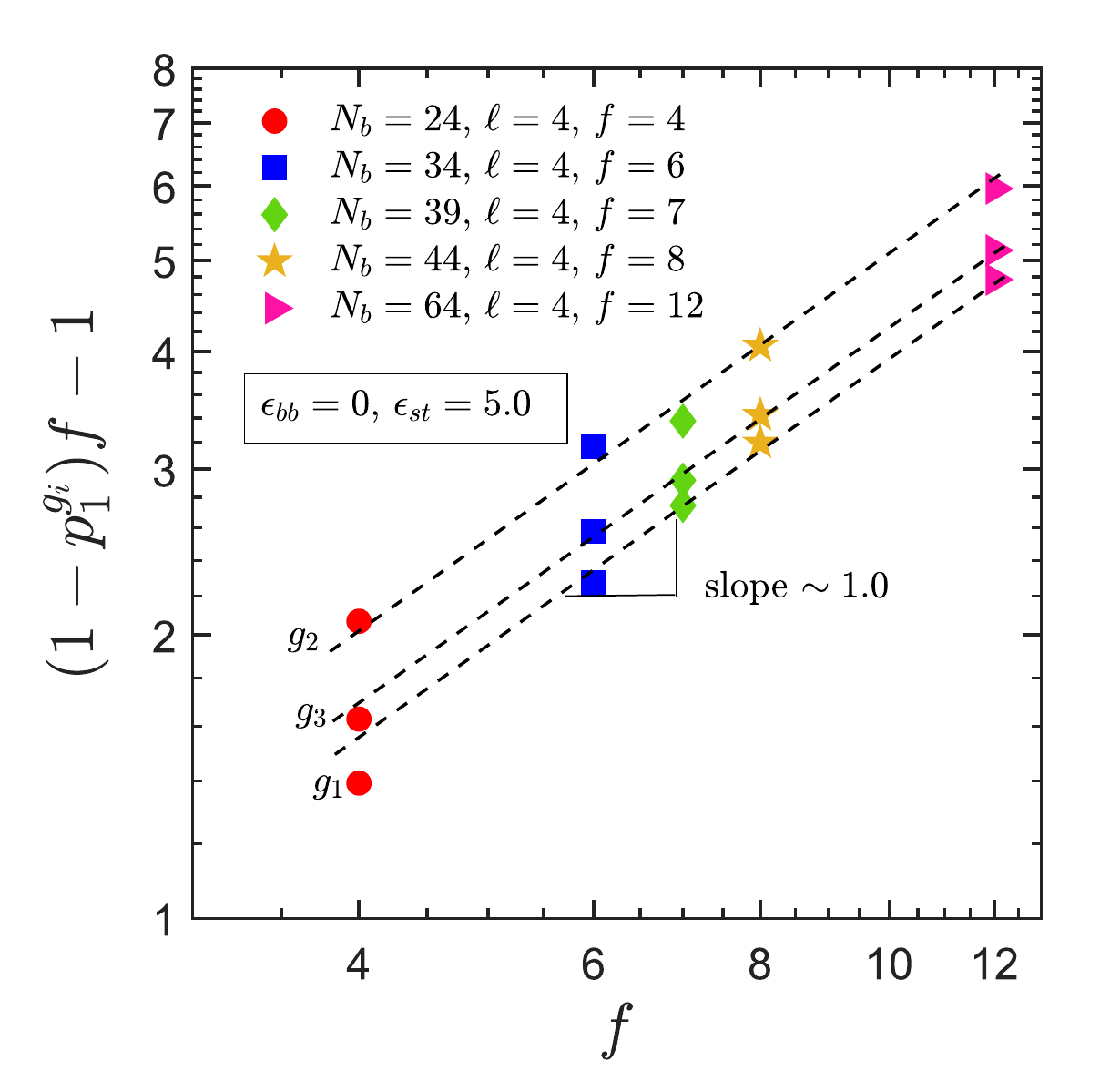} 
        & \includegraphics[width=71mm]{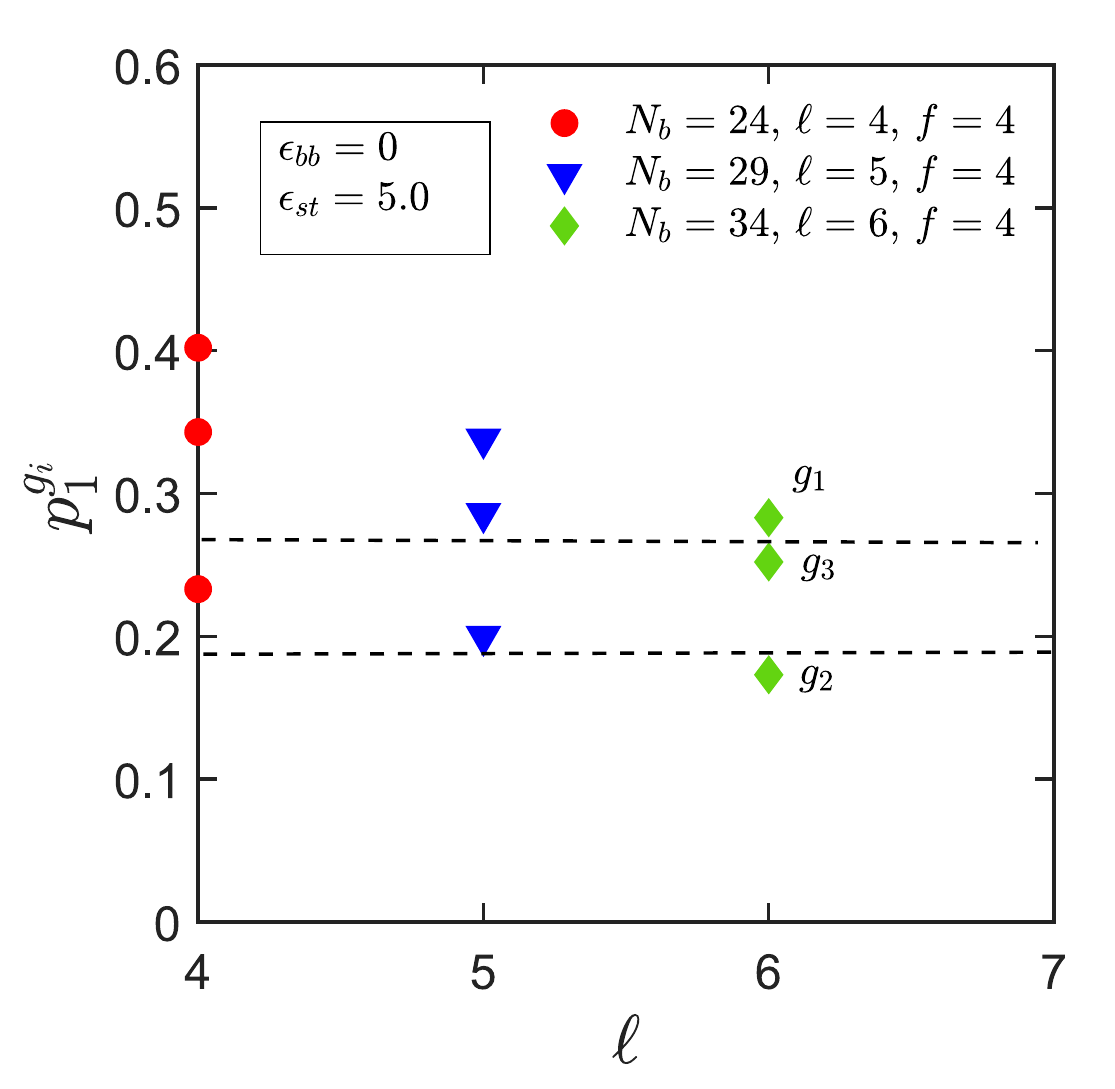} \\
        (c) & (d) \\
    \end{tabular}
    }
	\vskip-10pt
	\caption{Scaling with the number of stickers per chain, $f$, and the spacer length, $\ell$, in Regime~II, of the quantities $(1-p_{g_i})^2$ [(a) and (b), respectively], and $[(1-p_1^{g_i})f-1]$ [(c) and (d), respectively], that occur in Eq.~(\ref{Eq:gel_line}), for the different signatures of gelation.	
\label{fig:pgel_vs_fl}}
	\vskip-15pt
\end{figure*} 

Having determined the concentrations at the gel-point predicted by the different signatures of gelation, we can now verify if the dependence of $p_2^{g}$ on $p_1^{g}$ and $f$ coincides with the  prediction of the modified form of the Flory-Stockmayer theory~\cite{FloryBook,Stock2,Dob}. In other words, we can check if the dependence of $p_2^{g_i}$  on $[(1-p_1^{g_i})f-1]$ obeys Eq.~(\ref{FS-D}), where $g_i = g_1, g_2, g_3$, represents the three signatures of gelation. It is clear from Fig.~\ref{fig:p2_vs_f}~(a) that for all the three signatures of gelation, $p_2^{g}$ follows a linear scaling with the inverse of $((1-p_1^{g_i})f-1)$ for sufficiently long chains as predicted by Eq.~(\ref{FS-D}). Compared to gelation signatures $g_2$ and $g_3$, however, the approach to linear scaling for $g_1$ occurs at larger values of $N_b$. As mentioned earlier, the Flory-Stockmayer theory estimate of the gel-point matches well with the gel-point determined from the onset of bimodality in the chain-cluster size distribution ($g_2$).

\begin{figure}[ptbh]
    \centerline{
   {\includegraphics[width=80mm]{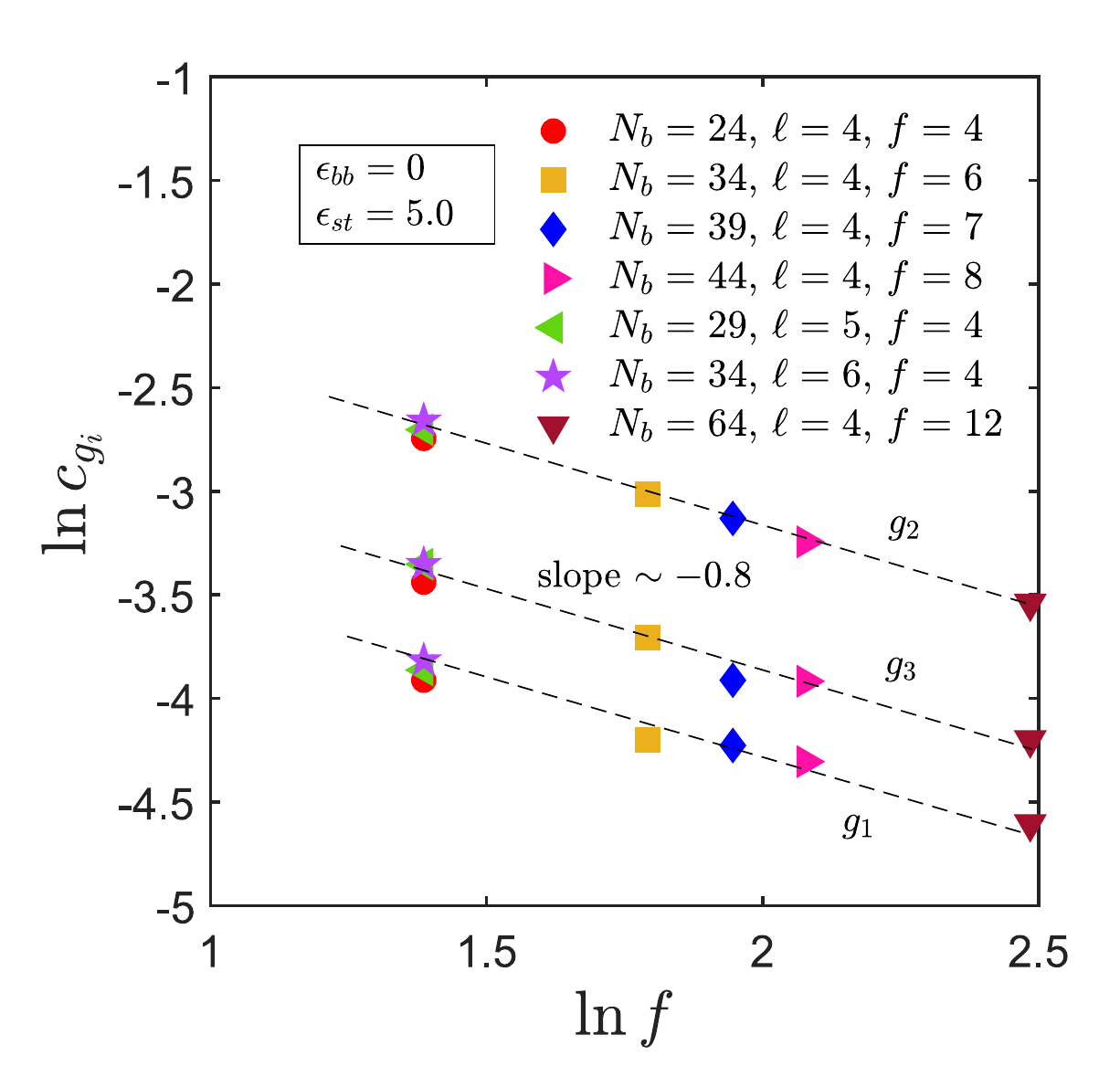}}     
    }
     \vskip-15pt
 \caption{Gelation concentration, $c_{g_i}$, as a function of the number of stickers in a chain (see Eq.~(\ref{Eq:gel_line2})), for the three different signatures of gelation in Regime~II. The sticker strength and backbone solvent quality are kept constant at $\epsilon_{st}=5.0$ and $\epsilon_{bb}=0$, respectively. Each symbol shape represents a system with a particular chain length $N_b$, spacer length $\ell$ and number of stickers per chain $f$ and the dashed lines are the scaling predictions for each of the gelation signatures.
\label{fig:gel_line}}
\end{figure}

The independence of the value of $c/c^*$ at the gel-point from chain length $N_b$ has been demonstrated for all the gelation signatures by keeping the spacer length $\ell$ fixed, while varying the number of stickers $f$ per chain. In Fig.~\ref{fig:p2_vs_f}~(b), the dependence of the fraction of inter-chain associations at the gel-point, $p_2^{g_i}$, on the spacer length, $\ell$ (and consequently, $N_b$), for the three signatures of gelation, is displayed for chains with a fixed number of stickers $f$.  Clearly,  $p_2^{g_i}$ is independent of $\ell$. Since $p_2^{g_i} = 1/[(1-p_1^{g_i})f-1]$, and $p_2^{g_i}$ is independent of $\ell$, this implies that $p_1^{g_i}$ should also be independent of $\ell$, at fixed values of $f$. This is demonstrated shortly below.  

The verification of the expression for the gelation line, Eq.~(\ref{Eq:gel_line}), as mentioned earlier, is examined here for the restricted case of constant $\hat \tau$ and $\epsilon_{st}$. Further, we wish to eliminate the quantities $(1-p_1^g)f-1$ and $(1-p_g)^2$ from Eq.~(\ref{Eq:gel_line}) so as to determine the dependence of $c_g$ on just the sticky chain properties, $\ell$ and $f$. In order to do so, the dependence of $(1-p_1^{g_i})f-1$ and $(1-p_{g_i})^2$  on $\ell$ and $f$ is displayed in Figs.~\ref{fig:pgel_vs_fl}. It is clear from Fig.~\ref{fig:pgel_vs_fl}~(a) that at a fixed value of $\ell$, $(1- p_{g_i})^2$ is independent of $f$ for all the three different signatures of gelation, while Fig.~\ref{fig:pgel_vs_fl}~(b) implies that for a fixed value of $f$, $(1- p_{g_i})^2$ scales linearly with $\ell$ in all three cases. Fig.~\ref{fig:pgel_vs_fl}~(c) suggests that when  $\ell$ is constant, $[(1-p_1^{g_i})f-1]$ scales linearly with $f$, for sufficiently long chains. Finally, as anticipated above, Fig.~\ref{fig:pgel_vs_fl}~(d) indicates that when $f$ is constant, $p_1^{g_i}$ tends to a constant value, independent of $\ell$, when $N_b$ becomes large. As observed earlier, the approach to asymptotic behaviour is slower in the case of $g_1$, compared to that of $g_2$ and $g_3$. 

Substituting the dependences on $\ell$ and $f$ for the quantities $[(1-p_1^g)f-1]$ and $(1-p_g^2)$, summarised in Figs.~\ref{fig:pgel_vs_fl}, into Eq.~(\ref{Eq:gel_line}), leads to the following expression for the monomer concentration along the gelation line,
\begin{equation}
\label{Eq:gel_line2}
c_{g_i} \sim \left[ \frac{\hat{\tau}^{\tfrac{\nu\theta_2}{(3\nu-1)}}}{f\,g_{ss}}\right]^{\tfrac{3\nu-1}{\nu(3+\theta_2)-1}} \sim f^{\,\, -\tfrac{4}{5}}
\end{equation}
where the assumptions of constant $\hat \tau$ and $\epsilon_{st}$, and the values, $\nu = 3/5$ and $\theta_2 = 1/3$, have been used to derive the second expression, which indicates that $c_{g_i}$ depends only on $f$ and not on $\ell$. It is clear from the results displayed in Fig.~\ref{fig:gel_line}, that simulations validate the revised expression for the gelation line, Eq.~(\ref{Eq:gel_line2}), for all the three different signatures of gelation. The overlapping of data corresponding to different values of $\ell$, for systems with $f=4$, also demonstrates the independence of $c_{g_i}$ from the number of spacer monomers between stickers. It is undoubtedly desirable to verify experimentally both the general and restricted forms of the dependence of $c_{g_i}$ on system parameters given in Eq.~(\ref{Eq:gel_line2}), as it would simultaneously permit an evaluation of the correctness of the scaling of $p_2$ predicted by Eq.~(\ref{FS-D}), the dependences revealed in Figs.~\ref{fig:pgel_vs_fl}, and the correct value of the des Cloizeaux exponent $\theta_2$.

\section{\label{sec:phase} Phase separation and the breakdown of scaling}

\begin{figure*}[t]
    \centerline{
    \begin{tabular}{cc}
        \multicolumn{2}{c}{\includegraphics[width=75mm]{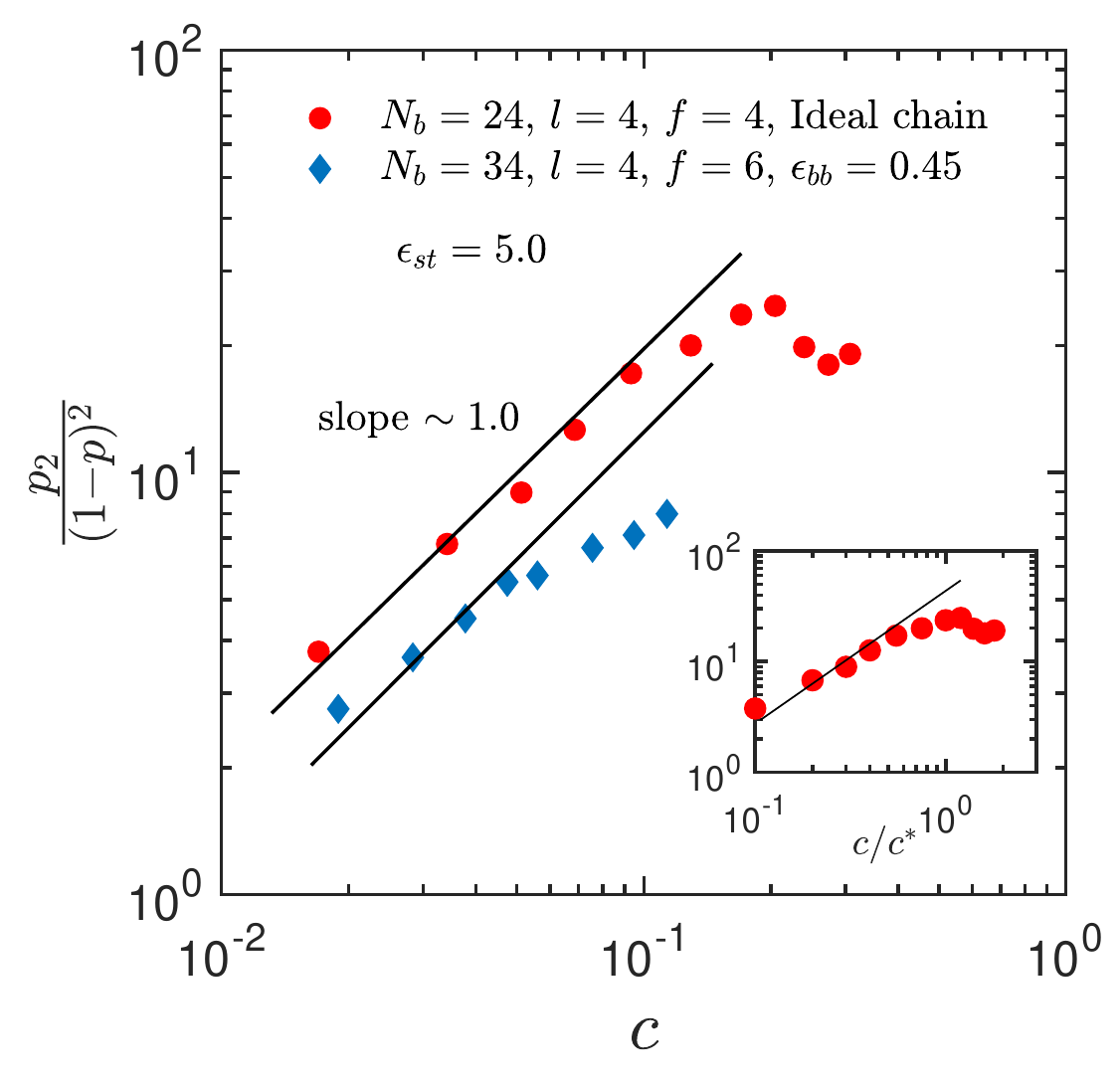}} \\
          \multicolumn{2}{c}{ (a) }\\[10pt]
         \includegraphics[width=63mm]{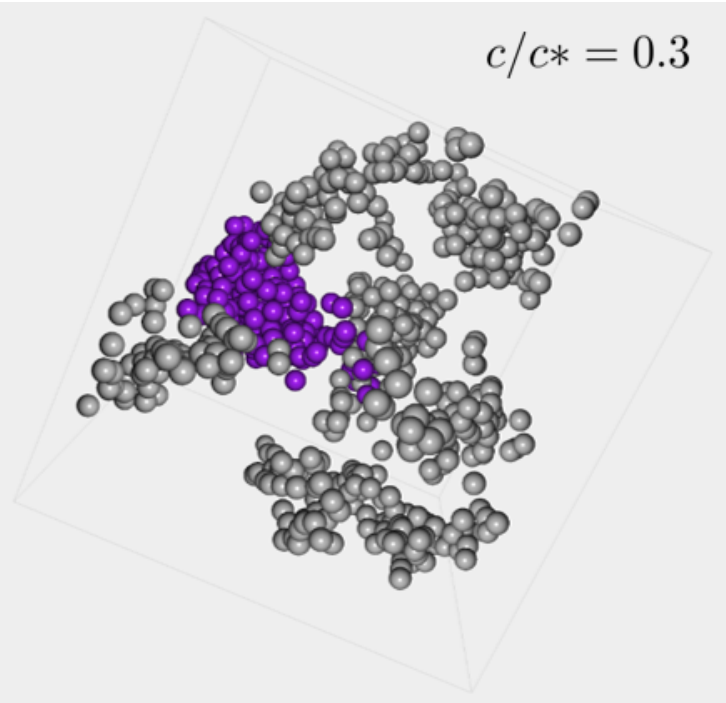} 
     &        \includegraphics[width=65mm]{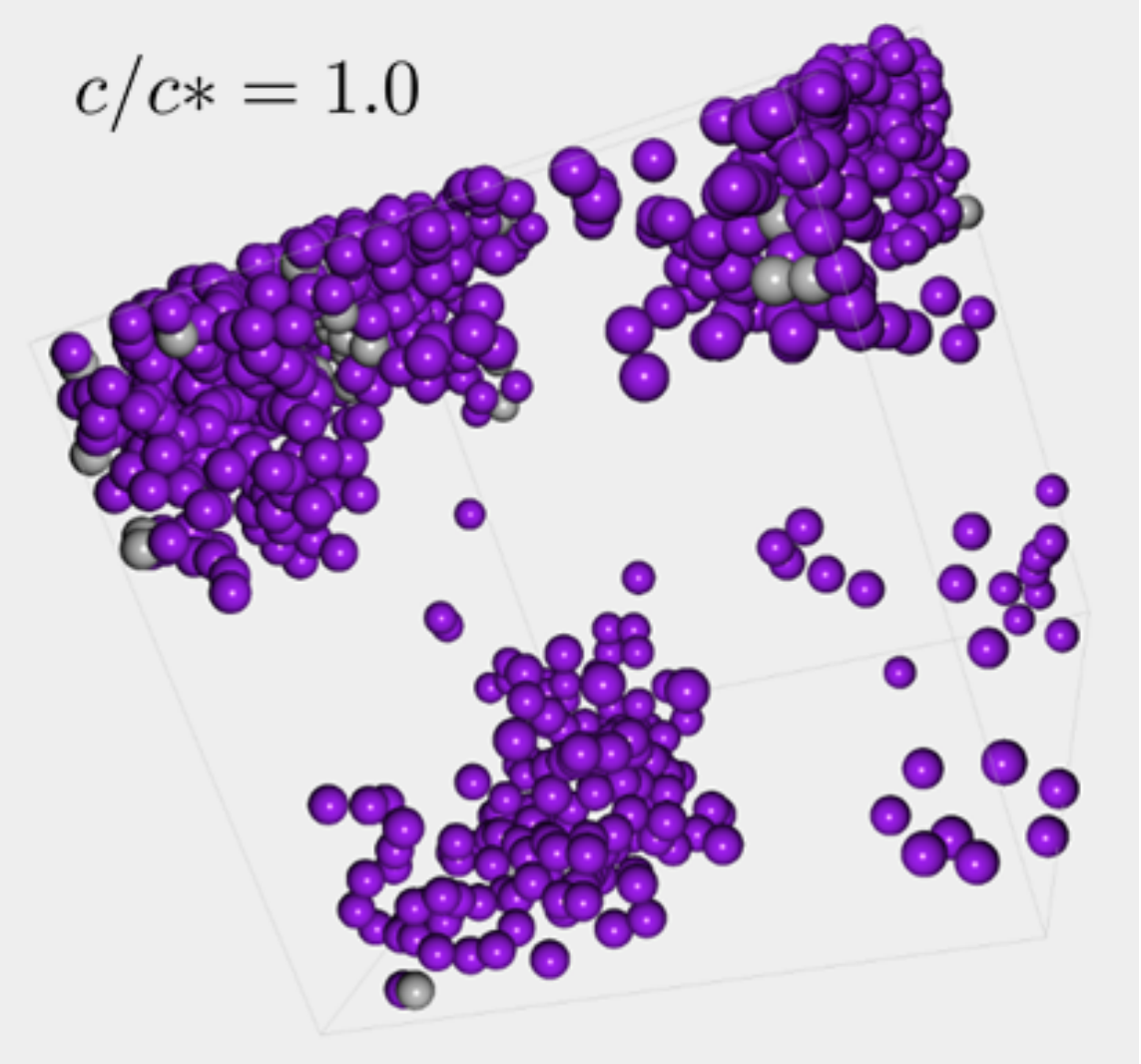} \\
         (b) &  (c) \\
    \end{tabular}
    }
    \caption{(a) Rescaled inter-chain degree of association as a function of monomer concentration, with backbone monomers under $\theta$-solvent conditions. For $N_b =24$, $\theta$-solvent conditions for the backbone are obtained by simulating ghost chains, while for $N_b = 34$, $\theta$-solvent conditions for the backbone are obtained by using the SDK potential with $\epsilon_{bb} = \epsilon_\theta = 0.45$. Inset is a plot of the ratio versus $c/c^*$, for the case $N_b = 24$. The solid lines are drawn with slopes equal to the prediction by scaling theory, while symbols represent simulation data. (b) and (c) Snapshots from the simulations for the system with $N_b=24$ in sub-figure~(a), at concentrations $c/c^*= 0.3$, and $c/c^*=1.0$, respectively. The purple coloured beads belong to chains that are all a part of a single cluster, while the colour grey is used to represent beads in chains that do not belong to this cluster.  \label{fig:phase_sep_theta}}
	\vskip-10pt
\end{figure*}

A solution of sufficiently long polymers under poor solvent conditions will phase separate with increasing monomer concentration. This applies both to homopolymer solutions~\cite{RubCol}, and to sticky polymer solutions~\cite{Dob}.  {In the case of sticky polymer solutions}, \citet{Aritra2019} {have calculated the second osmotic virial coefficient $B_2$ by determining the potential of mean force, $U(r)$, between a pair of polymer chains with their centres of mass separated by a distance $r$~\cite{Hall1994,Withers2003}. They have shown that for a chain of length $N_b$, with given values of the backbone solvent quality $\epsilon_{bb}$, and spacer length $\ell$, this procedure can be used to determine the value of sticker strength $\epsilon_{st}^\theta$ at which the sticky polymer chain as a whole behaves as a chain under $\theta$-solvent conditions, i.e., when $B_2$ becomes zero. For instance, for a chain with $N_b = 34, \ell = 4$, and $\epsilon_{bb} = 0.3$, the second virial coefficient $B_2 =0$ for $\epsilon_{st}^\theta \approx 3.2$. Note that even though the backbone monomers are under good solvent conditions (since $\epsilon_{bb} = 0.3 < \epsilon_{\theta} = 0.45$), the chain as a whole is under $\theta$-solvent conditions due to the affinity of the stickers for one another. It is clear then that a sticky polymer chain with $N_b = 34, \ell = 4, \epsilon_{bb} = 0.45$ and $\epsilon_{st} = 5$, will be under poor solvent conditions (i.e., the second virial coefficient $B_2 < 0$), since, firstly the backbone monomers are under $\theta$-solvent conditions, and secondly, the sticker strength, $\epsilon_{st} =5$, is greater than $\epsilon_{st}^\theta \approx 3.2$, determined for the case $\epsilon_{bb} = 0.3$. We can anticipate that a solution of such sticky polymers will phase separate with increasing concentration, and indeed this seems to be the case as discussed below}.

\begin{figure*}[t]
    \centerline{
    \begin{tabular}{c c}
        \includegraphics[width=72mm]{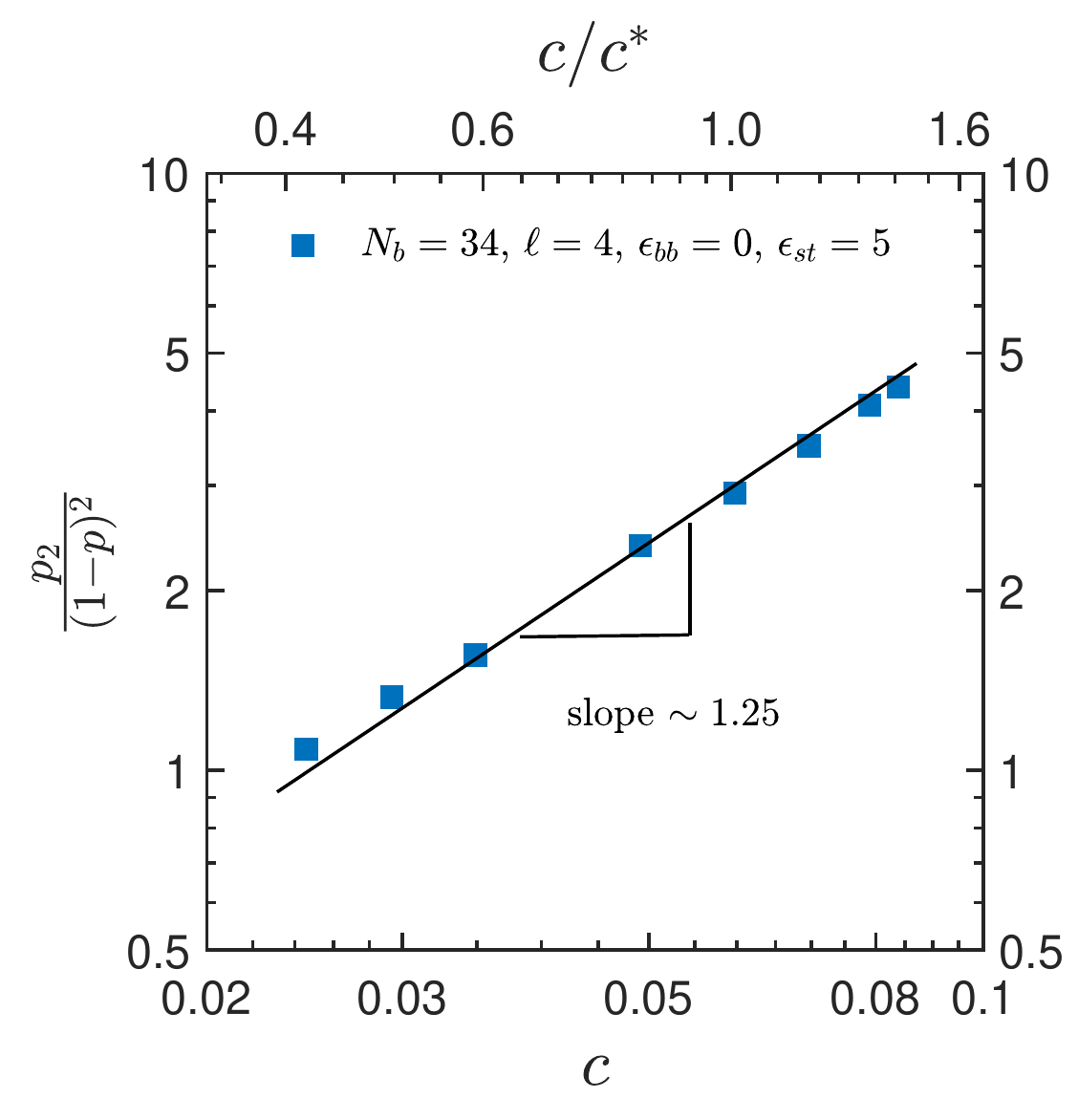} 
        & \includegraphics[width=68mm]{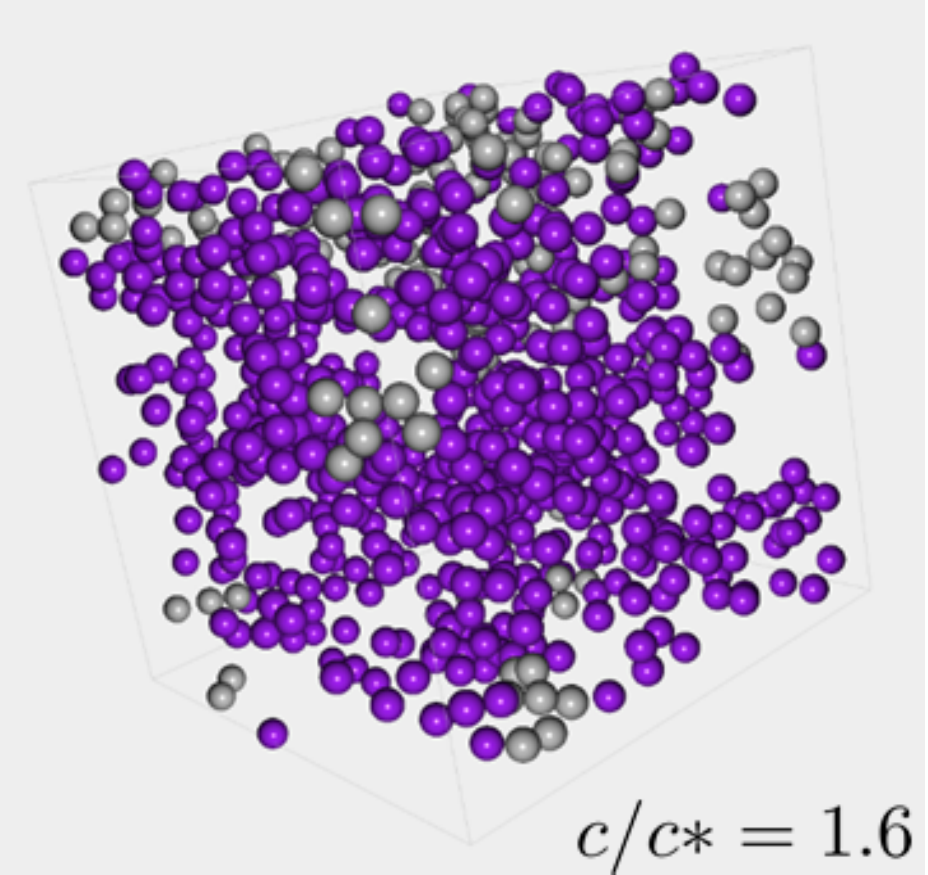} \\
        (a) & (b) \\
    \end{tabular}
    }
	\caption{(a) Rescaled inter-chain degree of association as a function of monomer concentration, with backbone monomers under  good solvent conditions corresponding to scaling regime~II. The solid line is drawn with slope equal to the prediction by scaling theory, while symbols represent simulation data. (b) Snapshot from the simulations for systems with $N_b=34$, $\ell=4$, $\epsilon_{bb} = 0$,  and $\epsilon_{st}=5.0$, at concentration $c/c^*= 1.6$.  Beads coloured purple are from chains that are all a part of the same system spanning cluster, while the grey coloured beads belong to chains that are not part of this cluster.  \label{fig:phase_sep_good} }
		\vskip-10pt
\end{figure*} 

We have previously shown that simulation results validate the predictions of scaling theory for sticky polymers, as displayed in Figs.~\ref{fig:RII_constStk} for a range of concentrations. The plot of the ratio $ \left[ p_2/(1-p^2) \right]$ is reproduced in Fig.~\ref{fig:phase_sep_theta}~(a), but this time at higher concentrations than shown previously. It is very clear that for sufficiently high concentrations, simulation data departs from the linear line representing the prediction of scaling theory, for both the approaches pursued here to simulate backbone monomers under $\theta$-solvent conditions. It seems likely that the breakdown of scaling theory coincides with the occurrence of phase separation, as indicated in the snapshots from simulations displayed in Figs.~\ref{fig:phase_sep_theta}~(b) and~(c).

These figures represent snapshots of a system with $N_b = 24, \ell = 4$, and $\epsilon_{st} = 5.0$, under $\theta$-solvent conditions for backbone monomers, at two different concentrations, $c/c^* = 0.3$ and $c/c^* = 1.0$. The purple coloured beads belong to chains that are all a part of the same cluster. The grey coloured beads belong to chains that are not part of the cluster represented by the purple beads. At the relatively low concentration of $c/c^* = 0.3$, Fig.~\ref{fig:phase_sep_theta}~(b) appears to suggest that there exist only small sized clusters, containing only a few chains, that are fairly homogeneously dispersed in the simulation cell. With increasing concentration, more free chains combine with existing clusters, along with the combination of clusters themselves, to give rise to increased cluster sizes, with more constituent chains in each cluster. At sufficiently high concentrations, such as at $c/c^* = 1.0$, the snapshot displayed in Fig.~\ref{fig:phase_sep_theta}~(c) suggests that most of the chains have clumped together to form a single large cluster. Note that since the simulation box has periodic images in the three coordinate directions, all the purple beads representing the single cluster are in fact in the neighbourhood of each of the corners of the box. The aggregation of chains in the cluster does not span the system homogeneously, suggesting that the solution has phase separated at some concentration, $0.3 < c/c^*< 1.0$, which is the range in which the scaling theory also breaks down.

The situation is very different for a sticky polymer system in which the chains have backbone monomers under very good solvent conditions. Under these circumstances, as indicated schematically in Fig.~\ref{fig:thetasurface}, the sticker strength $\epsilon^\theta_{st}$ required for the sticky chain as a whole to be under $\theta$-solvent conditions keeps increasing as  $\epsilon_{bb} \to 0$. The solvent quality for the sticky chain as a whole remains good in spite of the presence of stickers, and phase separation does not occur with increasing monomer concentration. As a consequence, it can be anticipated that unlike for chains with backbone monomers under $\theta$-solvent conditions, scaling predictions will remain valid even at high concentrations. This has already been commented upon in the context of Fig.~\ref{fig:RII_constStk} in section~\ref{sec:theta_and_II}, where it was pointed out the scaling relations remained valid even after the system is well into the gel phase. These observations are confirmed in Figs.~\ref{fig:phase_sep_good}, where in subfigure~(a) it can be seen that the ratio involving the inter-chain degree of association scales with monomer concentration according to the prediction of scaling theory even at the highest concentrations examined here, while subfigure~(b) indicates that at the scaled concentration $c/c^*= 1.6$, there exists a system spanning cluster, and that the chains are distributed homogeneously across the system, with no sign of phase separation.

\section{\label{sec:conclusions} Summary and conclusions}

A multi-particle Brownian dynamics simulation algorithm, with hydrodynamic interactions incorporated, which was formerly developed to describe semidilute polymer solutions~\cite{JainPRE}, has been extended to describe associative polymer solutions. Pairwise interactions between monomers that are on the chain backbone and between the stickers themselves, have been described with the SDK potential~\cite{SDK,Aritra2019}, which has advantages compared to other excluded volume potentials. 

The main static properties that have been evaluated here are the intra-chain and inter-chain degrees of conversion $p_1$ and $p_2$, respectively, and their dependence on system parameters such as the length of the chain, $N_b$, the number of stickers on a chain $f$, the distance between two stickers, $\ell$, the solvent quality parameter, $\hat \tau$, and the monomer concentration,  $c$.

Comparisons have been carried out with the predictions of a lattice-based mean-field theory~\cite{Dob} for ratios involving $p_1$, $p_2$, and the total fraction of associated stickers $p$. The scaling theory identifies different regimes of behaviour depending on the quality of the solvent for the backbone monomers, the monomer concentration, and the density of stickers on a chain. The use of the SDK potential allows a careful choice of parameter values such that simulations can be used to explore each of the different scaling regimes. The cluster computation algorithm of \citet{SevOtt1988} enables the calculation of the degrees of conversion, and the distribution of chain cluster sizes, along with their spatial extent.

The scaling theory of \citet{Dob} identifies two broad categories of behaviour based on whether the backbone monomers are under $\theta$ or good solvent conditions. The latter category is further divided into three regimes depending on the relative magnitude of the spacer segment, $\ell$, the number of monomers in a thermal blob, $g_T$, and the number of monomers in a correlation blob, $g_c$. In Regime~I, $\ell < g_T < g_c$, while in Regime~II, $g_T < \ell < g_c$, and in Regime~III, $g_T < g_c < \ell$. 

Simulation results are shown to validate the predictions of Dobrynin's mean-field theory~\cite{Dob} across a wide range of parameter values in all the scaling regimes, and  data is shown to collapse onto master plots when plotted in terms of suitable quantities. An important conclusion of this study is that the value of the des Cloizeaux exponent~\cite{desClo1980,Duplantier89,Hsu2004} proposed by \citet{Dob}, $\theta_2 = 1/3$, is accurate since it enables a collapse of the simulation data for all the scaling relations considered here.

The characterization of gelation in these systems has also been examined. Three different signatures of gelation are identified: (i) the concentration $c_{g_1}$ at which an incipient system-spanning network occurs, (ii) the concentration threshold $c_{g_2}$ at which the probability distribution of chain sizes becomes bimodal, and (iii) the monomer concentration $c_{g_3}$ at which there is a maximum in the free-chain concentration. Each of these three different sol-gel transition signatures is found to occur at a different concentration. The identification of the concentration at the sol-gel transition enables a verification of the modified Flory-Stockmayer expression~\cite{FloryBook,Stock2,Dob}, which relates the degree of inter-chain conversion, $p_2$, to the degree of intra-chain conversion, $p_1$, and the number of stickers on a chain, $f$. 

The only aspect of the phase behaviour of associative polymer solutions examined here is the gelation line, which separates the sol and gel phases. In this case as well, attention is restricted to the situation where the solvent quality  and sticker strength are constant, and the sticky chain is in scaling regime~II. This simplification leads to an expression for the dependence of the concentrations at gelation, $c_{g_i}; \, i=1, 2, 3$, on the number of stickers on a chain. Simulation results confirm the prediction of scaling theory when the modified Flory-Stockmayer expression is used for $p_2$. This is an experimentally testable prediction of scaling theory and simulations --- both of the des Cloizeaux exponent and the Flory-Stockmayer expression. 

Finally, it is shown that phase separation occurs with increasing concentration for systems in which the backbone monomers are under $\theta$-solvent conditions. Curiously, the predictions of scaling theory are found to breakdown  in the same range of concentrations in which phase separation is observed. On the other hand, for backbone monomer in good solvent conditions, there is no phase separation for the concentrations examined here, and scaling theory remains valid in both the sol and gel phases.

The success of the framework for the description of associative polymer solutions developed here in describing the predictions of static properties by scaling theory gives confidence that it can also be used to describe the equilibrium dynamics and the rheological behaviour of these solutions.

\appendix
\vskip30pt
\section*{Supplementary material}
Supporting information for this article contains four sections that discuss, (i) the equivalence of two different sticking rules for stickers within the cut-off radius, (ii) the influence of hydrodynamic interactions on the time taken to achieve a stationary state, (iii) the scaling of computational cost with chain size, and  (iii) all the data presented in section~\ref{sec:deg_of_conv} for the dependence of $R_g$, $p_1$ and $p_2$ on the various parameters $\{N_b,\ell, f, \epsilon_{bb}, \epsilon_{st}, c, c/c^*\}$, is given in tabular form in Table~S2 for comparison with future model predictions.

\linespread{1}\selectfont

\section*{\label{sec:acknwl}Acknowledgements}
This research was supported under Australian Research Council's Discovery Projects funding scheme (project number DP190101825). It was undertaken with the assistance of resources from the National Computational Infrastructure (NCI Australia), an NCRIS enabled capability supported by the Australian Government. We are grateful to Nathan Clisby for insightful discussions regarding the des Cloizeaux exponent $\theta_2$. The authors would like to thank the final year undergraduate research project students Declan Wain and Kyle Gibson for their assistance in generating the snapshots in Fig.~\ref{fig:SnapShot}.

\bibliography{Scalingref}
\bibliographystyle{JORnat}

\end{document}


\beginsupplement

\title{Supplementary Information for:\\
Universal scaling and characterisation of gelation in associative polymer solutions}


\author{Aritra Santra}
\affiliation{Department of Chemical Engineering, Monash University,
Melbourne, VIC 3800, Australia}
\author{B. D\"{u}nweg}
\affiliation{Max Plank Institute of Polymer Research, Ackermannweg 10, 55128 Mainz, Germany}
\affiliation{Department of Chemical Engineering, Monash University,
Melbourne, VIC 3800, Australia}
\author{J. Ravi Prakash}
\email{ravi.jagadeeshan@monash.edu}
\affiliation{Department of Chemical Engineering, Monash University,
Melbourne, VIC 3800, Australia}
 \homepage{https://users.monash.edu.au/~rprakash/}



 
\maketitle


\section{\label{sec:StickingRule} Equivalence of sticking rules}

\begin{figure*}[t]
    \centerline{
    \begin{tabular}{cc}
       \includegraphics[width=90mm]{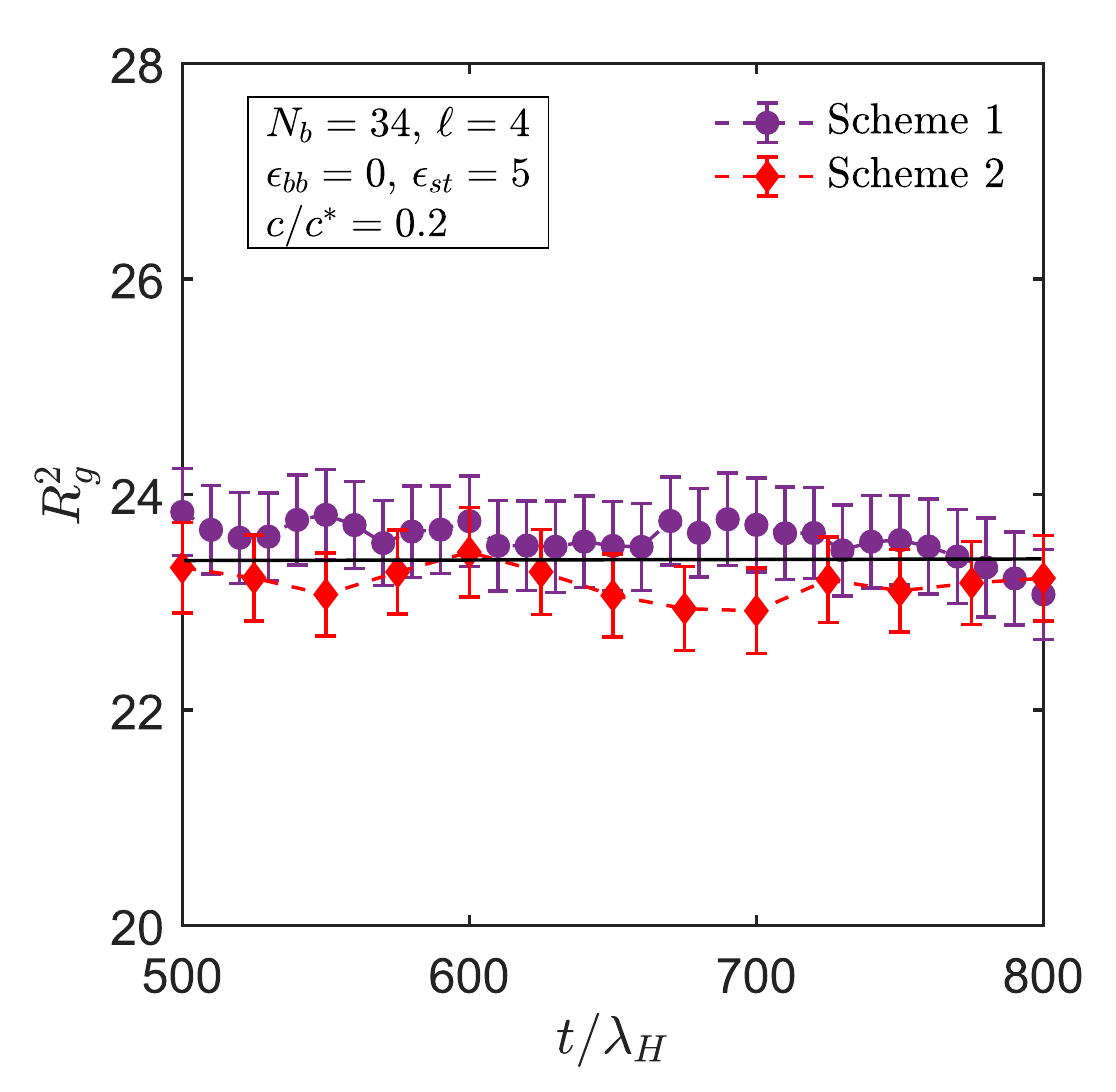} 
       &         \includegraphics[width=90mm]{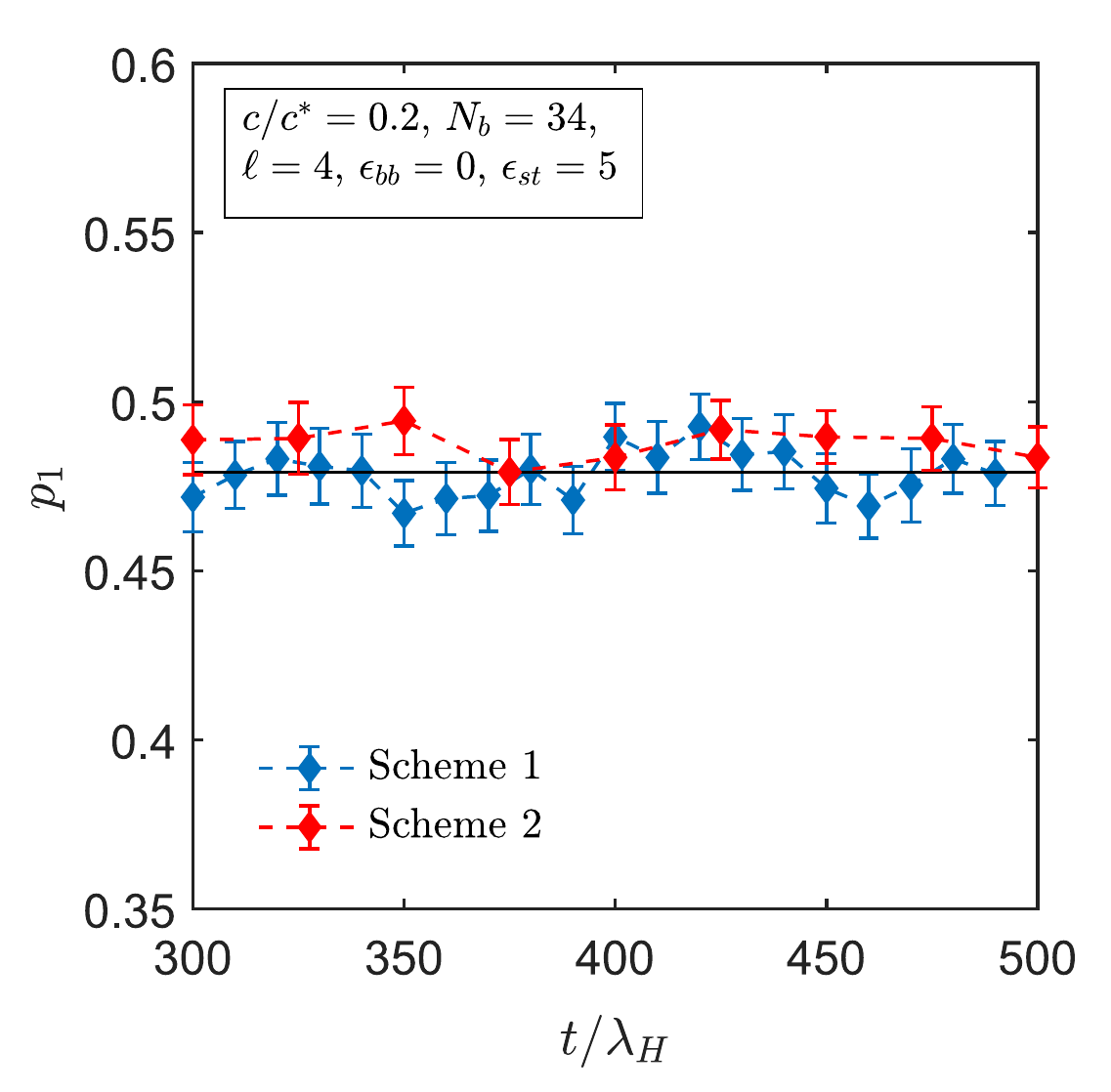} \\
          (a) &  (b) \\
        \multicolumn{2}{c}{\includegraphics[width=90mm]{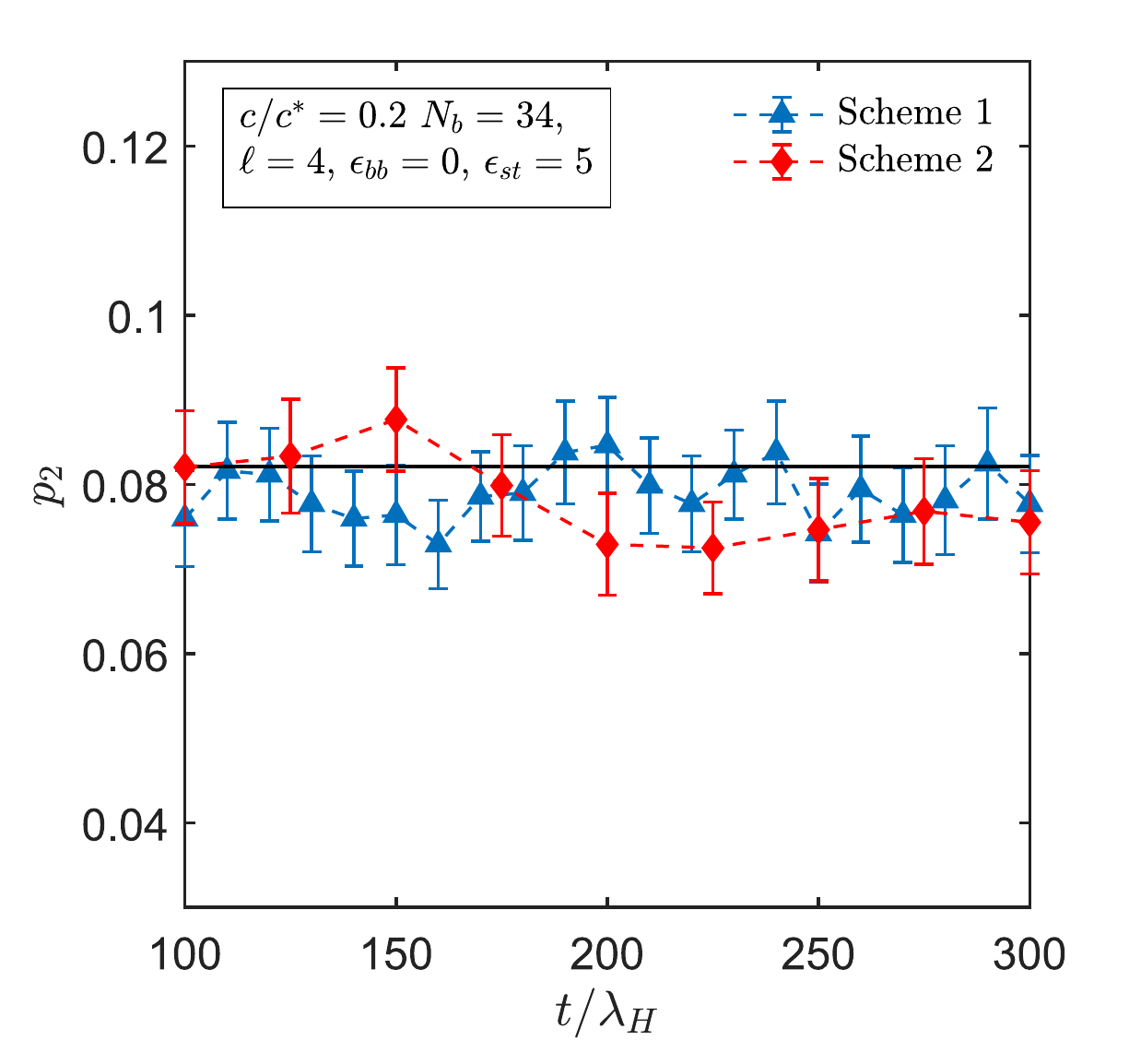}} \\
        \multicolumn{2}{c}{ (c)} \\
    \end{tabular}
    }
\caption{Comparison of the equilibrium values of (a) radius of gyration, $R_g^2$, (b) intra-chain association fraction, $p_1$, and (c) inter-chain association fraction, $p_2$, predicted by the two different sticking rules, scheme~1 and scheme~2. Simulations are carried out with chain length $N_b=34$, spacer length $\ell=4$, sticker strength $\epsilon_{st}=5$, and concentration $c/c^*=0.2$, for the backbone monomers under athermal solvent conditions, i.e., $\epsilon_{bb}=0$. The symbols are from simulations and the solid lines are equilibrium averaged values of the static properties.}
\label{fig:StkRule}
\end{figure*}

The rules implemented in the current algorithm for deciding when a pair of stickers associate with each other have been described in section~2.2 of the main paper. In particular, in cases where three or more stickers are within the cut-off radius, the sticking pairs are selected based on the order of labelling of the stickers. This rule  is denoted here as scheme~1. An alternative scheme could be to pick the sticking pairs at random when three or more stickers are within the interaction range (denoted here as scheme~2). We anticipate that since a large ensemble of chains are distributed randomly in a simulation box, bead labels of neighbouring beads are essentially random, and since the probability of three or higher body interactions for stickers is very low, the two schemes should effectively produce the same results. This is illustrated in Figs.~\ref{fig:StkRule} where the radius of gyration, and the intra-chain and inter-chain association fractions are plotted against time after reaching the stationary state. It can be seen that for each of the static properties, the results of both the schemes agree within error-bars, with the values fluctuating about the equilibrium averages.  

\section{\label{sec:HI2} Hydrodynamic interactions and time to equilibration}

As is well known, hydrodynamic interactions begin to get screened at the overlap concentration $c^*$, and get completely screened only in a polymer melt~\cite{RubCol,Jain2012}. Though they are confined at sufficiently long times~\cite{Ahlrichs2001} to length scales below the size of a correlation blob, which shrinks with increasing concentration, hydrodynamic interactions essentially determine the rich and complex dynamics of unentangled semidilute polymer solutions over a wide range of concentration~\cite{RubCol,Jain2012,Prakash2019}. It can therefore be anticipated that hydrodynamic interactions will also significantly influence dynamic properties of associative polymer solutions, such as (to name but a few), the on-and-off time scales of stickers, and relaxation times both on the level of single chains and of the network as a whole. It seems worthwhile therefore to examine the role of hydrodynamic interactions in associative polymer solutions, to determine the concentration beyond which they begin to get screened, and to establish whether there is a concentration beyond which their influence on dynamics can be safely ignored. 

The recent numerical investigations of associative polymer solutions by the J\"ulich group using MPCD~\cite{GompEq,GompFlo}, and by Castillo-Tejas et al.~\cite{Castillo} using NEMD, account explicitly for hydrodynamic interactions, and are promising approaches for investigating the rich behaviour of associative polymer solutions. While the published studies using these methods have drawn a number of important conclusions regarding the rheology of associative polymer solutions~\cite{GompFlo,Castillo}, to our knowledge, they have not so far explicitly examined the role of hydrodynamic interactions.  

While hydrodynamic interactions have no effect on equilibrium static properties, they do, nevertheless, play a role in determining the timescale over which equilibration is achieved. In this section, their influence on the time taken by the intra and inter-chain association fractions, and the radius of gyration, to reach their respective stationary values, is examined. 

The governing I{\^t}o stochastic differential equation for the time evolution of bead positions is given by Eq.~(4) of the main paper, with the various quantities that appear in the equation defined in section~2.1. As mentioned in the section, the regularized Rotne-Prager-Yamakawa (RPY) tensor is used to compute the hydrodynamic interaction tensor,
\begin{equation}
{\pmb{\Omega}_{\mu \nu}} = {\pmb{\Omega}} ( {\mathbf{r}_{\mu}} - {\mathbf{r}_{\nu}} )
\end{equation}
where 
\begin{equation}
\pmb{\Omega}(\mathbf{r}) =  {\Omega_1{ \pmb \delta} +\Omega_2\frac{\mathbf{r r}}{{r}^2}}
\end{equation}
with
\begin{equation*}
\Omega_1 = \begin{cases} \dfrac{3\sqrt{\pi}}{4} \dfrac{h^*}{r}\left({1+\dfrac{2\pi}{3}\dfrac{{h^*}^2}{{r}^2}}\right) & \text{for} \quad r\ge2\sqrt{\pi}h^* \\
 1- \dfrac{9}{32} \dfrac{r}{h^*\sqrt{\pi}} & \text{for} \quad r\leq 2\sqrt{\pi}h^* 
\end{cases}
\end{equation*}
and 
\begin{equation*}
\Omega_2 = \begin{cases} \dfrac{3\sqrt{\pi}}{4} \dfrac{h^*}{r} \left({1-\dfrac{2\pi}{3}\dfrac{{h^*}^2}{{r}^2}}\right) & \text{for} \quad r\ge2\sqrt{\pi}h^* \\
 \dfrac{3}{32} \dfrac{r}{h^*\sqrt{\pi}} & \text{for} \quad r\leq 2\sqrt{\pi}h^* 
\end{cases}
\end{equation*}
Here, the hydrodynamic interaction parameter $h^*$ is the dimensionless bead radius in the bead-spring model, defined as $h^* = a/(\sqrt{\pi k_BT/H})$. The decomposition of the diffusion tensor has been achieved with the help of Fixman's polynomial approximation based on the Chebyshev technique which has been widely used  earlier for both single chain~\cite{fixman1986implicit,PraRavi04,PrabhakarSFG} and multi-chain BD simulations~\cite{JainPRE,Stoltz,SaadatSemi}. The challenge of simulating semidilute solutions arises from the sum $\sum_{\nu}\mathbf D_{\mu\nu}\cdot(\mathbf F_\nu^{s}+ \mathbf F_\nu^{\textrm{SDK}})$ in Eq.~(4) of the main paper, which is conditionally convergent. Here it is evaluated using an optimized Ewald summation technique developed previously by Jain et al.~\cite{JainPRE}. For all the simulation results reported in this section, we use a value of $h^*=0.2$.

\begin{figure*}[t]
    \centerline{
    \begin{tabular}{c c}
        \includegraphics[width=90mm]{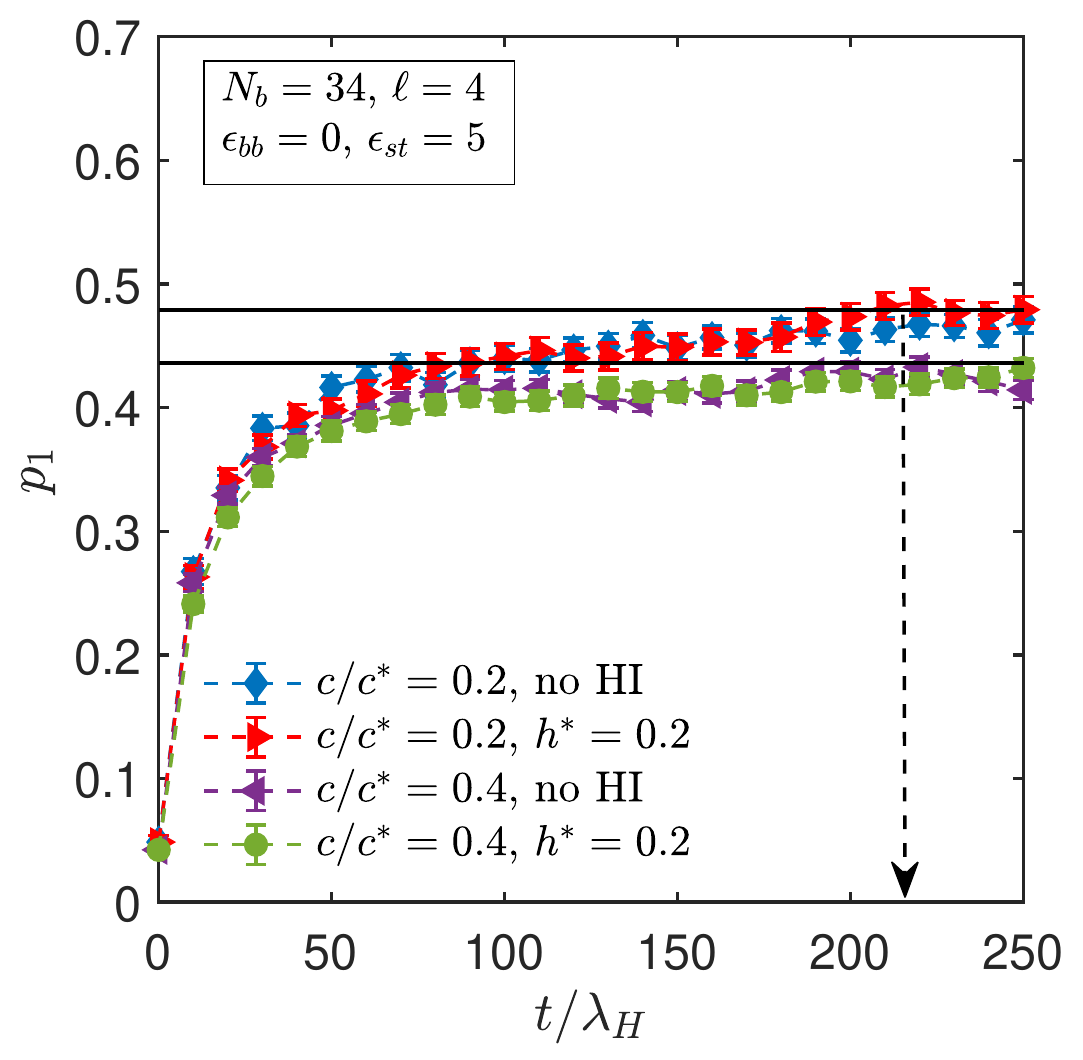} 
        & \includegraphics[width=91mm]{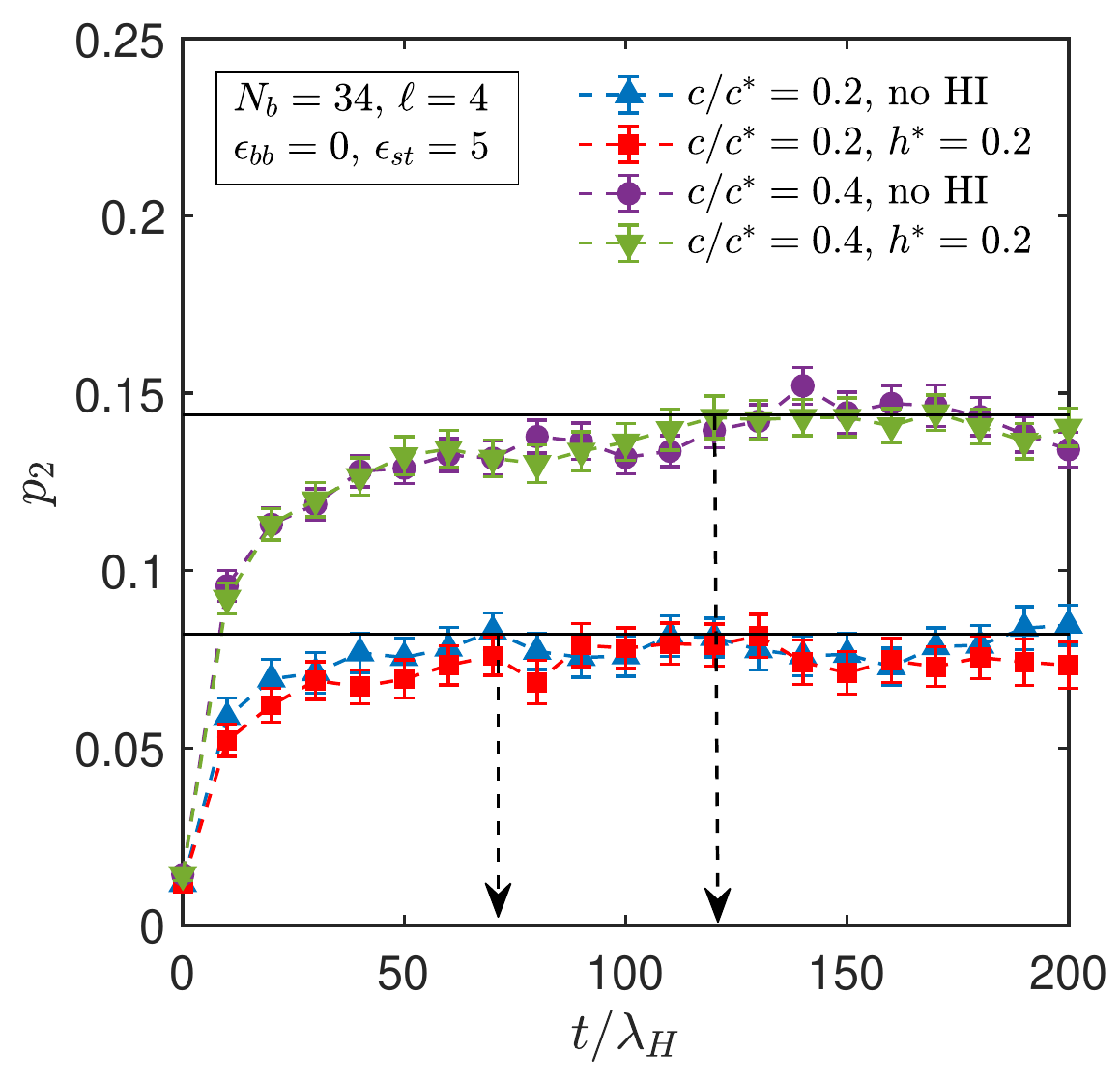} \\
        (a) & (b) \\
        \includegraphics[width=90mm]{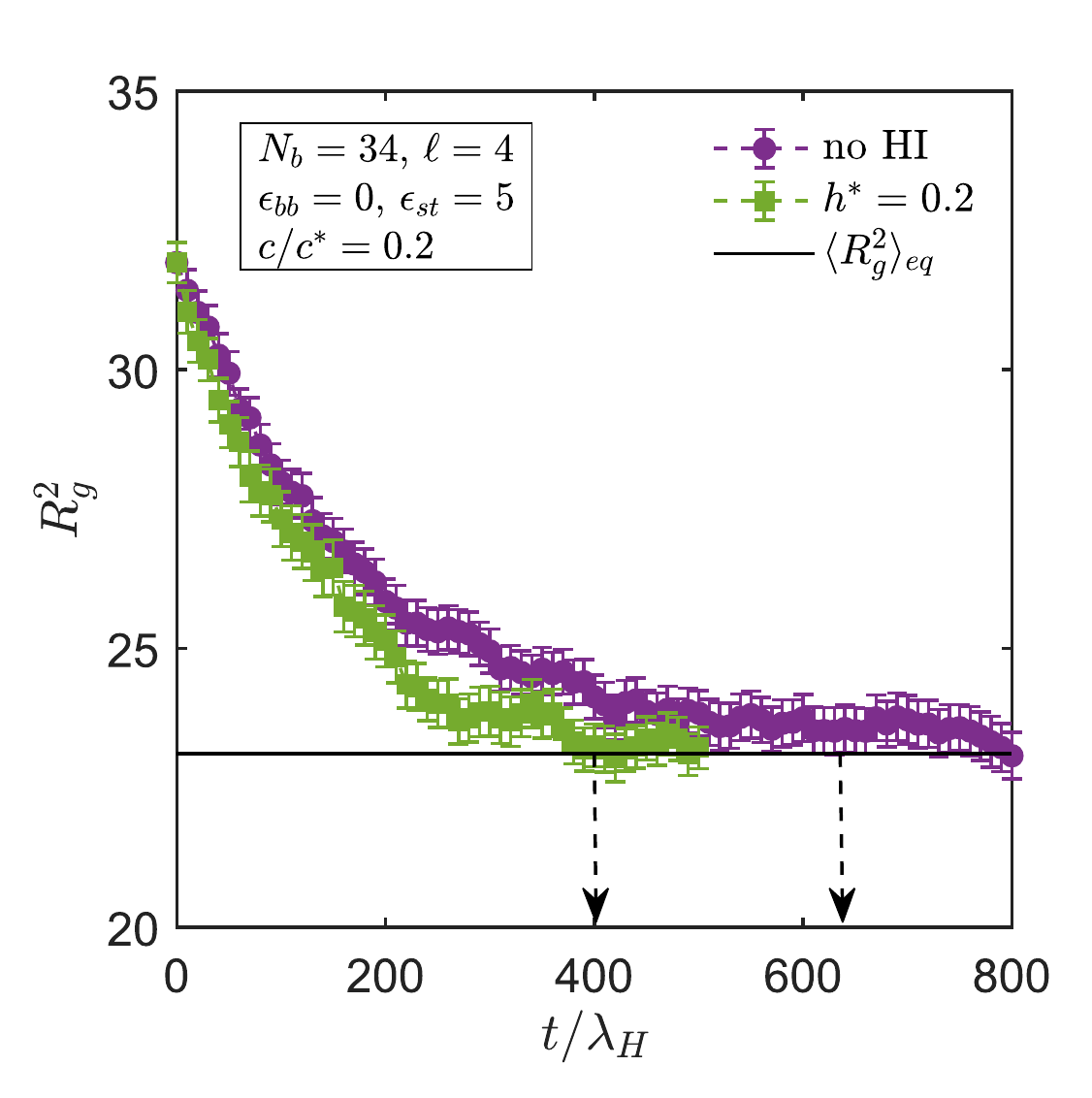} 
        & \includegraphics[width=87mm]{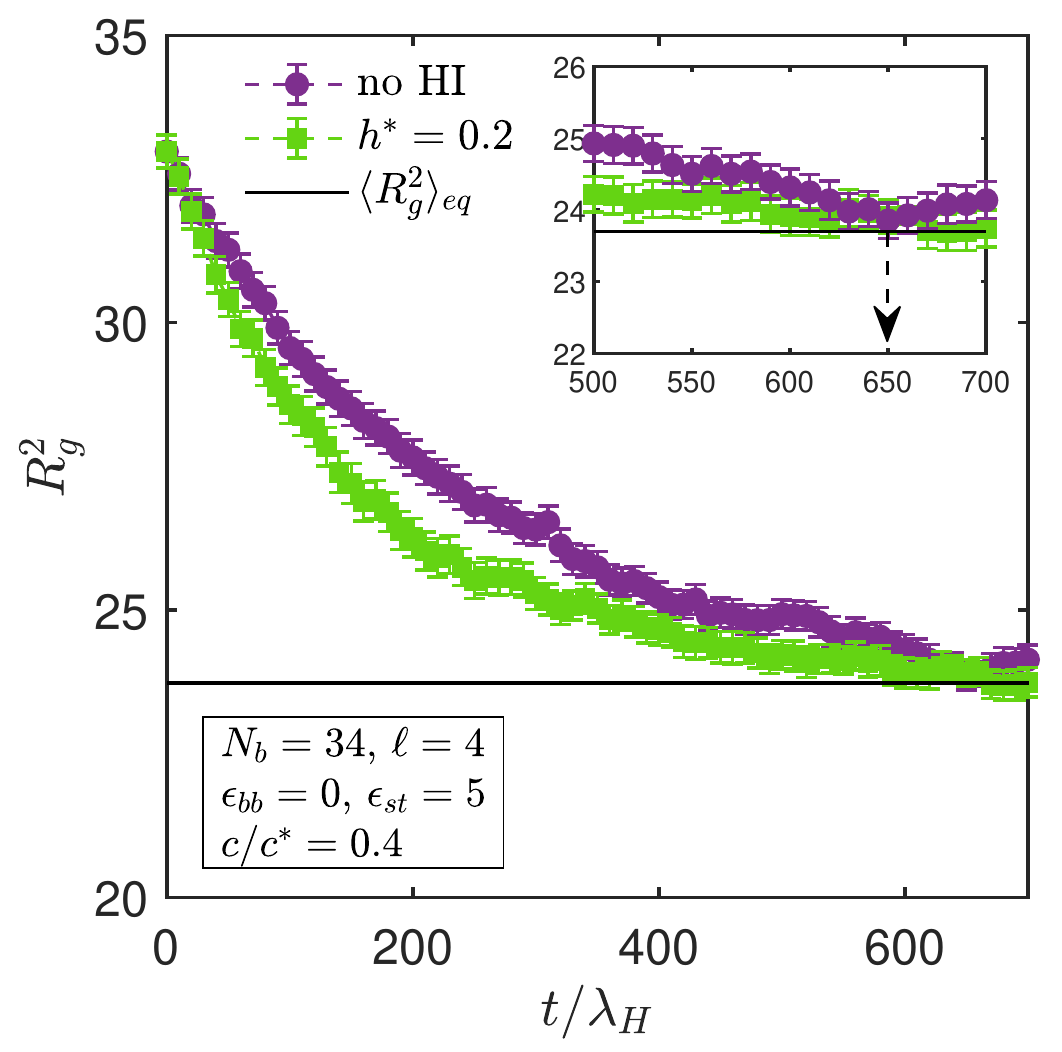} \\
        (c) & (d) \\
    \end{tabular}
    }
	\caption{The transient variation of (a) the intra-chain, and (b) the inter-chain association fractions, for two different concentrations, and the time evolution of the radius of gyration at the scaled concentrations, (c) $c/c^* = 0.2$, and (d) $c/c^* = 0.4$. Simulations with and without hydrodynamic interactions (HI) have been displayed for both properties, with the former carried out with the hydrodynamic interaction parameter, $h^* = 0.2$. The dashed vertical lines indicate the time required for equilibration. The choice of simulation parameters leads to a sticky polymer solution in scaling regime~II. 
\label{fig:HIequil}}
\end{figure*}

As mentioned in the main paper, prior to sampling equilibrium data, a typical simulation involves two equilibration steps. In the first, a run of about $3$ to $4$ Rouse relaxation times is carried out for a system of chains with only backbone monomers and no stickers. In the second step, stickers are introduced and a further run of about $5$ to $8$ Rouse relaxation times is carried out. For the results discussed in this section, hydrodynamic interactions are switched on at the end of the first step, simultaneously with the introduction of stickers. The transient trajectories during the second equilibration step are sampled in order to study the influence of hydrodynamic interactions. 

The transient variation of the intra and inter-chain association fractions are shown in Figs.~\ref{fig:HIequil}~(a) and~(b), respectively, for two different scaled concentrations, $c/c^* = 0.2$ and $c/c^* = 0.4$, and the time evolution of the radius of gyration at the same two values of concentration, are shown in Figs.~\ref{fig:HIequil}~(c) and~(d), respectively. The simulation parameters are such that, within the framework of scaling theory, the system lies in Regime~II. For both the degrees of conversion and the radius of gyration, simulations with and without hydrodynamic interactions have been displayed. Since both are static properties, their equilibrium values are unaffected by hydrodynamic interactions, and as can be clearly observed in Figs.~\ref{fig:HIequil}, this is indeed the case, with the results of simulations with and without hydrodynamic interactions being identical at sufficiently long times, when the systems have equilibrated. 

Interestingly, it appears from Fig.~\ref{fig:HIequil}~(a) and~(b), that both $p_1$ and $p_2$ are unaffected by hydrodynamic interactions for the entire period of observation, from the moment the stickers are turned on to the time at which equilibration is achieved (denoted by the dashed vertical lines). Both the degrees of conversion are seen to increase with increasing concentration. On the other hand, while the time to equilibration for $p_1$ is the same at $c/c^* = 0.2$ and $c/c^* = 0.4$, it takes longer for $p_2$ to equilibrate at the higher concentration. It is worth noting that for the sticker strength examined here, the time required for the equilibration of $p_1$ and $p_2$, which is a reflection of the diffusive timescale for stickers to find each other in space, is significantly shorter than the time required for the equilibration of the radius of gyration, which is a property of the chain as a whole.

In contrast to their lack of influence on the degrees of conversion, hydrodynamic interactions appear to have a pronounced influence on the time needed for the radius of gyration to equilibrate at $c/c^* = 0.2$. As can be seen from the two trajectories in Fig.~\ref{fig:HIequil}~(c) that correspond to simulations with (green squares) and without hydrodynamic interactions (purple circles), a stationary state is reached by a non-dimensional time, $t/\lambda_H \approx 400$, in the former case, while it requires $t/\lambda_H \approx 650$ in the latter case. At the higher value of scaled concentration, $c/c^* = 0.4$, however, even though the transient trajectories are different, both simulations with and without hydrodynamic interactions appear to reach a stationary state by a non-dimensional time, $t/\lambda_H \approx 650$ (see inset to Fig.~\ref{fig:HIequil}~(d)). This suggests that the influence of hydrodynamic interactions decreases with increasing concentration, as may be expected with the onset of screening in unentangled semidilute solutions. As observed from the results in the main paper, the value of the scaled concentration, $c/c^* = 0.2$, corresponds to a system in the sol phase, while according to at least one of the signatures of gelation considered in this work, the value $c/c^* = 0.4$, corresponds to a system in the gel phase. As a consequence, we may expect that dynamic properties such as relaxation times in the sol phase will be significantly affected by the presence of hydrodynamic interactions. A detailed study of the dynamics of sticky polymer solutions, using the present simulation algorithm, is however, outside the scope of the present study.

\section{\label{sec:CPU} Scaling of computational cost with chain size} 

The majority of the results reported in this work have been carried out on the supercomputer \textit{Gadi}, which is Australia's peak research supercomputer based at the National Computational Infrastructure in Canberra. Gadi, ranked 24 on the TOP500 list, is a 3,200 node supercomputer (with 155,000 CPU cores and 567 Terabytes of memory) comprising the latest generation Intel Cascade Lake and Nvidia V100 processors, with over 9 petaflops of peak performance. The technical specifications of the processors are: Primergy CX2570 M5, Thinksystem SD650, Xeon Platinum 8274/8268, Nvidia Tesla V100 SXM2, Mellanox HDR Infiniband cluster manufactured by FUJITSU. The computational cost estimates for simulating chains of various lengths $N_b$, spacer lengths $\ell$, and concentrations $c/c^*$ on this machine have been given in Table~\ref{tab:CPU_time}.

\begingroup
\begin{table}[t]
    \centering
\resizebox{\columnwidth}{!}
{\begin{tabular}{c|c|c|c|c}
        \hline
        & $N_b$ & Spacer length & $c/c^*$ & CPU time for 10 Rouse times\\
         &      &   ($\ell$)  &    &   (DD:HH:MM)  \\ 
        \hline
        \hline 
  	\multirow{3}{*}{Regime~II}    & $34$ & $6$  & $2.0$ & $00:03:07$\\
          & $154$ & $30$  & $2.0$ & $12:00:00$\\
           & $254$ & $50$  & $2.0$ & $54:07:00$\\
           \hline
     \multirow{4}{*}{Regime~III}    & $64$ & $12$  & $6.5$ & $01:19:00$\\
          & $79$ & $15$  & $6.5$ & $03:09:00$\\
           & $154$ & $30$  & $6.5$ & $25:00:00$\\   
           & $254$ & $50$  & $6.5$ & $112:00:00$\\              
 \hline
  \hline
    \end{tabular}}
\vskip10pt    
\caption{Estimates of CPU wall time required on the supercomputer \textit{Gadi} to simulate a single data point on systems with different chain lengths and concentrations in Regimes~II and~III.}
\label{tab:CPU_time}
\end{table}
\endgroup

\section{\label{sec:p1p2data} Tabulated values of simulation results for $R_g$, $p_1$ and $p_2$ as a function of the parameters $\{N_b,\ell, f, \epsilon_{bb}, \epsilon_{st}, c/c^*\}$}
 
The data presented in the main paper for the dependence of $R_g$, $p_1$ and $p_2$ on the various parameters $\{N_b,\ell, f, \epsilon_{bb}, \epsilon_{st}, c, c/c^*\}$, in the form of figures, is given here in tabular form in Table~\ref{tab:parameters}, so that they are readily available for comparison with any model predictions that may be made in the future.

\section*{\label{sec:acknwl}Acknowledgements}
This research was supported under Australian Research Council's Discovery Projects funding scheme (project number DP190101825). It was undertaken with the assistance of resources from the National Computational Infrastructure (NCI Australia), an NCRIS enabled capability supported by the Australian Government. We are grateful to Nathan Clisby for insightful discussions regarding the des Cloizeaux exponent $\theta_2$.


\newpage
\vspace{-20pt}
\begingroup
\begin{center}
\begin{longtable*}{ p{0.15\textwidth}p{0.1\textwidth}p{0.1\textwidth}p{0.1\textwidth}p{0.1\textwidth}p{0.15\textwidth}p{0.15\textwidth}p{0.15\textwidth}}\\[-5pt]      
\caption{\label{tab:parameters} {List of values for the radius of gyration ($R_g$), and the intra-chain ($p_1$) and inter-chain ($p_2$) degrees of conversion as a function of concentration for different sets of parameters used in this work.}}\\
\hline
         ($N_b,\,\ell,\,f$) & $\epsilon_{bb}$ & $\epsilon_{st}$ & $c/c^*$ & $c$ & $R_g$ & $p_1$ & $p_2$ \\
        \hline
        \endfirsthead
        \multicolumn{8}{c}%
       {\tablename\ \thetable\ -- \textit{Continued from previous page}} \\
        \hline
      ($N_b,\,\ell,\,f$) & $\epsilon_{bb}$ & $\epsilon_{st}$ & $c/c^*$ & $c$ & $R_g$ & $p_1$ & $p_2$ \\
        \hline
        \endhead
\hline \multicolumn{8}{r}{\textit{Continued on next page}} \\
\endfoot
\hline
\endlastfoot
        \hline 
	 \multirow{11}{*}{($24,\,4,\,4$)} & No EV  & No EV & $0.0$ & $0.0$ & $3.23\pm0.00$ & -- & -- \\  
           & No EV  & $5.0$ & $0.1$ & $0.017$ & $2.59\pm0.01$ & $0.548\pm0.004$ & $0.213\pm0.004$ \\
           & No EV & $5.0$ & $0.2$ & $0.034$ & $2.64\pm0.01$ & $0.448\pm0.003$ & $0.330\pm0.004$ \\
           & No EV & $5.0$ & $0.3$ & $0.051$ & $2.67\pm0.01$ & $0.401\pm0.004$ & $0.390\pm0.005$  \\
           & No EV  & $5.0$ & $0.4$ & $0.069$ & $2.72\pm0.01$ & $0.345\pm0.004$ & $0.464\pm0.006$  \\
           & No EV  & $5.0$ & $0.55$ & $0.093$ & $2.79\pm0.01$ & $0.295\pm0.004$ & $0.530\pm0.005$ \\
           & No EV  & $5.0$ & $0.75$ & $0.129$ & $2.91\pm0.01$ & $0.255\pm0.004$ & $0.576\pm0.005$\\
           & No EV  & $5.0$ & $1.0$ & $0.170$ & $3.00\pm0.01$ & $0.222\pm0.003$ & $0.617\pm0.003$  \\
           & No EV  & $5.0$ & $1.2$ & $0.205$ & $3.08\pm0.01$ & $0.206\pm0.002$ & $0.634\pm0.002$ \\
           & No EV  & $5.0$ & $1.4$ & $0.240$ & $3.18\pm0.01$ & $0.181\pm0.002$ & $0.639\pm0.002$\\
           & No EV  & $5.0$ & $1.6$ & $0.274$ & $3.22\pm0.02$ & $0.163\pm0.002$ & $0.648\pm0.002$\\
           \hline
     \multirow{6}{*}{($34,\,6,\,4$)} & No EV & No EV & $0.0$ & $0.0$ & $3.87\pm0.00$ & -- & -- \\
           & No EV & $2.5$ & $0.12$ & $0.017$ & $3.68\pm0.05$ & $0.338\pm0.005$ & $0.074\pm0.006$ \\
           & No EV & $2.5$ & $0.2$ & $0.030$ & $3.65\pm0.05$ & $0.308\pm0.005$ & $0.134\pm0.006$ \\
            & No EV  & $2.5$ & $0.3$ & $0.042$ & $3.64\pm0.03$ & $0.287\pm0.005$ & $0.193\pm0.006$  \\
           & No EV  & $2.5$ & $0.5$ & $0.068$ & $3.73\pm0.02$ &	$0.249\pm0.005$ & $0.263\pm0.005$ \\
           & No EV  & $2.5$ & $0.6$ & $0.085$ & $3.77\pm0.02$ & $0.225\pm0.005$ & $0.314\pm0.004$\\ 
            \hline
         \multirow{4}{*}{($34,\,4,\,6$)}   & No EV & No EV & $0.0$ & $0.0$ & $3.87\pm0.00$ & -- & -- \\
         & No EV & $5.0$ & $0.12$ & $0.017$ & $2.74\pm0.04$ & $0.643\pm0.004$ & $0.121\pm0.007$ \\
           & No EV & $5.0$ & $0.3$ & $0.042$ & $2.98\pm0.04$ & $0.505\pm0.004$ & $0.283\pm0.006$ \\
            & No EV  & $5.0$ & $0.5$ & $0.068$ & $3.06\pm0.03$ & $0.445\pm0.004$ & $0.360\pm0.005$  \\
            \hline
         \multirow{9}{*}{($34,\,4,\,6$)}   & $0.45$ & $0.45$ & $0.0$ & $0.0$ & $4.42\pm0.01$ & --&-- \\
           & $0.45$ & $5.0$ & $0.2$ & $0.019$ & $3.76\pm0.02$ & $0.468\pm0.004$ & $0.238\pm0.005$ \\
           & $0.45$ & $5.0$ & $0.3$ & $0.028$ & $3.86\pm0.02$ & $0.416\pm0.003$ & $0.298\pm0.003$ \\
            & $0.45$  & $5.0$ & $0.4$ & $0.038$ & $3.95\pm0.02$ & $0.367\pm0.002$ & $0.353\pm0.003$  \\   
            & $0.45$ & $5.0$ & $0.5$ & $0.047$ & $4.01\pm0.02$ & $0.334\pm0.003$ & $0.397\pm0.004$ \\
            & $0.45$  & $5.0$ & $0.6$ & $0.056$ & $4.09\pm0.02$ & $0.319\pm0.002$ & $0.412\pm0.003$  \\ 
            & $0.45$ & $5.0$ & $0.8$ & $0.076$ & $4.25\pm0.02$ & $0.275\pm0.003$ & $0.461\pm0.003$ \\
            & $0.45$  & $5.0$ & $1.0$ & $0.095$ & $4.29\pm0.02$ & $0.257\pm0.002$ & $0.482\pm0.002$  \\ 
            & $0.45$  & $5.0$ & $1.2$ & $0.113$ & $4.40\pm0.02$ & $0.228\pm0.002$ & $0.517\pm0.002$  \\
             \hline
         \multirow{7}{*}{($29,\,4,\,5$)}   & $0.45$ & $0.45$ & $0.0$ & $0.0$ & $4.07\pm0.02$ & --& -- \\
         & $0.45$ & $5.0$ & $0.2$ & $0.020$ & $3.54\pm0.02$ & $0.447\pm0.004$ & $0.250\pm0.005$ \\
            & $0.45$ & $5.0$ & $0.3$ & $0.032$ & $3.63\pm0.02$ & $0.376\pm0.004$ & $0.336\pm0.005$ \\
            & $0.45$  & $5.0$ & $0.4$ & $0.043$ & $3.72\pm0.01$ & $0.346\pm0.004$ & $0.370\pm0.004$  \\   
            & $0.45$ & $5.0$ & $0.5$ & $0.052$ & $3.80\pm0.02$ & $0.304\pm0.003$ & $0.421\pm0.003$ \\
            & $0.45$  & $5.0$ & $0.6$ & $0.063$ & $3.89\pm0.02$ & $0.281\pm0.003$ & $0.445\pm0.003$  \\
            & $0.45$  & $5.0$ & $0.8$ & $0.083$ & $3.98\pm0.02$ & $0.242\pm0.003$ & $0.490\pm0.003$  \\
             \hline
         \multirow{6}{*}{($34,\,6,\,4$)}   & $0.45$ & $0.45$ & $0.0$ & $0.0$ & $4.42\pm0.01$ & -- & -- \\
         & $0.45$ & $2.5$ & $0.1$ & $0.010$ & $4.25\pm0.06$ & $0.210\pm0.005$ & $0.042\pm0.005$ \\
           & $0.45$ & $2.5$ & $0.2$ & $0.019$ & $4.28\pm0.05$ & $0.204\pm0.005$ & $0.095\pm0.005$ \\
            & $0.45$  & $2.5$ & $0.3$ & $0.029$ & $4.28\pm0.03$ & $0.186\pm0.004$ & $0.135\pm0.005$  \\   
            & $0.45$ & $2.5$ & $0.48$ & $0.045$ & $4.34\pm0.04$ & $0.168\pm0.004$ & $0.185\pm0.004$ \\
            & $0.45$  & $2.5$ & $0.6$ & $0.054$ & $4.37\pm0.03$ & $0.155\pm0.003$ & $0.221\pm0.004$  \\
             \hline
         \multirow{7}{*}{($29,\,4,\,5$)}   & $0.3$ & $0.3$ & $0.0$ & $0.0$ & $4.41\pm0.02$ & -- & -- \\
     \multirow{8}{*}{Regime~I}  & $0.3$ & $2.5$ & $1.0$ & $0.082$ & $4.32\pm0.02$ & $0.185\pm0.001$ & 	$0.272\pm0.002$ \\     
         & $0.3$ & $2.5$ & $1.2$ & $0.095$ & $4.33\pm0.01$ & $0.176\pm0.001$ & $0.306\pm0.001$ \\
           & $0.3$  & $2.5$ & $1.4$ & $0.112$ & $4.33\pm0.01$ & $0.166\pm0.001$ & $0.339\pm0.001$  \\   
            & $0.3$ & $2.5$ & $1.8$ & $0.144$ & $4.34\pm0.01$ & $0.146\pm0.001$ & $0.406\pm0.001$ \\
            & $0.3$  & $2.5$ & $2.0$ & $0.161$ & $4.31\pm0.01$ & $0.140\pm0.001$ & $0.431\pm0.001$  \\
 	      & $0.3$ & $2.5$ & $2.2$ & $0.178$ & $4.33\pm0.01$ & $0.133\pm0.001$ & $0.459\pm0.001$ \\
	      & $0.3$  & $2.5$ & $2.4$ & $0.193$ & $4.31\pm0.01$ & $0.128\pm0.001$ & $0.478\pm0.001$  \\
            \hline
         \multirow{5}{*}{($34,\,4,\,6$)}   & $0.3$ & $0.3$ & $0.0$ & $0.0$ & $4.79\pm0.04$ & -- & -- \\
	\multirow{6}{*}{Regime~I} & $0.3$ & $2.5$ & $0.8$ & $0.059$ & $4.65\pm0.02$ & $0.219\pm0.002$ & 	$0.203\pm0.002$ \\
         & $0.3$ & $2.5$ & $1.0$ & $0.074$ & $4.67\pm0.02$ & $0.205\pm0.002$ & $0.245\pm0.002$ \\
            & $0.3$  & $2.5$ & $1.2$ & $0.088$ & $4.72\pm0.02$ & $0.190\pm0.002$ & $0.287\pm0.002$  \\   
            & $0.3$ & $2.5$ & $1.4$ & $0.103$ & $4.73\pm0.02$ & $0.178\pm0.001$ & $0.321\pm0.001$ \\
            & $0.3$  & $2.5$ & $1.6$ & $0.118$ & $4.71\pm0.02$ & $0.173\pm0.001$ & $0.349\pm0.001$  \\
            \hline
         \multirow{5}{*}{($34,\,6,\,4$)}   & $0.3$ & $0.3$ & $0.0$ & $0.0$ & $4.79\pm0.04$ & -- & -- \\
	\multirow{6}{*}{Regime~I}  & $0.3$ & $2.0$ & $0.8$ & $0.059$ & $4.78\pm0.02$ & $0.100\pm0.002$ & 	$0.140\pm0.002$ \\
          & $0.3$ & $2.0$ & $1.0$ & $0.074$ & $4.76\pm0.02$ & $0.097\pm0.001$ & $0.173\pm0.002$ \\
            & $0.3$  & $2.0$ & $1.2$ & $0.088$ & $4.72\pm0.02$ & $0.096\pm0.001$ & $0.204\pm0.001$  \\   
            & $0.3$ & $2.0$ & $1.4$ & $0.103$ & $4.76\pm0.02$ & $0.092\pm0.001$ & $0.236\pm0.001$ \\
            & $0.3$  & $2.0$ & $1.6$ & $0.118$ & $4.74\pm0.02$ & $0.088\pm0.001$ & $0.266\pm0.001$  \\
           \hline
         \multirow{4}{*}{($34,\,6,\,4$)}   & $0.35$ & $0.35$ & $0.0$ & $0.0$ & $4.72\pm0.02$ & --& -- \\
	\multirow{5}{*}{Regime~I} & $0.35$ & $2.5$ & $1.1$ & $0.085$ & $4.67\pm0.01$ & $0.120\pm0.001$ & 	$0.250\pm0.001$ \\
             & $0.35$ & $2.5$ & $1.3$ & $0.100$ & $4.67\pm0.01$ & $0.115\pm0.001$ & $0.284\pm0.001$ \\
	         & $0.35$  & $2.5$ & $1.5$ & $0.0115$ & $4.70\pm0.02$ & $0.107\pm0.001$ & $0.315\pm0.001$  \\   
 	        & $0.35$ & $2.5$ & $1.7$ & $0.131$ & $4.68\pm0.01$ & $0.103\pm0.001$ & $0.342\pm0.001$ \\
             \hline
         \multirow{11}{*}{($24,\,4,\,4$)}   & $0.0$ & $0.0$ & $0.0$ & $0.0$ & $4.47\pm0.05$ & -- & -- \\
	\multirow{12}{*}{Regime~II} & $0.0$ & $5.0$ & $0.1$ & $0.007$ & $3.97\pm0.02$ & $0.487\pm0.002$ & 	$0.057\pm0.002$ \\
          & $0.0$ & $5.0$ & $0.2$ & $0.013$ & $4.01\pm0.01$ & $0.437\pm0.002$ & $0.118\pm0.002$ \\
           & $0.0$  & $5.0$ & $0.3$ & $0.020$ & $4.04\pm0.01$ & $0.401\pm0.002$ & $0.168\pm0.002$  \\   
            & $0.0$ & $5.0$ & $0.4$ & $0.026$ & $4.09\pm0.01$ & $0.372\pm0.002$ & $0.207\pm0.002$ \\
            & $0.0$  & $5.0$ & $0.5$ & $0.032$ & $4.14\pm0.01$ & $0.342\pm0.002$ & $0.247\pm0.002$  \\   
             & $0.0$ & $5.0$ & $0.6$ & $0.039$ & $4.15\pm0.01$ & $0.316\pm0.001$ & $0.283\pm0.001$ \\
          & $0.0$  & $5.0$ & $0.8$ & $0.052$ & $4.20\pm0.01$ & $0.267\pm0.002$ & $0.356\pm0.002$  \\   
            & $0.0$ & $5.0$ & $1.0$ & $0.064$ & $4.24\pm0.01$ & $0.233\pm0.002$ & $0.409\pm0.002$ \\
            & $0.0$ & $5.0$ & $1.14$ & $0.073$ & $4.21\pm0.01$ & $0.219\pm0.002$ & $0.436\pm0.002$ \\
            & $0.0$ & $5.0$ & $1.28$ & $0.083$ & $4.25\pm0.01$ & $0.195\pm0.002$ & $0.473\pm0.002$ \\
            & $0.0$ & $5.0$ & $1.6$ & $0.102$ & $4.26\pm0.01$ & $0.163\pm0.001$ & $0.525\pm0.002$ \\
            \hline
         \multirow{11}{*}{($34,\,4,\,6$)}   & $0.0$ & $0.0$ & $0.0$ & $0.0$ & $5.48\pm0.02$ & -- & -- \\
	\multirow{12}{*}{Regime~II}  & $0.0$ & $5.0$ & $0.2$ & $0.010$ & $4.81\pm0.02$ & $0.487\pm0.002$ & 	$0.079\pm0.003$ \\
             & $0.0$ & $5.0$ & $0.35$ & $0.018$ & $4.87\pm0.02$ & $0.438\pm0.002$ & $0.141\pm0.003$ \\
     	         & $0.0$  & $5.0$ & $0.5$ & $0.025$ & $4.95\pm0.02$ & $0.404\pm0.002$ & $0.189\pm0.003$  \\   
  	          & $0.0$ & $5.0$ & $0.6$ & $0.030$ & $4.99\pm0.01$ & $0.385\pm0.002$ & $0.214\pm0.002$ \\
            & $0.0$  & $5.0$ & $0.8$ & $0.040$ & $5.01\pm0.1$ & $0.344\pm0.002$ & $0.273\pm0.002$  \\   
            & $0.0$ & $5.0$ & $0.9$ & $0.044$ & $5.07\pm0.02$ & $0.321\pm0.002$ & $0.301\pm0.002$ \\
            & $0.0$  & $5.0$ & $1.0$ & $0.049$ & $5.11\pm0.02$ & $0.305\pm0.002$ & $0.325\pm0.002$  \\   
            & $0.0$ & $5.0$ & $1.2$ & $0.060$ & $5.12\pm0.02$ & $0.278\pm0.001$ & $0.367\pm0.002$ \\
            & $0.0$  & $5.0$ & $1.4$ & $0.070$ & $5.15\pm0.02$ & $0.249\pm0.002$ & $0.410\pm0.002$  \\   
            & $0.0$ & $5.0$ & $1.6$ & $0.079$ & $5.17\pm0.02$ & $0.226\pm0.002$ & $0.445\pm0.002$ \\
            & $0.0$ & $5.0$ & $1.7$ & $0.084$ & $5.14\pm0.02$ & $0.217\pm0.002$ & $0.459\pm0.002$ \\
             \hline
         \multirow{9}{*}{($39,\,4,\,7$)}   & $0.0$ & $0.0$ & $0.0$ & $0.0$ & $5.98\pm0.04$ & -- & -- \\
   	\multirow{10}{*}{Regime~II}   & $0.0$ & $5.0$ & $0.2$ & $0.009$ & $5.16\pm0.02$ & $0.492\pm0.002$ & 	$0.080\pm0.003$ \\
            & $0.0$ & $5.0$ & $0.3$ & $0.014$ & $5.24\pm0.02$ & $0.474\pm0.002$ & $0.104\pm0.002$ \\
           & $0.0$  & $5.0$ & $0.4$ & $0.017$ & $5.26\pm0.02$ & $0.448\pm0.002$ & $0.137\pm0.002$  \\   
            & $0.0$ & $5.0$ & $0.5$ & $0.022$ & $5.30\pm0.02$ & $0.426\pm0.002$ & $0.165\pm0.002$ \\
             & $0.0$ & $5.0$ & $0.7$ & $0.031$ & $5.38\pm0.02$ & $0.384\pm0.002$ & $0.222\pm0.002$ \\
            & $0.0$  & $5.0$ & $0.9$ & $0.039$ & $5.45\pm0.02$ & $0.351\pm0.002$ & $0.267\pm0.002$  \\
             & $0.0$ & $5.0$ & $1.0$ & $0.044$ & $5.46\pm0.02$ & $0.334\pm0.002$ & $0.290\pm0.002$ \\
 	      & $0.0$  & $5.0$ & $1.1$ & $0.048$ & $5.48\pm0.02$ & 	$0.320\pm0.002$ & $0.312\pm0.002$  \\
	   & $0.0$  & $5.0$ & $1.2$ & $0.053$ & $5.52\pm0.02$ & $0.303\pm0.002$ & $0.336\pm0.002$  \\
            \hline
         \multirow{9}{*}{($44,\,4,\,8$)}   & $0.0$ & $0.0$ & $0.0$ & $0.0$ & $6.45\pm0.03$ & -- & -- \\
	\multirow{10}{*}{Regime~II}  & $0.0$ & $5.0$ & $0.2$ & $0.008$ & $5.49\pm0.03$ & $0.513\pm0.002$ & 	$0.058\pm0.003$ \\
             & $0.0$ & $5.0$ & $0.3$ & $0.012$ & $5.53\pm0.02$ & $0.483\pm0.002$ & $0.097\pm0.003$ \\
           & $0.0$  & $5.0$ & $0.5$ & $0.020$ & $5.61\pm0.02$ & $0.447\pm0.002$ & $0.146\pm0.002$  \\   
            & $0.0$ & $5.0$ & $0.7$ & $0.028$ & $5.72\pm0.02$ & $0.408\pm0.002$ & $0.196\pm0.002$ \\
             & $0.0$ & $5.0$ & $0.9$ & $0.035$ & $5.78\pm0.02$ & $0.375\pm0.002$ & $0.242\pm0.002$ \\
             & $0.0$ & $5.0$ & $1.0$ & $0.039$ & $5.86\pm0.02$ & $0.358\pm0.002$ & $0.264\pm0.002$ \\
             & $0.0$ & $5.0$ & $1.2$ & $0.047$ & $5.90\pm0.02$ & $0.330\pm0.002$ & $0.303\pm0.002$ \\
             & $0.0$ & $5.0$ & $1.4$ & $0.055$ & $5.93\pm0.02$ & $0.304\pm0.002$ & $0.340\pm0.002$ \\
             & $0.0$ & $5.0$ & $1.6$ & $0.063$ & $5.99\pm0.02$ & $0.277\pm0.001$ & $0.378\pm0.001$ \\
             \hline
	 \multirow{9}{*}{($34,\,6,\,4$)} & $0.0$ & $0.0$ & $0.0$ & $0.0$ & $5.48\pm0.02$ & -- & -- \\
	\multirow{10}{*}{Regime~II}  & $0.0$ & $5.0$ & $0.2$ & $0.010$ & $5.08\pm0.02$ & $0.334\pm0.002$ & 	$0.079\pm0.002$ \\
         & $0.0$ & $5.0$ & $0.4$ & $0.020$ & $5.12\pm0.02$ & $0.292\pm0.002$ & $0.156\pm0.003$ \\
       & $0.0$  & $5.0$ & $0.6$ & $0.029$ & $5.15\pm0.01$ & $0.259\pm0.002$ & $0.221\pm0.003$  \\   
        & $0.0$ & $5.0$ & $0.8$ & $0.039$ & $5.17\pm0.02$ & $0.234\pm0.002$ & $0.271\pm0.002$ \\
 	    & $0.0$ & $5.0$ & $1.0$ & $0.049$ & $5.18\pm0.01$ & $0.215\pm0.002$ & $0.323\pm0.002$ \\
	     & $0.0$ & $5.0$ & $1.4$ & $0.069$ & $5.17\pm0.02$ & $0.172\pm0.002$ & $0.415\pm0.002$ \\
         & $0.0$ & $5.0$ & $1.5$ & $0.075$ & $5.19\pm0.02$ & $0.163\pm0.002$ & $0.434\pm0.002$ \\
         & $0.0$ & $5.0$ & $1.6$ & $0.079$ & $5.20\pm0.02$ & $0.156\pm0.002$ & $0.449\pm0.002$ \\
         & $0.0$ & $5.0$ & $1.7$ & $0.084$ & $5.16\pm0.02$ & $0.149\pm0.002$ & $0.467\pm0.002$ \\
             \hline
         \multirow{8}{*}{($29,\,5,\,4$)}   & $0.0$ & $0.0$ & $0.0$ & $0.0$ & $5.01\pm0.02$ & -- & --\\
	\multirow{9}{*}{Regime~II}  & $0.0$ & $5.0$ & $0.35$ & $0.019$ & $4.61\pm0.02$ & $0.348\pm0.002$ & 	$0.151\pm0.002$ \\
        & $0.0$ & $5.0$ & $0.4$ & $0.022$ & $4.62\pm0.02$ & $0.328\pm0.002$ & $0.186\pm0.003$ \\
        & $0.0$ & $5.0$ & $0.6$ & $0.033$ & $4.67\pm0.01$ & $0.292\pm0.002$ & $0.247\pm0.002$  \\   
        & $0.0$ & $5.0$ & $0.8$ & $0.044$ & $4.70\pm0.01$ & $0.254\pm0.002$ & $0.311\pm0.002$ \\
        & $0.0$ & $5.0$ & $0.9$ & $0.049$ & $4.71\pm0.01$ & $0.240\pm0.002$ & $0.336\pm0.002$ \\
        & $0.0$ & $5.0$ & $1.0$ & $0.055$ & $4.72\pm0.01$ & $0.222\pm0.002$ & $0.367\pm0.002$ \\
        & $0.0$ & $5.0$ & $1.2$ & $0.067$ & $4.74\pm0.01$ & $0.199\pm0.001$ & $0.413\pm0.002$ \\
        & $0.0$ & $5.0$ & $1.4$ & $0.078$ & $4.72\pm0.01$ & $0.182\pm0.001$ & $0.451\pm0.002$ \\
             \hline
	\multirow{5}{*}{($34,\,6,\,4$)}   & $0.0$ & $0.0$ & $0.0$ & $0.0 $& $5.48\pm0.02$ & -- & -- \\
	\multirow{6}{*}{Regime~II} & $0.0$ & $4.0$ & $0.6$ & $0.029$ & $5.22\pm0.01$ & $0.187\pm0.001$ & 	$0.143\pm0.001$ \\
       & $0.0$ & $4.0$ & $0.8$ & $0.039$ & $5.23\pm0.01$ & $0.173\pm0.001$ & $0.192\pm0.001$ \\
	   & $0.0$  & $4.0$ & $0.9$ & $0.044$ & $5.23\pm0.01$ & $0.168\pm0.001$ & $0.214\pm0.001$  \\   
	   & $0.0$ & $4.0$ & $1.0$ & $0.049$ & $5.22\pm0.01$ & $0.163\pm0.001$ 	& $0.235\pm0.001$ \\
       & $0.0$ & $4.0$ & $1.1$ & $0.054$ & $5.21\pm0.01$ & $0.158\pm0.001$ & $0.256\pm0.001$ \\
            \hline            
               \multirow{11}{*}{($64,\,12,\,4$)}  & $0.0$ & $0.0$ & $0.0$ & $0.0$ & $8.03\pm0.05$ & -- & -- \\
	\multirow{12}{*}{Regime~II} & $0.0$ & $5.0$ & $0.33$ & $0.01$ & $7.78\pm0.07$ & $0.141\pm0.006$ & 	$0.051\pm0.004$ \\
\multirow{13}{*}{Regime~III}	& $0.0$ & $5.0$ & $0.5$ & $0.015$ & $7.78\pm0.06$ & $0.130\pm0.005$ & 	$0.079\pm0.004$ \\
	& $0.0$ & $5.0$ & $0.7$ & $0.021$ & $7.76\pm0.05$ & $0.129\pm0.003$ & 	$0.114\pm0.003$ \\
	& $0.0$ & $5.0$ & $1.0$ & $0.030$ & $7.59\pm0.04$ & $0.124\pm0.004$ & 	$0.160\pm0.003$ \\
	& $0.0$ & $5.0$ & $1.5$ & $0.045$ & $7.59\pm0.03$ & $0.107\pm0.003$ & 	$0.249\pm0.003$ \\
	& $0.0$ & $5.0$ & $2.0$ & $0.060$ & $7.51\pm0.03$ & $0.098\pm0.002$ & 	$0.318\pm0.003$ \\
	& $0.0$ & $5.0$ & $3.0$ & $0.090$ & $7.33\pm0.03$ & $0.084\pm0.002$ & 	$0.425\pm0.002$ \\
	& $0.0$ & $5.0$ & $4.0$ & $0.119$ & $7.21\pm0.03$ & $0.072\pm0.002$ & 	$0.500\pm0.002$ \\
	 & $0.0$ & $5.0$ & $4.8$ & $0.141$ & $7.08\pm0.02$ & $0.069\pm0.001$ & 	$0.538\pm0.001$ \\
       & $0.0$ & $5.0$ & $5.0$ & $0.148$ & $7.04\pm0.02$ & $0.068\pm0.001$ & $0.549\pm0.001$ \\
       & $0.0$ & $5.0$ & $5.5$ & $0.162$ & $7.00\pm0.02$ & $0.065\pm0.001$ & $0.567\pm0.001$ \\
       & $0.0$  & $5.0$ & $6.0$ & $0.177$ & $6.96\pm0.02$ & $0.059\pm0.001$ & $0.589\pm0.001$  \\   
         \hline      
	\multirow{4}{*}{($79,\,15,\,4$)}  & $0.0$ & $0.0$ & $0.0$ & $0.0$ & $9.11\pm0.06$ & -- & -- \\
	\multirow{5}{*}{Regime~III} & $0.0$ & $5.0$ & $5.0$ & $0.123$ & $7.96\pm0.03$ & $0.061\pm0.001$ & 	$0.484\pm0.002$ \\  
      & $0.0$ & $5.0$ & $5.5$ & $0.137$ & $7.91\pm0.03$ & $0.058\pm0.001$ & $0.510\pm0.001$ \\
      & $0.0$ & $5.0$ & $6.0$ & $0.149$ & $7.86\pm0.02$ & $0.056\pm0.001$ & $0.532\pm0.001$  \\   
      & $0.0$ & $5.0$ & $6.5$ & $0.162$ & $7.82\pm0.03$ & $0.054\pm0.001$ & $0.550\pm0.001$ \\
            \hline      
	\multirow{14}{*}{($34,\,6,\,4$)}   & $0.35$ & $0.35$ & $0.0$  & $0.0$ & $4.72\pm0.02$ &--& --\\
	\multirow{15}{*}{Sticky $\theta$  chain}  & $0.35$ & $3.6$ & $0.2$ & $0.016$ & $4.49\pm0.02$ & 		$0.266\pm0.003$ & $0.107\pm0.003$ \\
     & $0.35$ & $3.6$ & $0.3$ & $0.024$ & $4.52\pm0.02$ & $0.249\pm0.002$ & $0.146\pm0.003$ \\
     & $0.35$  & $3.6$ & $0.5$ & $0.039$ & $4.59\pm0.02$ & $0.217\pm0.003$ & $0.211\pm0.003$  \\  
	 & $0.35$ & $3.6$ & $0.7$ & $0.054$ & $4.62\pm0.02$ & $0.199\pm0.002$ & $0.259\pm0.003$ \\
	  & $0.35$ & $3.6$ & $0.9$ & $0.070$ & $4.62\pm0.01$ & $0.178\pm0.002$ & $0.310\pm0.002$ \\
     & $0.35$  & $3.6$ & $1.1$ & $0.085$ & $4.63\pm0.1$ & $0.162\pm0.002$ & $0.351\pm0.002$  \\   
     & $0.35$ & $3.6$ & $1.3$ & $0.100$ & $4.67\pm0.01$ & $0.148\pm0.001$ & $0.386\pm0.001$ \\
     & $0.35$ & $3.6$ & $1.5$ & $0.116$ & $4.70\pm0.01$ & $0.135\pm0.001$ & $0.419\pm0.001$ \\
     & $0.35$ & $3.6$ & $2.5$ & $0.193$ & $4.69\pm0.01$ & $0.098\pm0.001$ & $0.532\pm0.001$ \\
     & $0.35$  & $3.6$ & $3.5$ & $0.268$ & $4.66\pm0.01$ & $0.080\pm0.001$ & $0.596\pm0.001$  \\   
     & $0.35$ & $3.6$ & $4.0$ & $0.308$ & $4.64\pm0.02$ & $0.075\pm0.001$ & $0.620\pm0.001$ \\
     & $0.35$ & $3.6$ & $4.5$ & $0.348$ & $4.63\pm0.02$ & $0.069\pm0.001$ & $0.640\pm0.001$ \\
     & $0.35$ & $3.6$ & $5.0$ & $0.387$ & $4.63\pm0.01$ & $0.066\pm0.001$ & $0.656\pm0.001$ \\
     & $0.35$ & $3.6$ & $5.5$ & $0.425$ & $4.59\pm0.01$ & $0.064\pm0.001$ & $0.670\pm0.001$ \\   
\end{longtable*}
\end{center}
\endgroup  


\bibliography{Scalingref}